%% file: main.tex
\PassOptionsToPackage{names,dvipsnames}{xcolor}
\pdfoutput=1

\newif\ifdraft\draftfalse 
\newif\ifanon\anonfalse    
\newif\iffull\fullfalse   
\newif\iflongrefs\longrefsfalse 
\newif\ifbackref\backreffalse 
\newif\ifsooner\soonerfalse
\newif\iflater\laterfalse
\newif\iflncs\lncstrue
\newif\ifcamera\camerafalse 
\newif\ifallcites\allcitestrue
\newif\ifneedspace\needspacefalse
\newif\ifappendix\appendixfalse
\makeatletter \@input{texdirectives} \makeatother

\iflncs

\documentclass[runningheads,envcountsect,envcountsame]{llncs}

\usepackage{times}
\usepackage{DejaVuSansMono}
\usepackage{scalerel}
\usepackage{wrapfig}

\let\subparagraph\paragraph

\ifneedspace
\fi
\usepackage[compact]{titlesec}

\titlespacing*{\section}{0pt}{2ex plus .2ex minus .2ex}{1ex plus .2ex}
\titlespacing*{\subsection}{0pt}{2ex plus .2ex minus .2ex}{1ex plus .2ex}

\else 

\ifcamera
\documentclass[acmsmall]{acmart}
\else
\ifanon
\documentclass[acmsmall,review,anonymous,nonacm]{acmart}
\else
\documentclass[acmsmall,review=false,screen,nonacm]{acmart}
\fi
\fi

\acmJournal{PACMPL}
\acmYear{2018}
\acmMonth{1}
\acmVolume{2}
\acmNumber{POPL}
\acmPrice{}
\acmArticle{65}
\acmMonth{1}
\acmDOI{10.1145/3158153}
\ifcamera\else\copyrightyear{2017}\fi 

\ifanon
\setcopyright{none}             
\else
\setcopyright{rightsretained}   
\fi

\makeatletter
\def\@copyrightpermission{\ifcamera\\\\\\\fi This work is licensed under a \href{https://creativecommons.org/licenses/by/4.0/}{Creative Commons Attribution 4.0 International License}}
\makeatother

\ifcamera\else
\makeatletter
\def\@authorsaddresses{}
\fancypagestyle{firstpagestyle}{%
  \fancyhf{}%
  \renewcommand{\headrulewidth}{\z@}%
  \renewcommand{\footrulewidth}{\z@}%
    \fancyhead[L]{\ifanon\ACM@linecountL\fi}%
}
\fancypagestyle{standardpagestyle}{%
  \fancyhf{}%
  \renewcommand{\headrulewidth}{\z@}%
  \renewcommand{\footrulewidth}{\z@}%
    \fancyhead[LE]{\ifanon\ACM@linecountL\fi\@headfootfont\thepage}%
    \fancyhead[RO]{\@headfootfont\thepage}%
    \fancyhead[RE]{\@headfootfont\@shortauthors}%
    \fancyhead[LO]{\ifanon\ACM@linecountL\fi\@headfootfont\shorttitle}%
    \fancyfoot[RO,LE]{}%
}
\def\@mkbibcitation{}

\def\@titlefont{\sffamily\bfseries\fontsize{14}{17}\selectfont
}

\makeatother
\fi

\fi 

\usepackage{booktabs}   
\usepackage{subcaption} 

\usepackage[inference]{semantic}
\usepackage{amsmath}\allowdisplaybreaks
\usepackage{mathpartir}
\iflncs
\else
\usepackage{amsthm}
\fi
\usepackage{amssymb}
\usepackage{stmaryrd} 
\usepackage{xspace}
\usepackage{latexsym}
\usepackage{ifthen}
\usepackage{mathtools}
\usepackage{color}
\usepackage{listings}
\usepackage{verbatim}
\usepackage{tikz}
\usetikzlibrary{positioning,shadows,arrows,calc,backgrounds,fit,shapes,shapes.multipart,decorations.pathreplacing,shapes.misc,patterns,decorations.markings}
\tikzset{myiff/.style={double,implies-implies,,double equal sign distance}}
\tikzset{myimpl/.style={,double,-implies,double equal sign distance}}
\usepackage{tikzscale}
\usepackage{tikz-qtree}
\usepackage[T1]{fontenc}
\usepackage[scaled=.83]{beramono}
\usepackage{epigraph}
\usepackage{booktabs}
\usepackage{float}
\usepackage{etoolbox}
\usepackage{wrapfig}
\usepackage{scalerel}
\usepackage{nameref}
\usepackage[strict]{changepage}
\usepackage{coqed}
\usepackage{bm}
\usepackage{graphicx}
\usepackage[makeroom]{cancel}
\usepackage{centernot}
\usepackage{thm-restate}
\usepackage{enumitem}
\usepackage{fancyvrb}
\usepackage{varwidth}
\usepackage{bm}
\usepackage{combelow}
\usepackage{bussproofs}
\usepackage[normalem]{ulem} 
\usepackage{yhmath}
\usepackage[all]{xy}

\iflncs
\usepackage[numbers,sort]{natbib}
\fi

\newcommand\citepos[1]{\citeauthor{#1}'s\ \cite{#1}}

\definecolor{dkblue}{rgb}{0,0.1,0.5}
\definecolor{dkgreen}{rgb}{0,0.4,0}
\definecolor{dkred}{rgb}{0.6,0,0}
\definecolor{dkpurple}{rgb}{0.7,0,1.0}
\definecolor{purple}{rgb}{0.9,0,1.0}
\definecolor{olive}{rgb}{0.4, 0.4, 0.0}
\definecolor{teal}{rgb}{0.0,0.4,0.4}
\definecolor{azure}{rgb}{0.0, 0.5, 1.0}
\definecolor{gray}{rgb}{0.5, 0.5, 0.5}
\definecolor{dkgray}{rgb}{0.3, 0.3, 0.3}

\iflncs
\usepackage[
    pdftex,%
    pdfpagelabels,%
    colorlinks,%
    linkcolor=dkblue,%
    citecolor=dkblue,%
    filecolor=dkblue,%
    urlcolor=dkblue%
    \ifbackref,backref\fi%
]{hyperref}
\else
\usepackage{hyperref}
\hypersetup{
  breaklinks=true,
  citecolor=black,
  linkcolor=black,
  urlcolor=black,
}
\fi

\usepackage[capitalize]{cleveref}

\def\Snospace~{\S{}}

\def\Nnospace~{}

\usepackage{etoolbox}
\makeatletter
\patchcmd{\hyper@makecurrent}{%
    \ifx\Hy@param\Hy@chapterstring
        \let\Hy@param\Hy@chapapp
    \fi
}{%
    \iftoggle{inappendix}{
        \@checkappendixparam{chapter}%
        \@checkappendixparam{section}%
        \@checkappendixparam{subsection}%
        \@checkappendixparam{subsubsection}%
        \@checkappendixparam{paragraph}%
        \@checkappendixparam{subparagraph}%
    }{}%
}{}{\errmessage{failed to patch}}

\newcommand*{\@checkappendixparam}[1]{%
    \def\@checkappendixparamtmp{#1}%
    \ifx\Hy@param\@checkappendixparamtmp
        \let\Hy@param\Hy@appendixstring
    \fi
}
\makeatletter

\newtoggle{inappendix}
\togglefalse{inappendix}

\apptocmd{\appendix}{\toggletrue{inappendix}}{}{\errmessage{failed to patch}}


\AtBeginEnvironment{tikzpicture}{\catcode`$=3 } 

\input{./cmds}
\renewcommand{\comptd}[1]{\ensuremath{\bl{\left.\src{#1}\right\downarrow}}}

\newcommand{\comm}[3]{\ifdraft{{\color{#1}[#2: #3]}}\fi}
\newcommand{\marco}[1]{\comm{brown}{Marco}{#1}}

\newcommand{\ch}[1]{\comm{teal}{CH}{#1}}
\newcommand{\ca}[1]{\comm{olive}{CA}{#1}}

\newcommand{\jt}[1]{\comm{violet}{JT}{#1}}
\newcommand{\rb}[1]{\comm{orange}{RB}{#1}}
\newcommand{\et}[1]{\comm{dkpurple}{ET}{#1}}

\renewcommand{\cmp}[1]{\comptd{#1}} 

\iflncs
\else
\fi
\iflncs\else\fi

\newcommand{\mytitle}{Trace-Relating Compiler Correctness and Secure Compilation}

\iflncs
\makeatletter
\AtBeginDocument{
  \hypersetup{
    pdftitle = {\mytitle},
  }
}
\makeatother

\title{\LARGE
  Trace-Relating Compiler Correctness\\ and Secure Compilation
\iflncs\vspace{-0.8em}\else\ifanon\vspace{-2em}\fi\fi}
\titlerunning{\mytitle}
\else
\title{
\mytitle}
\fi

\ifanon
\author{}
\institute{}
\else

\iflncs
\author{\normalsize
       Carmine Abate$^1$ \quad
       Roberto Blanco$^1$ \quad
       \cb{S}tefan Ciob\^{a}c\u{a}$^2$ \quad
       Adrien Durier$^1$\\
       Deepak Garg$^3$\;
       C\u{a}t\u{a}lin Hri\cb{t}cu$^1$\;
       Marco Patrignani$^{4,5}$\;
       \'{E}ric Tanter$^{6,1}$\;
       J\'er\'emy Thibault$^1$\vspace{-0.3em}
}
\institute{\scriptsize
$^1$Inria Paris, France \;
$^2$UAIC Ia\c{s}i, Romania \;
$^3$MPI-SWS, Saarbr\"ucken, Germany \;
$^4$Stanford University, Stanford, USA \\
$^5$CISPA, Saarbr\"ucken, Germany \;
$^6$University of Chile, Santiago, Chile\vspace{-2.6em}
}

\authorrunning{C. Abate \ETAL}

\else 

\author{Carmine Abate}
\affiliation{
  \ifcamera\institution{Inria}\city{Paris}\country{France}
  \else\institution{Inria Paris}\fi}
\email{carmine.abate@inria.fr}

\author{Roberto Blanco}
\affiliation{
  \ifcamera\institution{Inria}\city{Paris}\country{France}
  \else\institution{Inria Paris}\fi}
\email{roberto.blanco@inria.fr}

\author{\cb{S}tefan Ciob\^{a}c\u{a}}
\affiliation{
  \ifcamera
    \institution{Alexandru Ioan Cuza University}\city{Ia\c{s}i}\country{Romania}
  \else\institution{Alexandru Ioan Cuza University, Ia\cb{s}i}\fi}
\email{stefan.ciobaca@info.uaic.ro}

\author{Deepak Garg}
\affiliation{
  \institution{MPI-SWS}
  \ifcamera\city{Saarbr\"ucken}\country{Germany}\fi}
\email{dg@mpi-sws.org}

\author{C\u{a}t\u{a}lin Hri\cb{t}cu}
\affiliation{
  \ifcamera\institution{Inria}\city{Paris}\country{France}
  \else\institution{Inria Paris}\fi}
\email{catalin.hritcu@gmail.com}

\author{Marco Patrignani}
\orcid{0000-0003-3411-9678}
\affiliation{
  \institution{Stanford University}
  \ifcamera\city{Palo Alto}\country{USA}\fi}
\affiliation{
  \institution{CISPA Helmholz Center for Information Security}
  \ifcamera\city{Saarbr\"ucken}\country{Germany}\fi}
\email{mp@cs.stanford.edu}

\author{\'{E}ric Tanter}
\affiliation{
  \institution{University of Chile}
  \ifcamera\city{Santiago}\country{Chile}\fi}
\affiliation{
  \ifcamera\institution{Inria}\city{Paris}\country{France}
  \else\institution{Inria Paris}\fi}
\email{etanter@dcc.uchile.cl}

\author{J\'er\'emy Thibault}
\affiliation{
  \ifcamera\institution{Inria}\city{Paris}\country{France}
  \else\institution{Inria Paris}\fi}
\email{jeremy.thibault@inria.fr}

\author{}

\makeatletter
\renewcommand{\@shortauthors}{C. Abate, R. Blanco, \cb{S}. Ciob\^{a}c\u{a}, A. Durier, D. Garg, C. Hri\cb{t}cu, M. Patrignani, \'{E}. Tanter, and J. Thibault}
\makeatother

\fi 

\fi 

\numberwithin{theorem}{section} 
\begin{document}
\setlength{\abovedisplayskip}{2pt}
\setlength{\belowdisplayskip}{1pt}
\setlength{\abovedisplayshortskip}{2pt}
\setlength{\belowdisplayshortskip}{1pt}

\makeatletter
\def\thm@space@setup{\thm@preskip=0pt \thm@postskip=0pt}
\def\def@space@setup{\def@preskip=0pt \def@postskip=0pt}
\def\lem@space@setup{\lem@preskip=0pt \lem@postskip=0pt}
\makeatother


\iflncs\maketitle\fi

\begin{abstract}
\input{abstract.tex}
\end{abstract}

\iflncs\else\maketitle\fi

\ifcamera
\fi

\input{intro}

\section{Trace-Relating Compiler Correctness}
\label{sec:compiler-correctness}




In this section, we start by generalizing the trace property preservation
definitions at the end of the introduction to $\tpsigma$ and
$\tptau$, which depend on two {\em arbitrary} mappings
$\sigma$ and $\tau$ (\autoref{sec:correct-trace-props}).
We prove that, whenever $\sigma$ and $\tau$ form a Galois connection,
$\tpsigma$ and $\tptau$ are equivalent
(\autoref{thm:galois}).
%
%
%
%
%
%
%
%
%
We then exploit a bijective correspondence between trace relations and Galois
connections to close the trinitarian view (\autoref{sec:trinity}),
with two main benefits:
first, it helps us assess the meaningfulness of a given trace relation
by looking at the property mappings it induces; second, it allows
us to construct new compiler correctness definitions starting from a
desired mapping of properties.
Finally, we generalize the classic result that compiler correctness (\IE
\ccequal) is enough to preserve not just trace properties but also all
subset-closed hyperproperties~\cite{ClarksonS10}.
For this, we show that \cctilde is also equivalent to subset-closed
hyperproperty preservation, for which we also define both a version in
terms of $\tilde{\sigma}$ and a version in terms of $\tilde{\tau}$
(\autoref{sec:subset-closed}).

\subsection{Property Mappings}
\label{sec:correct-trace-props}
As explained in \autoref{sec:intro}, trace-relating compiler correctness
\cctilde, by itself, lacks a crisp description of which trace properties
are preserved by compilation.
Since even the syntax of traces can differ between source and target, one can
either look at trace properties of the source (but then one needs to interpret
them in the target), or at trace properties of the target (but then one needs to
interpret them in the source).
Formally we need two property mappings, $\tau : \propS \to \propT$ and
$~\sigma : \propT \to \propS$, which lead us to the following
generalization of trace property preservation ($\tp$).
%
%
\begin{definition}[\tpsigma and \tptau] \label{defn:TP}
 Given two property mappings, $\tau : \propS \rightarrow 2^\trg{Trace_T}$ and
  $\sigma : \propT \rightarrow\propS$, for a compilation chain $\cmp{\cdot}$ we define:
{\small
\[
    \tptau \equiv
    \forall \piS{\pi}. ~\forall \src{W}. ~ \src{W} ~\satS~ \piS{\pi} \Rightarrow 
                                                      \cmp{W} ~\satT~ \tau(\piS{\pi}); 
\;\;\;
    \tpsigma \equiv
    \forall \piT{\pi}. ~\forall\src{W}.~ \src{W} ~\satS~ \sigma(\piT{\pi}) \Rightarrow 
                                                   \cmp{W} ~\satT~ \piT{\pi}.
\]
}
\end{definition}
For an arbitrary source program $\src{W}$, $\tau$ interprets a source
property $\piS{\pi}$ as the \emph{target guarantee} for $\cmp{W}$.
Dually, $\sigma$ defines a \emph{source obligation} sufficient for the
satisfaction of a target property $\piT{\pi}$ after
compilation. Ideally:
\begin{itemize}[nosep]
\item Given $\piT{\pi}$, the target interpretation of
  the source obligation $\sigma(\piT{\pi})$ should actually guarantee
  that $\piT{\pi}$ holds,
  \IE~$\tau(\sigma(\piT{\pi})) \subseteq \piT{\pi}$;
\item Dually for $\piS{\pi}$, we would not want the source obligation for
  $\tau(\piS{\pi})$ to be harder than $\piS{\pi}$ itself,
  \IE~$\sigma(\tau(\piS{\pi})) \supseteq \piS{\pi}$.
\end{itemize}
These requirements are satisfied when the two maps form a {\em Galois
  connection} between the posets of source and target properties
ordered by inclusion. 
We briefly recall the definition and the characteristic property of
Galois connections~\cite{CousotCousot77-1, Melton:1986}.
 
\begin{definition}[Galois connection] Let $(X, \preceq)$ and
  $(Y, \sqsubseteq)$ be two posets.  A pair of maps,
  $\alpha: X \to Y$, $\gamma: Y \to X$ is a Galois connection \ii{iff}
  it satisfies the \emph{adjunction law}:
  $\forall x \in X. ~ \forall y \in Y. ~ \alpha(x) \sqsubseteq y \iff
  x \preceq \gamma(y)$.
  $\alpha$ (resp. $\gamma$)
  is the lower (upper) adjoint or
  abstraction (concretization) function and $Y$ ($X$)
  the abstract (concrete) domain.
\end{definition}
We will often write
$\alpha: (X, \preceq) \leftrightarrows (Y, \sqsubseteq) :\gamma$ to
denote a Galois connection, or simply
$\alpha: X \leftrightarrows Y :\gamma$, or even
$\alpha \leftrightarrows \gamma$ when the involved posets are clear
from context.
\begin{lemma}[Characteristic property of Galois connections] \label{lem:chG}
  If $\alpha{:} (X, \preceq) \leftrightarrows {(Y, \sqsubseteq) {:}\gamma}$ is a Galois connection, then $\alpha, \gamma$ are monotone and
  they satisfy these properties:
  \begin{align*}
   i)& \quad \forall x \in X.  ~ x \preceq \gamma(\alpha (x)); 
   &
   ii)& \quad \forall y \in Y. ~ \alpha(\gamma (y)) \sqsubseteq y.
  \end{align*}
If $X,Y$ are complete lattices, then $\alpha$ is continuous,
\IE~$\forall F \subseteq X.  ~\alpha(\bigsqcup F) = \bigsqcup \alpha(F)$.
\end{lemma}
%
If two property mappings, $\tau$ and $\sigma$, form a Galois connection on trace
properties ordered by set inclusion, \autoref{lem:chG} (with
$\alpha = \tau$ and $\gamma = \sigma$) tells us that they
satisfy the ideal conditions we discussed above, ~\IE
$\tau(\sigma(\piT{\pi})) \subseteq \piT{\pi}$ and
$\sigma(\tau(\piS{\pi})) \supseteq \piS{\pi}$.\footnote{While
  target traces are often \emph{``more concrete''} than source ones,
  trace properties $\prop$ (which in Coq we represent as
  the function type $\text{Trace} {\to} \text{Prop}$)
  are contravariant in $\text{Trace}$ and thus
  target properties correspond to the \emph{abstract domain}.
}

The two ideal conditions on $\tau$ and $\sigma$ are sufficient to
show the equivalence of the criteria they define, respectively
$\tptau$ and $\tpsigma$.

\begin{theorem}[\tptau and \tpsigma coincide \coqhref{Def.v}]
\label{thm:galois}
Let $\tau : 2^\src{Trace_S} \rightleftarrows \propT : \sigma$ be a
Galois connection, with $\tau$ and $\sigma$ the lower and upper
adjoints (resp.).  Then $ \tptau \iff \tpsigma $.
\end{theorem}

\subsection{Trace Relations and Property Mappings}
\label{sec:trinity}
We now investigate the relation between $\cctilde$,
\tptau and \tpsigma.
%
%
We show that for a trace relation and its corresponding Galois
connection (\autoref{lem:correspond}), the three criteria are
equivalent (\autoref{thm:correspondcriteria}).
This equivalence offers interesting insights for both verification and
design of a correct compiler. For a $\cctilde$ compiler, the
equivalence makes explicit both the guarantees one has after
compilation ($\tilde{\tau}$) and source proof obligations to ensure
the satisfaction of a given target property ($\tilde{\sigma}$).  On
the other hand, a compiler designer might first determine the target
guarantees the compiler itself must provide, ~\IE $\tau$, and then
prove an equivalent statement, $\cctilde$, for which more convenient
proof techniques exist in the literature~\cite{BeringerSDA14,TanMKFON19}.

\begin{definition}[Existential and Universal Image \cite{gardiner1994algebraic}]\label{defn:eximage}
Given any two sets $X$ and $Y$ and a relation ${\sim} \subseteq A \times B$, 
define its existential or direct image, $\tilde{\tau}: 2^X \to 2^Y$ and
its universal image, $\tilde{\sigma}: 2^Y \to 2^X$ as follows:
$$
\tilde{\tau} = \lambda ~\pi \in 2^X. ~ \myset{y}{\exists x \ldotp x \sim y \wedge x \in \pi};
\tilde{\sigma} = \lambda ~\pi \in 2^Y. ~ \myset{x}{\forall y \ldotp x \sim y \Rightarrow y \in \pi}.
$$
\end{definition}
When trace relations are considered, the existential and
universal images can be used to instantiate \autoref{defn:TP} leading
to the trinitarian view already mentioned in \autoref{sec:intro}.

\begin{theorem}[Trinitarian View \coqhref{TraceCriterion.v}] \label{thm:trinitarianview}
  For any trace relation $\sim$ and its existential and universal images
  $\tilde{\tau}$ and $\tilde{\sigma}$, we have:
  \(\tptautilde \iff \cctilde \iff \tpsigmatilde\).
\end{theorem}
This result relies both on \autoref{thm:galois} and on the fact that
the existential and universal images of a trace relation form a Galois
connection (\coqhref{Galois.v}). Below we further generalize this result
(\autoref{thm:correspondcriteria}) relying on a bijective
correspondence between trace relations and Galois connections on
properties.
\begin{lemma}[Trace relations $\cong$ Galois connections on trace properties]
 \label{lem:correspond}
%
%
 The function
 $\sim \;\mapsto \tilde{\tau} \leftrightarrows \tilde{\sigma}$ that
 maps a trace relation to its existential and universal images is a
 bijection between trace relations
 $2^{\src{Trace_S} \times \trg{Trace_T}}$ and Galois connections on
 trace properties $\propS \leftrightarrows \propT$. 
 Its inverse is $\tau \leftrightarrows \sigma \mapsto \hat{\sim}$,
 where
 $\tS{s} \,\hat{\sim}\, \tT{t} \equiv \tT{t} \in \tau(\{ \tS{s} \})$.
\end{lemma}
\begin{proof}
  \citet{gardiner1994algebraic} show that the existential image is a
  functor from the category of sets and relations to the category of
  predicate transformers, mapping a set $X \mapsto 2^X$ and a relation
  $\sim \;\subseteq X \times Y \mapsto \tilde{\tau} : 2^X \to 2^Y$.
  They also show that such a functor is an isomorphism -- hence
  bijective -- when one considers only monotonic predicate
  transformers that 
  have a -- unique -- upper adjoint.
  The universal image of $\sim$, $\tilde{\sigma}$, is the unique
  adjoint of $\tilde{\tau}$ (\coqhref{Galois.v}), hence
  $\sim \;\mapsto \tilde{\tau} \leftrightarrows \tilde{\sigma}$ is
  itself bijective.  \qed
\end{proof}
The bijection just introduced allows us to generalize
\autoref{thm:trinitarianview} and switch between the three views of
compiler correctness described earlier at will.

\begin{theorem}[Correspondence of Criteria] \label{thm:correspondcriteria} 
  For any trace relation $\sim$ and corresponding Galois connection
  $\tau \leftrightarrows \sigma$, we have:
  \(\tptau \iff \cctilde \iff \tpsigma\).
\end{theorem}
\begin{proof}

  For a trace relation $\sim$ and the Galois connection
  $\tilde{\tau} \leftrightarrows \tilde{\sigma}$, the result follows
  from \autoref{thm:trinitarianview}.
  For a Galois connection $\tau \leftrightarrows \sigma$ and
  $\hat{\sim}$, use \autoref{lem:correspond} to conclude that the
  existential and universal images of $\hat{\sim}$ coincide with $\tau$
  and $\sigma$, respectively; the goal then follows from \autoref{thm:trinitarianview}.
  %
  \qed
\end{proof} 
%
%
We conclude by explicitly noting that sometimes the lifted properties may be
trivial: the target guarantee can be the true property (the set of all traces),
or the source obligation the false property (the empty set of traces).
This might be the case when source observations abstract away too much
information (\autoref{sec:example-resources} presents an example).
\subsection{Preservation of Subset-Closed Hyperproperties}\label{sec:subset-closed}
%

A $\ccequal$ compiler ensures the preservation not only of trace
properties, but also of all subset-closed hyperproperties, which are known
to be preserved by refinement~\cite{ClarksonS10}.
An example of a subset-closed hyperproperty is {\em noninterference}
\cite{ClarksonS10}; a $\ccequal$ compiler thus guarantees that if $\src{W}$ is
noninterfering with respect to the inputs and outputs in the trace then so is
$\cmp{W}$.
To be able to talk about how (hyper)properties such as noninterference
are preserved, in this section we propose another trinitarian view involving
$\cctilde$ and preservation of subset-closed hyperproperties
(\autoref{thm:ssch}),
slightly weakened in that source and target property mappings will need to be
closed under subsets.

First, recall that a program satisfies a hyperproperty when its complete
set of traces, which from now on we will call its {\em behavior}, is a member of
the hyperproperty~\cite{ClarksonS10}.
\begin{definition}[Hyperproperty Satisfaction]
  A program $W$ satisfies a hyperproperty $H$, written
  $W \models H$,{}
  iff $\ii{beh}(W) \in H$, where
  $\ii{beh}(W) = \myset{t}{W {\rightsquigarrow} t}$.
\end{definition}
Hyperproperty preservation is a strong requirement in general.
Fortunately, many interesting hyperproperties are {\em subset-closed} ($\ii{SCH}$
for short), which simplifies their preservation 
%
%
since it suffices to show that the behaviors of the compiled program refine the
behaviors of the source one, which coincides with the statement of $\ccequal$.

To talk about hyperproperty preservation in the trace-relating setting, we need
an interpretation of source hyperproperties into the target and vice versa.
The one we consider builds on top of the two trace property mappings $\tau$ and
$\sigma$, which are naturally lifted to hyperproperty mappings.
This way we are able to extract two hyperproperty mappings from a trace relation
similarly to \autoref{sec:trinity}:

\begin{definition}[Lifting property mappings to hyperproperty mappings] 
  Let $\tau: \propS \to \propT$ and $\sigma: \propT \to \propS$ be arbitrary property mappings.          
  The images of $\hS{H} \in 2^{\propS}, \hT{H} \in 2^{\propT}$ under $\tau$ and $\sigma$ are, respectively:
  \begin{align*}
    \tau(\hS{H}) & = \myset{\tau(\piS{\pi})}{\piS{\pi} \in \hS{H}}; 
    &
    \sigma(\hT{H}) & = \myset{\sigma(\piT{\pi})}{\piT{\pi} \in \hT{H}}.
  \end{align*}
\end{definition}
Formally we are defining two new mappings, this time on
hyperproperties, but by a small abuse of notation we still denote them
by $\tau$ and $\sigma$.

Interestingly, it is not possible to apply the argument used for $\ccequal$
to show that a $\cctilde$ compiler guarantees
$\cmp{W} ~\satT~ \tilde{\tau}(\hS{H})$ whenever $\src{W} ~\satS~ \hS{H}$.
This is in fact not true because direct images do not necessarily
preserve subset-closure  \cite{mastroeni2018verifying, naumann2019whither}. 
%
To fix this we close the image of $\tilde{\tau}$
and $\tilde{\sigma}$ under subsets (denoted as $\mi{Cl_\subseteq}$)
and obtain:
\begin{theorem}[Preservation of Subset-Closed Hyperproperties
  \coqhref{SSCHCriterion.v}] \label{thm:ssch}
  For any trace relation
  $\sim$ and its existential and
  universal images lifted to hyperproperties, $\tilde{\tau}$ and $\tilde{\sigma}$, and for
  $\ii{Cl}_{\subseteq} (H) = \{\pi \mid \exists \pi' \in H. ~ \pi
  \subseteq \pi'\}$,
  we have: \[\schptautilde \iff \cctilde \iff \schpsigmatilde, \text{ where}\]
  \begin{align*}
    \schptautilde & \equiv
      \forall \src{W} \forall \hS{H} \in \hS{SCH}.
     \src{W} ~\satS~ \hS{H} \Rightarrow \cmp{W} ~\satT~ \ii{Cl}_{\subseteq}(\tilde{\tau}(\hS{H}));\\
    \schpsigmatilde & \equiv
      \forall \src{W} \forall \hT{H} \in \hT{SCH}.
     \src{W} ~\satS~ \ii{Cl}_{\subseteq}(\tilde{\sigma}(\hT{H})) \Rightarrow \cmp{W} ~\satT~ \hT{H} .
  \end{align*}
\end{theorem}
\autoref{thm:ssch} makes us aware of the potential loss of precision when
interested in preserving subset-closed hyperproperties through compilation.
In \autoref{sec:example-noninterference} we focus on a security
relevant subset-closed hyperproperty, noninterference, and show that
such a loss of precision can be intended as a declassification of
noninterference.  

\section{Instances of Trace-Relating Compiler Correctness}
\label{sec:instances}

The trace-relating view of compiler correctness above can serve as a unifying
framework for studying a range of interesting compilers.
This section provides several representative instantiations of the framework:
source languages with undefined behavior that compilation can turn into
arbitrary target behavior (\autoref{sec:example-undef}),
target languages with 
resource exhaustion that cannot
happen in the source (\autoref{sec:example-resources}),
changes in the representation of values (\autoref{sec:example-diff-values}),
and differences in the granularity of data and observable events
(\autoref{sec:instance-marco}).

\subsection{Undefined Behavior}
\label{sec:example-undef}
\input{undefbeh}

\subsection{Resource Exhaustion}
\label{sec:example-resources}
\input{resource-exhaustion}

\subsection{Different Source and Target Values}
\label{sec:example-diff-values}



\input{diff_values}

\subsection{Abstraction Mismatches} \label{sec:example-split-io}
\label{sec:instance-marco}\label{SEC:M}


\input{instance-more-target-statements}

\section{Trace-Relating Compilation and Noninterference Preservation}
\label{sec:example-noninterference}
\input{ANI}

\section{Trace-Relating Secure Compilation}
\label{sec:secure-compilation}
\input{secure-compilation-motivation}

\subsection{Trace-Relating Secure Compilation: A Spectrum of Trinities}\label{sec:sec-comp-trini}
\input{sec-comp-explain}
\input{expand-sec-diagram}
\subsection{Instance of Trace-Relating Robust Preservation of Trace Properties}\label{sec:sec-comp-traces}
\input{robust-trace}

\subsection{Instances of Trace-Relating Robust Preservation of Safety and Hypersafety}\label{sec:comp-summ}
\input{rw-sec-crits-short}

\section{Related Work}
\label{sec:related}

\ca{
\begin{itemize}
\item TODO: if $\sim$ is a partial function then the subset closure is
            preserved \cite{naumann2019whither} 
            (future work may be a sufficient condition for (Hyper)Safety) 
\item Given \autoref{lem:correspond} a logical relation may
      correspond to a Galois connection between types (see also \cite{backhouse2004safety}).
      What does this mean in terms of compiler correctness?
\end{itemize}
}

We already discussed how our results relate to some existing work in
correct compilation~\cite{Leroy09,TanMKFON19} and secure
compilation~\cite{AbateBGHPT19,PatrignaniG17,PatrignaniG18}. 
We also already mentioned that most of our definitions and results
make no assumptions about the structure of traces.
One result that relies on the structure of traces is
\autoref{thm:rsp-sec-trinity}, which involves some \emph{finite prefix} $m$,
suggesting traces should be some sort of sequences of events (or states),
as customary when one wants to refer to safety
properties~\cite{ClarksonS10}. It is however sufficient to fix
a topology on properties where safety properties coincide with
closed sets \cite{PasquaM17}.
Even for reasoning about safety, hypersafety, or arbitrary
hyperproperties, traces can therefore be values, sequences of program
states, or of input output events, or even the recently proposed
\emph{interaction trees}~\cite{XiaZHHMPZ20}. In the latter case
we believe that the compilation from IMP to ASM proposed by
\citet{XiaZHHMPZ20} can be seen as an instance of \hctilde, for
the relation they call ``trace equivalence.''

%
%
%
%
\myparagraph{Compilers Where Our Work Could Be Useful}
%
%
%
Our work should be broadly applicable to understanding the guarantees provided
by many verified compilers.
For instance, \citet{WangWS19} recently proposed a CompCert variant that
compiles all the way down to machine code, and it would be interesting to see if 
the model at the end of \autoref{sec:example-undef} applies there too.
This and many other verified compilers~\cite{SevcikVNJS13, KangHMGZV15,
  CarbonneauxHRS14, MullenZTG16} beyond CakeML~\cite{TanMKFON19} deal with
resource exhaustion and it would be interesting to also apply the ideas of
\autoref{sec:example-resources} to them.
%
%
%
\citet{HurD11} devised a correct compiler from an ML language to assembly using
a cross-language logical relation to state their \ccbasic theorem.
They do not have traces, though were one to add them, the logical relation on
values would serve as the basis for the trace relation and therefore their
result would attain \cctilde.\ch{Theirs is a {\em compositional} compiler
  result though.}

Switching to more informative traces capturing the interaction between the
program and the context is often used as a proof technique for secure
compilation~\cite{JeffreyR05, PatrignaniC15, AbateBGHPT19}.
Most of these results consider a cross-language relation, so they probably could
be proved to attain one of the criteria from
\autoref{fig:diagram-preserve-props-sec}.


\myparagraph{Generalizations of Compiler Correctness}
The compiler correctness definition of \citet{Morris73a} was already general
enough to account for trace relations, since it considered a translation between
the semantics of the source program and that of the compiled program, which
he called ``decode'' in his diagram, reproduced in \autoref{fig:morris} (left).
And even some of the more recent compiler correctness definitions preserve this
kind of flexibility \cite{PattersonA19}.
While \cctilde can be seen as an instance of
a definition by \citet{Morris73a}, we are not aware of any prior work
that investigated the preservation of properties
when the ``decode translation'' is neither the identity nor a bijection,
and source properties need to be re-interpreted as target ones and vice versa.

\begin{figure}[!t]
  \begin{center}
	  \begin{minipage}{0.6\textwidth}
		  \begin{tikzpicture}[scale = 0.70,every node/.style={align=center,font=\small}]
		    \node  (W) at (0, 2)    {source language};
		    \node  (Wcmp) at (0, 0) {target language};
		    \node  (Z) at (8, 0)    {target meanings};
		    \node  (Zdec) at (8, 2) {source meanings};
		    \draw [->] (W) -- (Wcmp)  node[midway,left,fill = white] {compile};
		    \draw [->] (Wcmp) -- (Z)  node[midway,above] {target semantics};
		    \draw [->] (Z) -- (Zdec)  node[midway,right,fill = white] {decode};
		    \draw [->] (W) -- (Zdec)  node[midway,above] {source semantics};
		  \end{tikzpicture}
	  \end{minipage}
	  \hspace{3em}
	  \vrule
	  \hspace{1em}
	  \begin{minipage}{0.2\textwidth}
	  	\begin{tikzpicture}[scale = 0.70]
	    \node  (W) at (0, 1.2)    {\src{W}};
	    \node  (Wcmp) at (0, 0) {\cmp{W}};
	    \node  (Z) at (2, 0)    {\trg{Z}};
	    \node  (Zdec) at (2, 1.2) {$\trg{Z}^{\#}$};
	    \draw [->] (W) to (Wcmp);
	    \draw [->>] (Wcmp) -- (Z) node[midway,above] {\tiny{\trg{T}}};
	    \draw [->] (Z) to (Zdec);
	    \draw [->>] (W) -- (Zdec) node[midway,above] {\tiny{\src{S}}} ;
	  \end{tikzpicture}
	  \end{minipage}
  \end{center}
    \caption{\citepos{Morris73a} (left) and \citepos{Melton:1986} and \citepos{sabry1997reflection} (right) \vspace{1em}}\label{fig:morris}
\end{figure}

\myparagraph{Correct Compilation and Galois Connections}
\citet{Melton:1986} and \citet{sabry1997reflection} expressed a strong
variant of compiler correctness using the diagram of \autoref{fig:morris} (right)
\cite{sabry1997reflection, Melton:1986}.
%
%
They require that compiled programs \emph{parallel} the computation
steps of the original source programs, which can be proven showing
the existence of a \emph{decompilation} map $\#$ that makes the
diagram commute, or equivalently, the existence of an adjoint for
$\downarrow$ ~($W \leq W' \iff W \twoheadrightarrow W'$ for both source and target). 
The ``parallel'' intuition can be formalized as an instance of \cctilde.
Take source and target traces to be finite or infinite sequences of
program states (maximal trace semantics
\cite{cousot2002constructive}), and relate them exactly like
\citet{Melton:1986} and \citet{sabry1997reflection}\ifappendix; more details can
be found in \autoref{sec:relatedwork}\fi.
\ca{I think they do the following assumption in order to have a proof
  (\citet{Melton:1986}) or because of their goals
  (\citet{sabry1997reflection})}
\ca{ While \citet{Melton:1986} requires $\downarrow$ to be injective
  (insertion), for their goals \citet{sabry1997reflection} assume a
  reflection, so that target expressions can be embedded in the source
  language.  }

\myparagraph{Translation Validation}
Translation validation is an important alternative to proving that all runs of a
compiler are correct.
A variant of \cctilde for translation validation can simply be obtained by
specializing the definition to a particular $\src{W}$, and one can obtain again
the same trinitarian view.
Similarly for our other criteria, including our extensions of the secure
compilation criteria of \citet{AbateBGHPT19}, which \citet{BusiDG19} seem to
already be considering in the context of translation validation.

\section{Conclusion and Future Work}
\label{sec:conclusion}
We have extended the property preservation view on compiler correctness to arbitrary
trace relations, and believe that this will be useful for understanding 
the guarantees various compilers provide.
An open question is whether, given a compiler, there exists a most precise
$\sim$ relation for which this compiler is correct.
As mentioned in \autoref{sec:intro}, every compiler is $\cctilde$ for some
$\sim$, but under which conditions is there a most precise relation?
In practice, more precision may not always be better though, as it may be at
odds with compiler efficiency and may not align with more subjective notions of
usefulness, leading to tradeoffs in the selection of suitable relations.
Finally, another interesting direction for future work is studying whether using
the relation to Galois connections allows to more easily compose trace
relations for different purposes, say, for a compiler whose target language has
undefined behavior, resource exhaustion, and side-channels.
In particular, are there ways to obtain complex relations by combining
simpler ones in a way that eases the compiler verification burden?




{

\iflncs
\medskip
\myparagraph{Acknowledgements}
\else
\begin{acks}
\fi
We thank Akram El-Korashy and Amin Timany for participating in an early discussion
about this work
%
and the anonymous reviewers
for their valuable feedback.
%
This work was in part supported by the
\iflncs
\href{https://erc.europa.eu}{European Research Council}
\else
\grantsponsor{1}{European Research Council}{https://erc.europa.eu/}
\fi
under \href{https://secure-compilation.github.io/}{ERC Starting Grant SECOMP}
(\iflncs 715753\else\grantnum{1}{715753}\fi),
by the German Federal Ministry of
Education and Research (BMBF) through funding for the CISPA-Stanford
Center for Cybersecurity (FKZ: 13N1S0762),
by DARPA grant SSITH/HOPE (FA8650-15-C-7558) and by UAIC internal grant 07/2018.
\iflncs\else
\end{acks}
\fi
}

\ifappendix
\appendix

\input{appendixintro}

\input{proofindex}

\input{appendix-other}
\MP{
	commented here is the forward simulation stuff.
}
\MP{
	stuff from the next file should go in the RW, then comment the input
}
\input{other_works}
\input{app-proofs-marco-statements}

\input{composition-appendix}
\section{In-Depth Trace-Relating Secure Compilation}\label{sec:in-depth-trsec}
We expand \autoref{sec:comp-summ} and delve more in-depth on how to interpret related work as instance of our framework.

\subsection{Trace-Relating Robust Preservation of Safety Properties}\label{sec:sec-comp-safe}
\input{rob-pres-safety}

\subsection{Trace-Relating Robust Preservation of Hypersafety Properties}\label{sec:sec-comp-hypersafe}
\input{instance-marcodeepak}

\fi

\iflncs
\bibliographystyle{abbrvnaturl}
\footnotesize
\else 

\ifcamera
\bibliographystyle{ACM-Reference-Format}
\citestyle{acmauthoryear}   
\else
\bibliographystyle{abbrvnaturl}
\fi

\fi 

\bibliography{mp,safe}


\vfill

{\small\medskip\noindent{\bf Open Access} This chapter is licensed under the terms of the Creative Commons\break Attribution 4.0 International License (\url{http://creativecommons.org/licenses/by/4.0/}), which permits use, sharing, adaptation, distribution and reproduction in any medium or format, as long as you give appropriate credit to the original author(s) and the source, provide a link to the Creative Commons license and indicate if changes were made.}

{\small \spaceskip .28em plus .1em minus .1em The images or other third party material in this chapter are included in the chapter's Creative Commons license, unless indicated otherwise in a credit line to the material.~If material is not included in the chapter's Creative Commons license and your intended\break use is not permitted by statutory regulation or exceeds the permitted use, you will need to obtain permission directly from the copyright holder.}

\medskip\noindent\includegraphics{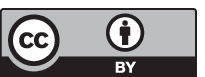}

\end{document}

%% file: cmds.tex
\newcommand{\MP}[1]{\marco{#1}}

\newcommand*{\EG}{e.g.,\xspace}
\newcommand*{\IE}{i.e.,\xspace}
\newcommand*{\ETAL}{et al.\xspace}
\newcommand*{\ETC}{etc.\xspace}

\newcommand{\mi}[1]{\ensuremath{\mathit{#1}}}
\newcommand{\ii}[1]{\mi{#1}}
\newcommand{\mr}[1]{\ensuremath{\mathrm{#1}}}

\newcommand{\mtt}[1]{\ensuremath{\mathtt{#1}}}
\newcommand{\mf}[1]{\ensuremath{\mathbf{#1}}}

\newcommand{\ms}[1]{\ensuremath{\mathsf{#1}}}
\newcommand{\mb}[1]{\ensuremath{\mathbb{#1}}}

\newcommand{\R}{\mathcal{R}}
\newcommand{\RR}{\mathrel{\R}}
\newcommand{\WT}{\trg{W_T}}
\newcommand{\WS}{\src{W_S}}

\newcommand{\relmiddle}[1]{\mathrel{}\middle#1\mathrel{}}

\newcommand\bnfdef{\ensuremath{\mathrel{::=}}}

\newcommand{\OB}[1]{\ensuremath{\overline{#1}}}

\newcommand{\myset}[2]{\ensuremath{\left\{#1 \,\relmiddle|\, #2\right\}}}

\newcommand*{\QEDA}{\hfill\ensuremath{\blacksquare}}%

\iflncs
\newcommand*{\qedhere}{\hfill\qed}
\fi

\AtEndEnvironment{problem}{\null\hfill\QEDA}
\AtEndEnvironment{example}{}

\Crefname{lstlisting}{Listing}{Listings}
\Crefname{problem}{Problem}{Problems}

\Crefname{equation}{Rule}{Rules}

\newcommand{\compskel}[3]{\ensuremath{\bl{\left.\src{#1}\right\downarrow^{}_{}}}} 

\newcommand{\cmp}[1]{\compskel{#1}{}{}} 

\newcommand{\funname}[1]{\mtt{#1}}
\newcommand{\fun}[2]{\ensuremath{{\bl{\funname{#1}\left(#2\right)}}}\xspace}

\renewcommand{\hole}[1]{\ensuremath{\left[#1\right]}}

\newcommand{\evalctx}[0]{\ensuremath{\mb{E}}}
\newcommand{\evalctxs}[1]{\src{\evalctx\hole{#1}}\xspace}

\newcommand{\skips}[0]{\src{skip}\xspace}
\newcommand{\skipt}[0]{\trg{skip}\xspace}

\newcommand{\neutcol}[0]{black}
\newcommand{\stlccol}[0]{RoyalBlue}
\newcommand{\ulccol}[0]{RedOrange}
\newcommand{\commoncol}[0]{black}    
\newcommand{\othercol}[0]{Emerald}

\newcommand{\col}[2]{\ensuremath{{\color{#1}{#2}}}}

\newcommand{\src}[1]{\ms{\col{\stlccol}{#1}}}
\newcommand{\trg}[1]{\mf{\col{\ulccol }{#1}}}
\newcommand{\trgb}[1]{\ensuremath{\bm{\col{\ulccol }{#1}}}}
\newcommand{\bl}[1]{\col{\neutcol }{#1}}
\newcommand{\com}[1]{\mi{\col{\commoncol }{#1}}}

\newcommand{\ooth}[1]{\mtt{\col{\othercol }{#1}}}

\newcounter{typerule}
\crefname{typerule}{rule}{rules}

\newcommand{\typeruleInt}[5]{
	\def\thetyperule{#1}%
	\refstepcounter{typerule}%
	\label{tr:#4}%
  \ensuremath{\begin{array}{c}#5 \inference{#2}{#3}\end{array}}
}
\newcommand{\typerule}[4]{
  \typeruleInt{#1}{#2}{#3}{#4}{\textsf{\scriptsize ({#1})} \\      }
}

\newcounter{criteria}
\crefname{criteria}{}{}

\pgfdeclarelayer{background}
\pgfdeclarelayer{veryback}
\pgfdeclarelayer{veryback2}
\pgfdeclarelayer{veryback3}
\pgfdeclarelayer{back2}
\pgfdeclarelayer{foreground}
\pgfsetlayers{veryback3,veryback2,veryback,background,back2,main,foreground}

\newcommand{\myparagraph}[1]{ \smallskip \noindent\noindent\textbf{#1.}~}

\newcounter{line}

\newcommand{\xto}[1]{\ensuremath{~\mathrel{\xrightarrow{~#1~}}~}}

\newcommand{\xtos}[1]{\src{\xto{#1}}}

\newcommand{\xtot}[1]{\trg{\xto{#1}}}

\definecolor{mygreen}{rgb}{0,0.6,0}
\definecolor{mygray}{rgb}{0.5,0.5,0.5}
\definecolor{mymauve}{rgb}{0.58,0,0.82}

\lstdefinelanguage{Java} 
{morekeywords={abstract, all, and, as, assert, but, check, disj, else, exactly, extends, fact, for, fun, iden, if, iff, implies, in, Int, int, let, lone, module, no, none, not, one, open, or, part, pred, run, seq, set, sig, some, sum, then, univ, package, class, public, private, null, return, new, interface, extern, object, implements, System, static, super, try , catch, throw, throws, Unit, var, val, of, principal, trust},
sensitive=true,
keywordstyle=\bfseries\color{green!40!black},
commentstyle=\itshape\color{purple!40!black},
morecomment=[l][\small\itshape\color{purple!40!black}]{//},
identifierstyle=\color{blue},
stringstyle=\color{orange},
basicstyle=\small,
basicstyle={\small\ttfamily},
numbers=left,
numberstyle=\tiny\color{mygray},
tabsize=2,
numbersep=3pt,
breaklines=true,
lineskip=-2pt,
stepnumber=1,
captionpos=b,
breaklines=true,
breakatwhitespace=false,
showspaces=false,
showtabs=false,
float=!h,
columns=fullflexible,escapeinside={(*@}{@*)},
moredelim=**[is][\color{red!60}]{@}{@},
literate={->}{{$\to$}}1 {^}{{$\mspace{-3mu}\widehat{\quad}\mspace{-3mu}$}}1
{<}{$<$ }2 {>}{$>$ }2 {>=}{$\geq$ }2 {=<}{$\leq$ }2
{<:}{{$<\mspace{-3mu}:$}}2 {:>}{{$:\mspace{-3mu}>$}}2
{=>}{{$\Rightarrow$ }}2 {+}{$+$ }2 {++}{{$+\mspace{-8mu}+$ }}2
{<=>}{{$\Leftrightarrow$ }}2 {+}{$+$ }2 {++}{{$+\mspace{-8mu}+$ }}2
{\~}{{$\mspace{-3mu}\widetilde{\quad}\mspace{-3mu}$}}1
{!=}{$\neq$ }2 {*}{${}^{\ast}$}1 
{\#}{$\#$}1
}

\DeclareMathOperator\ceq{\ensuremath{\mathrel{\simeq_{\mi{ctx}}}}}

\def\teqaux#1{\vcenter{\hbox{\ooalign{\hfil
       \raise6pt \hbox{\scriptsize{T}}\hfil\cr\hfil
       $=$}}}}

\def\beqaux#1{\vcenter{\hbox{\ooalign{\hfil
       \raise6pt \hbox{\scriptsize{\ensuremath{\beta}}}\hfil\cr\hfil
       $=$}}}}

\def\ceqwaux#1{\vcenter{\hbox{\ooalign{\hfil
       \raise6pt \hbox{\scriptsize{w-b}}\hfil\cr\hfil
       $\ceq$}}}}

\def\praux#1{\vcenter{\hbox{\ooalign{\hfil
       \raise6pt \hbox{$\sqsubset$}\hfil\cr\hfil
       $\sim$}}}}

\newcommand{\tick}[0]{\trg{\surd}}

\iflncs
\else
\theoremstyle{definition}
\newtheorem{definition}{Definition}[section]

\newtheorem{theorem}[definition]{Theorem}

\newtheorem{example}[definition]{Example}

\newtheorem{lemma}[definition]{Lemma}
\fi
\Crefname{corollary}{Corollary}{Corollaries}
\Crefname{informal}{Definition}{Definition}
\Crefname{assumption}{Assumption}{Assumptions}
\crefname{assumption}{Assumption}{Assumptions}
\Crefname{property}{Property}{Properties}
\crefname{property}{Property}{Properties}

\newcommand{\pair}[1]{\ensuremath{\left\langle#1\right\rangle}}
\newcommand{\projone}[1]{\ensuremath{#1.1}}
\newcommand{\projtwo}[1]{\ensuremath{#1.2}}

\iflncs

\else

\fi

\newcommand{\op}[0]{\ensuremath{\, \oplus\, }}

\newcommand{\ifte}[3]{{if}~#1~{then}~#2~{else}~#3}
\newcommand{\iflete}[4]{{if}~#1~\trgb{\leq}~#2~{then}~#3~{else}~#4}
\newcommand{\iftes}[3]{\src{if}~#1~\src{then}~#2~\src{else}~#3}

\newcommand{\ifztet}[3]{\trg{ifz}~#1~\trg{then}~#2~\trg{else}~#3}
\newcommand{\ifztes}[3]{\src{ifz}~#1~\src{then}~#2~\src{else}~#3}

\def\formatCompilers#1{\ms{#1}\xspace}

\def\porc{P}
\def\porcc{C}
\def\criterion#1{\formatCompilers{#1\porc}}

\def\pf#1{\ensuremath{\let\porc\porcc #1}}

\newcommand{\rschp}[0]{\criterion{RSCH}}

\newcommand{\rtp}[0]{\criterion{RT}}
\newcommand{\tp}{\criterion{T}}

\newcommand{\rhsp}[0]{\criterion{RHS}}

\newcommand{\rsp}[0]{\criterion{RS}}

\newcommand{\red}[0]{\ensuremath{\ \hookrightarrow\ }}

\newcommand{\redapp}[1]{\ensuremath{\!\!\!^{#1}\ }}
\newcommand{\redstar}[0]{\redapp{*}}

\newcommand{\redstars}[0]{\src{\redstar}}

\AtEndEnvironment{example}{\null\hfill$\boxdot$}

\makeatletter
\xdef\@thefnmark{\@empty}

\makeatother

\renewcommand{\emptyset}[0]{\varnothing}

\newcounter{hps}
\crefname{hps}{}{}

\newcommand{\comptd}[1]{\compskel{#1}{\Lt}{\Ld}}

\newcommand{\sem}{\boldsymbol{\rightsquigarrow}}

\newcommand{\tS}[1]{\src{#1}}
\newcommand{\tT}[1]{\trg{#1}}

\newcommand{\prop}[0]{2^{\text{Trace}}}
\newcommand{\propS}[0]{2^{\src{{Trace}_S}}}
\newcommand{\propT}[0]{2^{\trg{{Trace}_T}}}

\newcommand{\piS}[1]{\ensuremath{\src{#1_S}}}
\newcommand{\piT}[1]{\ensuremath{\trgb{#1_{\mr{T}}}}}
\newcommand{\hS}[1]{\ensuremath{\src{#1_S}}}
\newcommand{\hT}[1]{\ensuremath{\trg{#1_T}}}

\newcommand{\satS}{\src{\models}}
\newcommand{\satT}{\trgb{\models}}

\newcommand{\critdiff}[2]{\ensuremath{#1^{#2}}\xspace}

\newcommand{\eundefS}{\ensuremath{\src{Goes\_wrong}}}
\newcommand{\eundef}{\ensuremath{\mi{Goes\_wrong}}}
\newcommand{\oomT}{\ensuremath{\trg{Resource\_Limit\_Hit}}}

\newcommand{\ani}[2]{\ensuremath{\mathit{ANI}_{#1} ^{#2}}}
\newcommand{\uco}[1]{\ensuremath{\mathit{uco}(#1)}}

\newcommand{\traceP}[0]{\criterion{T}}

\newcommand{\hyperP}[0]{\criterion{H}}

\newcommand{\tildePF}[1]{\critdiff{\pf{#1}}{\sim}}

\newcommand{\wtmkt}[2]{\cmp{#1} \trg{\sem}\tT{#2}}
\newcommand{\wsmkt}[2]{\src{#1} \src{\sem}\tS{#2}}

\newcommand{\TypeNat}{N}
\newcommand{\TypeBool}{B}
\newcommand{\emptylist}{\emptyset}
\newcommand{\booltag}{\xspace} 
\newcommand{\nattag}{\xspace} 

\newcommand{\langlettm}[0]{N}
\newcommand{\Sm}[0]{\src{\langlettm^{p}_{}}\xspace}
\newcommand{\Tm}[0]{\trg{\langlettm^{n}_{}}\xspace}
\newcommand{\cmpm}[1]{\compskel{#1}{\Sm}{\Tm}}
\newcommand{\compm}[1]{\cmpm{\left(#1\right)}}

\newcommand{\send}[0]{send}
\newcommand{\sends}[1]{\src{\send~#1}}
\newcommand{\sendt}[1]{\trg{\send~#1}}

\newcommand{\ift}[0]{\text{ if }}
\newcommand{\thent}[0]{\text{ then }}
\newcommand{\andt}[0]{\text{ and }}

\newcommand{\gensends}[1]{\fun{gensend}{#1}}

\DeclareMathOperator\simbase{\sim^0}
\newcommand{\simnat}[0]{\ensuremath{\simbase}}
\DeclareMathOperator\simmess{\sim} 
\newcommand{\simm}[0]{\ensuremath{\simmess}}
\DeclareMathOperator\simqueue{\sim} 
\newcommand{\simq}[0]{\ensuremath{\simqueue}}

\newcommand{\listconcat}[0]{\ensuremath{\cdot}}

\newcommand{\langlettr}[0]{R}
\newcommand{\Sr}[0]{\src{\langlettr^{S}_{}}\xspace}
\newcommand{\Tr}[0]{\trg{\langlettr^{T}_{}}\xspace}

\newcommand{\SrPar}[0]{\src{P}\xspace}
\newcommand{\SrCtx}[0]{\src{C}\xspace}
\newcommand{\SrPrg}[0]{\src{W}\xspace}

\newcommand{\SrOutL}[0]{\src{out_S}\xspace}
\newcommand{\TrOutL}[0]{\trg{out_S}\xspace}
\newcommand{\TrOutH}[0]{\trg{out_T}\xspace}

\newcommand{\myfilter}[1]{|_{#1}}

\newcommand{\cmpr}[1]{\compskel{#1}{\Sr}{\Tr}}

\newcommand{\tcol}[0]{blue}
\newcommand{\sccol}[0]{red}
\newcommand{\trtcol}[0]{green}

\newcommand{\ccbasic}[0]{\formatCompilers{CC}}
\newcommand{\ccequal}[0]{\formatCompilers{CC^=}}
\newcommand{\cctilde}[0]{\formatCompilers{CC^{\sim}}}

\newcommand{\cctildei}[1]{\formatCompilers{CC^{#1}}}

\newcommand{\scequal}[0]{\formatCompilers{SC^=}}
\newcommand{\sctilde}[0]{\formatCompilers{SC^{\sim}}}

\newcommand{\hcequal}[0]{\formatCompilers{HC^=}}
\newcommand{\hctilde}[0]{\formatCompilers{HC^{\sim}}}

\newcommand{\tptau}[0]{\formatCompilers{TP^{\tau}}}
\newcommand{\tptautilde}[0]{\formatCompilers{TP^{\tilde{\tau}}}}
\newcommand{\tpsigma}[0]{\formatCompilers{TP^{\sigma}}}
\newcommand{\tpsigmatilde}[0]{\formatCompilers{TP^{\tilde{\sigma}}}} 

\newcommand{\sptautilde}[0]{\formatCompilers{TP^{\ii{Safe}\circ\tilde{\tau}}}}
\newcommand{\spsigmatilde}[0]{\formatCompilers{SP^{\tilde{\sigma}}}}

\newcommand{\schptautilde}[0]{\formatCompilers{SCHP^{\ii{Cl}_{\subseteq}\circ\tilde{\tau}}}}
\newcommand{\schpsigmatilde}[0]{\formatCompilers{SCHP^{\ii{Cl}_{\subseteq}\circ\tilde{\sigma}}}}

\newcommand{\hptautilde}[0]{\formatCompilers{HP^{\tilde{\tau}}}}
\newcommand{\hpsigmatilde}[0]{\formatCompilers{HP^{\tilde{\sigma}}}}

\newcommand{\rtctilde}[0]{\formatCompilers{RTC^{\sim}}}        
\newcommand{\rtctildei}[1]{\formatCompilers{RTC^{#1}}}         
\newcommand{\rsctilde}[0]{\formatCompilers{RSC^{\sim}}}        
\newcommand{\rhsctilde}[0]{\formatCompilers{RHSC^{\sim}}}      
\newcommand{\rschptilde}[0]{\formatCompilers{RSCHC^{\sim}}}    
\newcommand{\rhctilde}[0]{\formatCompilers{RHC^{\sim}}}        
\newcommand{\rtworeltctilde}[0]{\formatCompilers{R2rTC^{\sim}}}    
\newcommand{\rtworelsctilde}[0]{\formatCompilers{R2rSC^{\sim}}}    
\newcommand{\rtworelschpctilde}[0]{\formatCompilers{R2rSCHC^{\sim}}} 

\newcommand{\rtptautilde}[0]{\formatCompilers{RTP^{\tilde{\tau}}}}
\newcommand{\rtpsigmatilde}[0]{\formatCompilers{RTP^{\tilde{\sigma}}}}

\newcommand{\rsptautilde}[0]{\formatCompilers{RTP^{\ii{Safe}\circ\tilde{\tau}}}}

\newcommand{\rspsigmatilde}[0]{\formatCompilers{RSP^{\tilde{\sigma}}}}

\newcommand{\rschptautilde}[0]{\formatCompilers{RSCHP^{\ii{Cl}_{\subseteq}\circ\tilde{\tau}}}}
\newcommand{\rschpsigmatilde}[0]{\formatCompilers{RSCHP^{\ii{Cl}_{\subseteq}\circ\tilde{\sigma}}}}

\newcommand{\rhsptautilde}[0]{\formatCompilers{RSCHP^{\ii{HSafe}\circ\tilde{\tau}}}}
\newcommand{\rhspsigmatilde}[0]{\formatCompilers{RHSP^{\ii{Cl}_{\subseteq}\circ\tilde{\sigma}}}}

\newcommand{\rhptautilde}[0]{\formatCompilers{RHP^{\tilde{\tau}}}}
\newcommand{\rhpsigmatilde}[0]{\formatCompilers{RHP^{\tilde{\sigma}}}}

\newcommand{\rtworelschptautilde}[0]{\formatCompilers{R2rSCHP^{\ii{Cl}_{\subseteq}\circ\tilde{\tau}}}}
\newcommand{\rtworelschpsigmatilde}[0]{\formatCompilers{R2rSCHP^{\ii{Cl}_{\subseteq}\circ\tilde{\sigma}}}}

\newcommand{\rtworelsptautilde}[0]{\formatCompilers{R2rTP^{\ii{2rSafe}\circ\tilde{\tau}}}}
\newcommand{\rtworelspsigmatilde}[0]{\formatCompilers{R2rSP^{\tilde{\sigma}}}}

\newcommand{\rtworeltptautilde}[0]{\formatCompilers{R2rTP^{\tilde{\tau}}}}
\newcommand{\rtworeltpsigmatilde}[0]{\formatCompilers{R2rTP^{\tilde{\sigma}}}}

\newcommand{\aniS}[0]{\ani{\src{\phi}}{\src{\rho}}}
\newcommand{\aniT}[0]{\ani{\trgb{\phi^{\#}}}{\trgb{\rho^{\#}}}}

\newcommand{\inp}{^{\scaleobj{0.7}{\circ}}}
\newcommand{\out}{^{\scaleobj{0.7}{\bullet}}}

\makeatletter
\newcommand{\oset}[2]{%
  \mathrel{\mathop{#2}\limits^{\vbox to -.5\ex@{\kern-\tw@\ex@
   \hbox{\tiny #1}\vss}}}}
\makeatother
\newcommand{\simin}{\oset{$\scaleobj{0.7}{\circ}$}{\sim}}

\newcommand{\simout}{\oset{$\scaleobj{0.7}{\bullet}$}{\sim}}

\newcommand{\lowcls}[1]{[#1]_{\mathit{low}}}

\newcommand{\repo}{https://github.com/secure-compilation/different_traces/blob/esop2020-camera-ready}
\newcommand{\coqhref}[1]{\href{\repo/#1}{\CoqSymbol}}
\newcommand{\coqedhref}[1]{\href{\repo/#1}{\Coqed}}

%% file: abstract.tex

Compiler correctness is, in its simplest form, defined as the
inclusion of the set of traces of the compiled program into the set of
traces of the original program, which is equivalent to the
preservation of all trace properties.
Here traces collect, for instance, the externally observable events of each
execution.
This definition requires, however, the set of traces of the source and target
languages to be exactly the same, which is not the case when the languages are
far apart or when observations are fine-grained.
To overcome this issue, we study a generalized compiler correctness
definition, which uses source and target traces drawn from potentially
different sets and connected by an arbitrary relation.
We set out to understand what guarantees this generalized compiler correctness
definition gives us when instantiated with a non-trivial relation on traces.
When this trace relation is not equality, it is no longer possible to preserve
the trace properties of the source program unchanged.
Instead, we provide a generic characterization of the target trace property
ensured by correctly compiling a program that satisfies a given source property,
and dually, of the source trace property one is required to show in order to
obtain a certain target property for the compiled code.
We show that this view on compiler correctness can naturally account for
undefined behavior, resource exhaustion, different source and target values,
side-channels, and various abstraction mismatches.
Finally, we show that the same generalization also applies to many 
secure compilation definitions, which characterize the
protection of a compiled program against linked adversarial code.

%% file: intro.tex
\section{Introduction}\label{sec:intro}



Compiler correctness is an old idea~\cite{McCarthyP67, MilnerW72, Morris73a}
that has seen a significant revival in recent times.
This new wave was started by the creation of the CompCert verified C
compiler~\cite{Leroy09} and continued by the proposal of many significant
extensions and variants of CompCert~\cite{SevcikVNJS13, MullenZTG16,
  StewartBCA15, KangKHDV15, KangHMGZV15, BessonBW19, WangWS19, GuSKWKS0CR18,
  RamananandroSWKF15, CarbonneauxHRS14, BoldoJLM15}
and the success of many other milestone compiler verification projects,
including Vellvm~\cite{ZhaoNMZ12},
Pilsner~\cite{NeisHKMDV15},
CakeML~\cite{TanMKFON19},
CertiCoq~\cite{CertiCoq17}, \ETC
%
%
Yet, even for these verified compilers, the precise statement of
correctness matters.
%
Since proof assistants are used to conduct the verification, an external
observer does not have to understand the proofs in order to trust them, but one
still has to deeply understand the statement that was proved.
And this is true not just for correct compilation, but also for secure
compilation, which is the more recent idea that our compilation chains should 
do more to 
also ensure security of our programs
\cite{SecureCompilationSIGPLAN19, dagstuhl-sc-2018}.


\myparagraph{Basic Compiler Correctness}
The gold standard for compiler correctness is {\em semantic preservation}, which
intuitively says that the semantics of a compiled program (in the target language)
is compatible with the semantics of the original program (in the source language).
For practical verified compilers, such as CompCert \cite{Leroy09} and CakeML
\cite{TanMKFON19}, semantic preservation is stated extrinsically, by referring
to {\em traces}.
In these two settings,
a trace is an ordered sequence of events---such as inputs from and outputs to
an external environment---that are produced by the execution of a program.

A basic definition of compiler correctness can be given by the set inclusion of
the traces of the compiled program into the traces of the original program. Formally~\cite{Leroy09}:

\begin{definition}[Basic Compiler Correctness ($\ccbasic$)]
\label{def:bcc}
A compiler $\downarrow$ is {\em correct} iff
$$
\forall \src{W}~t.~ \trg{\cmp{W}} \trg{\sem} t \Rightarrow \src{W} \src{\sem} t.
$$
\end{definition}
This definition says that for any whole\footnote{For simplicity, for now
  we ignore separate compilation and linking, returning to it in
  \autoref{sec:secure-compilation}.
}
source program \src{W}, if we compile it (denoted \cmp{W}), execute it with
respect to the semantics of the target language, and observe a trace
$t$, then the original $\src{W}$ can produce {\em the same} trace $t$ with
respect to the semantics of the source language.%
\footnote{%
  Typesetting convention~\cite{patrignani2020use}: we use a \src{blue}, \src{sans\text{-}serif} font for \src{source} elements,
  an \trg{orange}, \trg{bold} font for \trg{target} ones
  and a \com{\commoncol}, \com{italic} font for elements common to both languages.
}
%
%
This definition is simple and easy to understand,
since it only references a few familiar concepts:
a compiler between a source and a target language,
each equipped with a trace-producing 
semantics (usually nondeterministic).
%

\myparagraph{Beyond Basic Compiler Correctness}
This basic compiler correctness definition assumes that any trace produced by
a compiled program can be produced by the source program.
This is a very strict requirement, and in particular implies that the source and
target traces are drawn from the same set and that the same source trace
corresponds to a given target trace.
%
These assumptions are often too strong, and hence in practice verified compiler
efforts use different formulations of compiler correctness:
\begin{description}[nosep]
\item[CompCert~\cite{Leroy09}]
The original compiler correctness theorem of CompCert~\cite{Leroy09} can be seen
as an instance of basic compiler correctness, but it does not provide any
guarantees for programs that can exhibit undefined
behavior~\cite{Regehr10}.
As allowed by the C standard, such unsafe programs are not even considered to be
in the source language, so are not quantified over.
%
This has important practical implications, since undefined behavior 
%
often leads to exploitable security vulnerabilities~\cite{Heartbleed,
  memory-unsafety,HallerJPPGBK16} and serious confusion even among experienced C
and C++ developers~\cite{Regehr10, Lattner11, WangZKS13, WangCCJZK12}.
As such, since 2010, CompCert provides an additional top-level correctness
theorem\footnote{%
  Stated at the top of the CompCert file
  \href{http://compcert.inria.fr/doc/html/compcert.driver.Complements.html\#transf_c_program_preservation}{\tt
    driver/Complements.v} and discussed by \citet{Regehr10}.}
that better accounts for the presence of unsafe programs by
providing guarantees for them up to the point when they encounter
undefined behavior~\cite{Regehr10}.
This new theorem goes beyond the basic correctness definition above, as
a target trace need only correspond to a source trace \emph{up to the
  occurrence} of undefined behavior in the source trace.

\item[CakeML~\cite{TanMKFON19}] 
Compiler correctness for CakeML accounts for
  memory exhaustion in target executions. Crucially, memory exhaustion events
  cannot occur in source traces, only in target traces.
  Hence, dually to CompCert, compiler correctness only requires source and
  target traces to coincide up to the occurrence of a memory exhaustion event
  in the target trace.
\end{description}

\myparagraph{Trace-Relating Compiler Correctness}
Generalized formalizations of compiler correctness like the 
ones above can
be naturally expressed as instances of a uniform definition, which we call
{\em trace-relating compiler correctness}.
This generalizes basic compiler correctness by
(a) considering that source and target traces belong to {\em possibly distinct}
sets $\src{Trace_S}$ and $\trg{Trace_T}$, and (b) being parameterized
by an arbitrary {\em trace relation} $\sim$.


\begin{definition}[Trace-Relating Compiler Correctness
  (\cctilde)] \label{defn:tracepreservation}
A compiler $\downarrow$ is {\em correct} with respect to a trace relation
$\sim \,\subseteq \src{Trace_S} \times \trg{Trace_T}$ iff
$$
\forall\src{W}. \forall \tT{t}.~ \wtmkt{W}{t}
         \Rightarrow \mathrel{\exists\tS{s} \sim \tT{t}} \ldotp \wsmkt{W}{s}.
$$
\end{definition}
This definition requires that, for any target trace $\tT{t}$ produced by the
compiled program $\cmp{W}$, there exist a source trace $\tS{s}$ that can be
produced by the original program $\src{W}$ and is {\em related} to $\tT{t}$
according to $\sim$ (\IE $\tS{s} \sim \tT{t}$).
By choosing the trace relation appropriately, one can recover the different
notions of compiler correctness presented above:
\begin{description}[nosep]
\item[Basic CC] Take $\tS{s} \sim \tT{t}$ to be $\tS{s} =
  \tT{t}$. Trivially, the basic CC of \autoref{def:bcc} is $\ccequal$.

\item[CompCert] Undefined behavior is modeled in CompCert
  as a trace-terminating event \eundef\xspace
  that can occur in any of its languages (source,
  target, and all intermediate languages), so for a given
  phase (or composition thereof), we have $\src{Trace_S} = \trg{Trace_T}$.
  Nevertheless, the relation between source and target traces with which to
  instantiate $\cctilde$ to obtain CompCert's current theorem is:
$$
\tS{s} \sim \tT{t} \quad\equiv\quad \tS{s} = \tT{t} \vee
    (\exists m \leq \tT{t}.~ \tS{s} = m{\cdot}\eundef).
$$
A compiler satisfying $\cctilde$ for this trace relation
can turn a source trace ending in undefined behavior
$m{\cdot}\eundef$ (where ``${\cdot}$'' is concatenation) either into
the same trace in the target (first disjunct), or into a target trace that
starts with the prefix $m$ but then continues {\em arbitrarily}
(second disjunct, ``${\leq}$'' is the prefix relation).

\item[CakeML] Here, target traces are sequences of symbols from an alphabet
$\trg{\Sigma_T}$ that has a specific trace-terminating event,
\trg{Resource\_limit\_hit},
which is not available in the source alphabet $\src{\Sigma_S}$ 
(\IE $\trg{\Sigma_T} = \src{\Sigma_S} \cup \{ \trg{Resource\_limit\_hit} \}$.
Then, the compiler correctness theorem of CakeML can be obtained by
instantiating  $\cctilde$  with the following $\sim$ relation:
$$
\tS{s} \sim \tT{t} \quad\equiv\quad \tS{s} = \tT{t} \vee
    (\exists m.~ m\leq \tS{s}.~ \tT{t} = m{\cdot}\trg{Resource\_limit\_hit}).
$$
The resulting $\cctilde$ instance relates a target trace ending in
\trg{Resource\_limit\_hit} after executing $m$ to a source trace that first
produces $m$ and then continues in a way given by the semantics of the source
program.
%
\end{description}

\smallskip
Beyond undefined behavior and resource exhaustion, there are many other
practical uses for \cctilde: in this paper we show that it also
accounts for differences between source and target values, for a single
source output being turned into a series of 
target outputs, and for side-channels.


On the flip side, the compiler correctness statement and its implications can be
more difficult to understand for $\cctilde$ than for $\ccequal$.
The full implications of choosing a particular $\sim$ relation can be subtle.
%
In fact, using a bad relation can make the compiler correctness statement trivial or unexpected.
%
For instance, it should be easy to see that if one uses the total relation, which
relates all
source traces to all target ones, the $\cctilde$ property holds for every
compiler, yet it might take one a bit more effort to understand that the
same is true even for the following relation:%
\begin{align*}
\tS{s} \sim \tT{t} \quad\equiv\quad \exists \src{W}. \wsmkt{W}{s} \wedge \wtmkt{W}{t}.
\end{align*}

\myparagraph{Reasoning About Trace Properties}
%
%
To understand more about a particular $\cctilde$ instance, we propose to also
look at how it preserves {\em trace properties}---defined as sets of allowed
traces~\cite{LamportS84}---from the source to the target.
For instance, it is well known that $\ccequal$ is equivalent to the
preservation of all trace properties (where $W \models \pi$ reads
``$W$ satisfies $\pi$'' and stands for
$\forall t.~ W{\sem} t \Rightarrow t \in \pi$):
$$
\ccequal \quad\equiv\quad \forall \pi \in \prop ~\forall \src{W}. ~
  \src{W} \src{\models} \pi \Rightarrow \cmp{W} \trgb{\models} \pi.
$$
However, to the best of our knowledge, similar results have not been formulated for
trace relations beyond equality, when it is no longer possible to preserve
the trace properties of the source program unchanged.
%
%
%
For trace-relating compiler correctness, where source and target traces can be
drawn from different sets and related by an arbitrary trace relation, there are
two crucial questions to ask:
%
\begin{enumerate}[nosep]
\item For a source trace property $\piS{\pi}$ of a program---established for
  instance by formal verification---what is the strongest
  target property that any  $\cctilde$  compiler is guaranteed to ensure for
  the produced target program?
\item For a target trace property ${\piT{\pi}}$, what is the weakest source
  property we need to show of the original source program to obtain
  $\piT{\pi}$ for the result of any  $\cctilde$  compiler?
%
%
%
\end{enumerate}
Far from being mere hypothetical questions, they can help the developer of a
verified compiler to better understand the compiler correctness theorem they
are proving, and we expect that any user of such a compiler will need to ask
either one or the other if they are to make use of that theorem.
%
In this work we provide a simple and natural answer to these
questions, for any instance of \cctilde. Building upon a bijection
between relations and
Galois connections \cite{gardiner1994algebraic, backhouse2004safety,
  naumann1998categorical}, 
%
we observe that any trace relation $\sim$ corresponds to two {\em property
  mappings} $\tilde{\tau}$ and $\tilde{\sigma}$, which are functions
mapping source properties to target ones ($\tilde{\tau}$ standing for
``to target'') and target properties to source ones ($\tilde{\sigma}$
standing for ``to source''): 
$$
\tilde{\tau}(\piS{\pi}) = \myset{\tT{t}}{\exists \tS{s} \ldotp \tS{s} \sim \tT{t} \wedge \tS{s} \in \piS{\pi}};
\qquad
\tilde{\sigma}(\piT{\pi}) = \myset{\tS{s}}{\forall \tT{t} \ldotp \tS{s} \sim \tT{t} \Rightarrow \tT{t} \in \piT{\pi}}.
$$
The \emph{existential image} of $\sim$, $\tilde{\tau}$, answers the
first question above by mapping a given source property $\piS{\pi}$ to
the target property that contains all target traces for which {\em
  there exists a related source trace} that satisfies $\piS{\pi}$.
Dually, the \emph{universal image} of $\sim$, $\tilde{\sigma}$,
answers the second question by mapping a given target property
$\piT{\pi}$ to the source property that contains all source traces for
which {\em all related target traces} satisfy $\piT{\pi}$.
%
%
%
%
%
We introduce two new correct compilation definitions in terms of {\em
  trace property preservation} (\tp):
\tptautilde quantifies over all source trace properties
and uses $\tilde{\tau}$ to obtain the corresponding target properties.
\critdiff{\traceP}{\tilde{\sigma}} quantifies over all target trace
properties and uses $\tilde{\sigma}$ to obtain the corresponding source
properties.
We prove that these two definitions are equivalent to \cctilde,
yielding a novel trinitarian view of compiler correctness (\autoref{fig:diagram-cc}).
%



\input{diagram-cc-only}



\newpage
\myparagraph{Contributions}
\begin{itemize}[leftmargin=0em,nosep,label=$\blacktriangleright$]

\item We propose a new trinitarian view of compiler correctness that
accounts for non-trivial \iffull
\else trace relations\fi.
While, as discussed above, specific instances of the $\cctilde$ definition
have already been used in practice,
we seem to be the first to propose assessing the
meaningfulness of $\cctilde$ instances in terms of how properties are
preserved between the source and the target, and in particular by looking at the
property mappings $\tilde{\sigma}$ and $\tilde{\tau}$ induced by the trace
relation $\sim$.
We prove that \cctilde, $\tpsigmatilde$, and
$\tptautilde$ are equivalent for any trace relation (\autoref{sec:trinity}),
as illustrated in \autoref{fig:diagram-cc}.
%
%
%
In the opposite direction, we show that for every trace relation
corresponding to a given Galois connection
\cite{gardiner1994algebraic}, an analogous equivalence holds.
Finally, we extend these results (\autoref{sec:subset-closed})
from the preservation of trace properties to
the larger class of subset-closed hyperproperties (\EG noninterference).%

\item
We use $\cctilde$ compilers of various complexities
to illustrate that our view on compiler
correctness naturally accounts
for undefined behavior (\autoref{sec:example-undef}),
resource exhaustion (\autoref{sec:example-resources}),
different source and target values (\autoref{sec:example-diff-values}),
and differences in the granularity of data and observable events
(\autoref{sec:example-split-io}).
%
We expect these ideas to apply to any other discrepancies between source
and target traces.
For each compiler we show how to choose the relation
between source and target traces and how the induced property mappings preserve
interesting trace properties and subset-closed hyperproperties. 
We look at the way particular $\tilde{\sigma}$ and $\tilde{\tau}$ work on
different kinds of properties and how the produced properties can be expressed for
different kinds of traces.

\item We analyze the impact of correct compilation on
  noninterference~\cite{GoguenM82}, showing what can still be preserved
  (and thus also what is lost)
  when target observations are finer than source ones, \EG~side-channel
  observations (\autoref{sec:example-noninterference}). We formalize
  the guarantee obtained by correct compilation of a noninterfering
  program as {\em abstract
    noninterference}~\cite{giacobazzi2018abstract}, a weakening of
  target noninterference. Dually, we identify a family of
  declassifications of target noninterference for which source
  reasoning is possible.


\item Finally, we show that the trinitarian view also extends to a
large class of {\em secure compilation} definitions \cite{AbateBGHPT19},
formally characterizing the protection of the compiled program against linked
adversarial code (\autoref{sec:secure-compilation}).
For each secure compilation definition we again
propose both a property-free characterization in the style of \cctilde,
and two characterizations in terms of preserving a class of source or target
properties satisfied against arbitrary adversarial contexts.
The additional quantification over contexts allows for finer
distinctions when considering different property classes, so we
study mapping classes not only of trace properties and hyperproperties,
but also of relational hyperproperties~\cite{AbateBGHPT19}.
%
%
An example secure compiler accounting for a target that can produce additional
trace events that are not possible in the source illustrates this approach.
\end{itemize}
The paper closes with discussions of
related~(\autoref{sec:related}) and future work~(\autoref{sec:conclusion}).
\ifappendix The \else An online \fi appendix contains omitted technical details\ifappendix\else:
{\small \url{https://arxiv.org/abs/1907.05320}}\fi.

The traces considered in our examples are structured, usually as sequences of
events.
We notice however that unless explicitly mentioned, all our definitions and
results are more general and make no assumption whatsoever about the structure
of traces.
Most of the theorems formally or informally mentioned in the paper
were mechanized in the Coq proof assistant and are marked with \CoqSymbol.
This development has around 10k
lines of code, is described in the online appendix, and is available
at the following address:\\
{\small \url{https://github.com/secure-compilation/different_traces}.}

%% file: diagram-cc-only.tex

\begin{figure}[!ht]
\vspace{2em}
\begin{center}




\begin{tikzpicture}
  \node[](cctilde){
    \phantom{$\sim$}
    \cctilde
  };
  \node[below right =.4 and 1 of cctilde](tctau){
    \tptautilde
  };
  \node[below left =.4 and 1 of cctilde](tcsigma){
    \tpsigmatilde
  };

  \draw[myiff] (cctilde.-30) to (tctau.150);
  \draw[myiff] (cctilde.-150) to (tcsigma.30);
  \draw[myiff] (tctau.180) to (tcsigma.0);

  \node[above of=cctilde, yshift=-1em](cctildedef){
    $\forall\src{W}.~ \forall \tT{t}.~ \wtmkt{W}{t} \Rightarrow \mathrel{\exists\tS{s} \sim \tT{t}} \ldotp \wsmkt{W}{s}$
  };
  \node[left =of tcsigma.north east](tcsigmadef){
    $
    \begin{aligned}[b]
      &
      \forall \piT{\pi}. ~\forall\src{W}.~ \src{W} ~\satS~ \tilde{\sigma}(\piT{\pi}) 
      \\
      &\ \Rightarrow \cmp{W} ~\satT~ \piT{\pi}      
    \end{aligned}
    \equiv$
  };
  \node[right =of tctau.north west](tctaudef){
    $\equiv
    \begin{aligned}[b]
      &
      \forall \piS{\pi}. ~\forall \src{W}. ~ \src{W} ~\satS~ \piS{\pi} 
      \\
      &\
      \Rightarrow \cmp{W} ~\satT~ \tilde{\tau}(\piS{\pi})
    \end{aligned}
    $
  };
  \path (cctilde.90) -- (cctildedef.-90) node[midway,sloped] (eq) {$\equiv$};
\end{tikzpicture}
\end{center}
\caption{The equivalent compiler correctness definitions forming our trinitarian view.}
\label{fig:diagram-cc}
\vspace{1em}
\end{figure}

%% file: undefbeh.tex
We start by expanding upon the discussion of undefined behavior
in \autoref{sec:intro}.
%
%
%
%
We first study the model of CompCert, where source and target alphabets are the same,
including the event for undefined behavior.
The trace relation weakens equality by allowing undefined behavior to be
replaced with an arbitrary sequence of events.
%
\begin{example}[CompCert-like Undefined Behavior Relation]\label{ex:undef}
Source and target traces are sequences of events drawn from \(\Sigma\),
where \(\eundef \in \Sigma\) is a terminal event that represents an undefined
behavior.
We then use the trace relation from the introduction:
\[
\tS{s} \sim \tT{t} \quad\equiv\quad \tS{s} = \tT{t} \vee \exists m \leq \tT{t}.~ \tS{s} = {m}\cdot\eundef.
\]
%
Each trace of a target program produced by a \cctilde compiler is either also a
trace of the original source program or it has a finite prefix that the source
program also produces, immediately before encountering undefined behavior.
%
As explained in \autoref{sec:intro}, one of the correctness theorems in CompCert
can be rephrased as this variant of $\cctilde$.


We proved that the property mappings induced by the relation can be written as (\coqhref{UndefBehaviorCompCert.v}):
{\small\begin{align*}
  \tilde{\sigma}(\piT{\pi}) =&\ 
    \myset{\tS{s}}{\tS{s} {\in} \piT{\pi} \wedge \tS{s} \neq m{\cdot}\eundef}
  \cup
    \myset{{m}{\cdot}\eundef}{\forall \tT{t}.~ m {\leq} \tT{t} {\implies} \tT{t} {\in} \piT{\pi}};
    \\
  \tilde{\tau}(\piS{\pi}) =&\ 
    \myset{\tT{t}}{\tT{t} {\in} \piS{\pi}}
                             \cup \myset{\tT{t}}{\exists
                                  m \leq \tT{t}.~ {m} {\cdot} \eundef \in \piS{\pi}}.
\end{align*}}%
These two mappings explain what a \cctilde compiler ensures
for the $\sim$ relation above.
The target-to-source mapping \(\tilde{\sigma}\) states that to prove that
a compiled program has a property \(\trgb{\pi_T}\) using source-level reasoning,
one has to prove that any trace produced by the source program must either be a target
trace satisfying $\trgb{\pi_T}$ 
or have undefined behavior, but only provided that
{\em any continuation} of the trace
substituted for the undefined behavior satisfies $\trgb{\pi_T}$.
%
The source-to-target mapping \(\tilde{\tau}\) states that by compiling a
program satisfying a property \(\piS{\pi}\) we obtain a program that
produces traces that satisfy the same property or that extend a source
trace that ends in undefined behavior.

These definitions can help us reason about programs. For instance,
\(\tilde{\sigma}\) specifies that, to prove that an event does not happen in the
target, it is not enough to prove that it does not happen in the source: it is
also necessary to prove that the source program is 
does not have any undefined behavior (second disjunct).
Indeed, if it had an undefined behavior, its continuations could exhibit
the unwanted event.
%
\end{example}



This relation can be easily generalized to other settings.
%
For instance, consider the setting in which we compile down to a low-level language like
machine code.
Target traces can now contain new events that cannot occur in the source: indeed,
in modern architectures like x86 a compiler typically uses only a fraction of
the available instruction set.
Some instructions might even perform dangerous operations, such as writing to the
hard drive.
%
Formally, the source and target do not have the same events any more.
Thus, we consider a source alphabet \(\src{\Sigma_S} = \Sigma \cup \{\eundefS\}\), and
a target alphabet \(\trg{\Sigma_T} = \Sigma \cup \Sigma'\).
The trace relation is defined in the same way and we obtain the same property
mappings as above, except that since target traces now have more events (some of
which may be dangerous), and the arbitrary continuations of target traces get
more interesting.
For instance, consider a new event that represents writing data on the hard
drive, and suppose we want to prove that this event cannot happen for a
compiled program.
Then, proving this property requires exactly proving that the source
program exhibits no undefined behavior~\cite{CaoBGDA18}.
%
More generally,
what one can prove about target-only events can only be either that they
cannot appear (because there is no undefined behavior) or that any of them can
appear (in the case of undefined behavior).

In \autoref{sec:sec-comp-traces} we study a similar example, showing that even
in a safe language linked adversarial contexts can cause dangerous target events
that have no source correspondent.



%% file: resource-exhaustion.tex

Let us return to the discussion about resource exhaustion in
\autoref{sec:intro}.
%
%

\begin{example}[Resource Exhaustion]\label{ex:res-exh}
We consider traces made of events drawn from \src{\Sigma_S} in the source, and
\(\trg{\Sigma_T} = \src{\Sigma_S} \cup \{\oomT\}\) in the target.
Recall the trace relation for resource exhaustion:
\[
\tS{s} \sim \tT{t} \quad\equiv\quad \tS{s} = \tT{t} \vee \exists {m} \leq \tS{s}.~ \tT{t} = {m}\cdot\oomT.
\]
Formally, this relation is similar to the one for undefined behavior, except
this time it is the target trace that is allowed to end early instead of the
source trace.

The induced trace property
mappings \(\tilde{\sigma}\) and \(\tilde{\tau}\) are the following
(\coqhref{ResourceExhaustion.v}):
\begin{align*}
  \tilde{\sigma}(\trgb{\pi_T}) &= \{ \tS{s} \mid \tS{s} \in \trgb{\pi_T} \} \cap
    \{\tS{s} \mid \forall {m} \leq \tS{s}.~{m}\cdot\oomT \in \trgb{\pi_T}\};
    \\
  \tilde{\tau}(\tS{\pi_S}) &= \tS{\pi_S} \cup
    \{ {m}\cdot\oomT \mid \exists \tS{s} \in \tS{\pi_S}.~{m} \leq \tS{s}\}.
\end{align*}
These capture the following intuitions.
  The target-to-source mapping \(\tilde{\sigma}\) states that
  to prove a property of the compiled program
  one has to show that the traces of the source program
  satisfy two conditions: (1) they must also satisfy the target property; and
  (2) the termination of every one of their prefixes by a resource
  exhaustion error must be allowed by the target property.
%
This is rather restrictive:
any property that prevents resource exhaustion cannot be proved
using source-level reasoning. Indeed, if \(\piT{\pi}\) does not allow
resource exhaustion, then \(\tilde{\sigma}(\piT{\pi}) = \src{\emptyset}\).
This is to be expected since resource exhaustion is simply not accounted for
at the source level.
  %
%
The \iffull\else other \fi
mapping \(\tilde{\tau}\) states that a compiled program produces traces that
either belong to the same properties as the traces of the source program or
end early due to resource exhaustion.

In this example, safety properties~\cite{LamportS84} are mapped
(in both directions) to other safety properties (\coqhref{ResourceExhaustion.v}).
This can be desirable for a relation: since safety properties are usually easier
to reason about, one interested only in safety properties at the target can
reason about them using source-level reasoning tools for safety properties.
\ifappendix
\autoref{sec:safety-preservation} provides a detailed account of safety
properties and how one can restrict the notion of compiler correctness if one
is only interested in them.
\fi
%


The compiler correctness theorem in CakeML is an instance of \cctilde
for the $\sim$ relation above.
We have also implemented two small compilers that are correct for this relation.
The full details can be found in the Coq development in the supplementary
materials. 
%
The first compiler (\coqhref{ResourceExhaustionExample.v}) goes from a simple
expression language (similar to the one in \autoref{sec:example-diff-values} but
without inputs)
to the same language except that execution is bounded by some amount of fuel: each execution
step consumes some amount of fuel and execution immediately halts when it
runs out of fuel. The compiler is the identity.

The second compiler (\coqhref{ResourceExhaustion/CompilerStackLimited.v}) is
more interesting: we proved this \cctilde instance
for a variant of a compiler from a
\textsc{while} language to a simple stack machine by Xavier Leroy~\cite{DeepSpecLeroy}.
We enriched the two languages with outputs and modified the semantics of the stack machine so that
it falls into an error state if the stack reaches a certain size.
The proof uses a standard
forward simulation modified to account for
failure.
\end{example}

We conclude this subsection by noting that the resource exhaustion relation and
the undefined behavior relation from the previous subsection can easily be combined.
Indeed, given a relation \(\sim_{\text{UB}}\) and a
relation \(\sim_{\text{RE}}\) defined as above on the same sets of traces,
we can build a new relation \(\sim\) that allows both refinement of undefined
behavior and resource exhaustion by taking their union:
\(\src{s} \sim \trg{t} \equiv \src{s} \sim_{\text{UB}} \trg{t} \vee \src{s} \sim_{\text{RE}} \trg{t}\).
A compiler that is \formatCompilers{CC^{\sim_{\text{UB}}}} or \formatCompilers{CC^{\sim_{\text{RE}}}} is
trivially \cctilde, though the converse is not true.

%% file: diff_values.tex
%

We now illustrate trace-relating compilation for a translation
mapping source-level booleans to target-level natural numbers. Given
the simplicity of this compiler, most of the details of the
formalization are deferred to the online appendix.

The source language is a pure, statically typed expression language whose
expressions \src{e} include naturals \src{n}, booleans \src{b}, conditionals,
arithmetic and relational operations, boolean inputs \src{in_b} and natural
inputs \src{in_n}.
A trace \src{s} is a list of inputs \src{is} paired with a
result \src{r}, which can be a natural, a boolean, or an
error. Well-typed programs never produce error (\coqhref{TypeRelationExample.v}).
Types $\src{ty}$ are either $\src{\TypeNat}$ (naturals) or $\src{\TypeBool}$
(booleans); typing is standard.
The source language has a standard big-step operational semantics
($\src{e \sem \pair{is,r}}$) which tells how an expression \src{e} generates a
trace \src{\pair{is,r}}.
%
%
%
%
The target language is analogous, except that it is untyped, only
has naturals \trg{n} and its only inputs are naturals \trg{in_n}.
%
%
%
%
The semantics of the target language is also given in big-step style. Since
we only have naturals and all expressions operate on them, no error
result is possible in the target.

%
The compiler is homomorphic, translating a source expression to the same
target expression; the only differences are natural numbers (and conditionals),
as noted below.
%
\begin{align*}
\cmp{\booltag true} =
	&\ \trg{\nattag 1}
		&
		\cmp{in_b} =
			&\ \trg{in_n}
				&
				\cmp{e_1 \leq e_2} =
					&\ \trg{\iflete{\cmp{e_1}}{\cmp{e_2}}{1}{0}}
\\
\cmp{\booltag false} =
		&\ \trg{\nattag 0}
		&
		\cmp{in_n} =
			&\ \trg{in_n}
				&
				\cmp{\ifte{e_1}{e_2}{e_3}} =
					&\ \trg{\iflete{\cmp{e_1}}{0}{\cmp{e_3}}{\cmp{e_2}}}
\end{align*}
%
When compiling an \emph{if-then-else} the
target condition \src{\cmp{e_1}}~\trgb{\leq}~\trg{0} is used to check
that \src{e_1} is false, and therefore the \emph{then} and \emph{else}
branches of the source are swapped in the target.

\myparagraph{Relating Traces}
We relate basic values (naturals and booleans) in a non-injective fashion as
noted below. Then, we extend the relation to lists of inputs pointwise
(\Cref{tr:ste-em,tr:ste-con}) and lift that relation to traces
(\Cref{tr:ste-n,tr:ste-b}).
\begin{align*}
\src{n} \sim&\ \trg{n}
&
\src{true} \sim&\ \trg{n} \quad\text{if $\trg{n} > 0$}
&
\src{false} \sim&\ \trg{0}
\end{align*}
\vspace{-1.5em}
%
\begin{center}
	\typerule{Empty}{
	}{
        \src{\emptylist} \sim \trg{\emptylist}
	}{ste-em}
	\typerule{Cons}{
        \src{i} \sim \trg{i} &
        \src{is} \sim \trg{is}
	}{
	\src{i\listconcat is} \sim \trg{i\listconcat is}
	}{ste-con}
	\vrule
	\typerule{Nat}{
        \src{is} \sim \trg{is}
        &
        \src{n} \sim \trg{n}
	}{
        \src{\pair{is, \nattag n}} \sim \trg{\pair{is, \nattag n}}
	}{ste-n}
	\typerule{Bool}{
        \src{is} \sim \trg{is} &
        \src{b} \sim \trg{n}
	}{
        \src{\pair{is, \booltag b}} \sim \trg{\pair{is, \nattag n}}
	}{ste-b}
\end{center}
%
%

\myparagraph{Property mappings}
The property mappings
$\tilde{\sigma}$ and $\tilde{\tau}$ induced by the trace relation $\sim$ defined
above capture the intuition behind encoding booleans as naturals:
\begin{itemize}[nosep]
\item the source-to-target mapping allows $\src{true}$ to be
  encoded by any non-zero number;
\item the target-to-source mapping requires that
  $\trg{0}$ be replaceable by \emph{both} $\src{0}$ and $\src{false}$.
\end{itemize}

\myparagraph{Compiler correctness}
With the relation above, the compiler is proven to satisfy \cctilde.
\begin{theorem}[\cmp{\cdot} is correct \coqhref{TypeRelationExampleInput.v}]\label{thm:inst-dv-cc}
	\cmp{\cdot} is \cctilde.
\end{theorem}

\myparagraph{Simulations with different traces}
The difficulty in proving \autoref{thm:inst-dv-cc} arises from the
trace-relating compilation setting:
For compilation chains that have the same source and target traces, it is 
customary to prove compiler correctness using a forward 
simulation (i.e., a simulation between source and target transition system); 
then, using determinacy~\cite{Engelfriet85, milner82} of the target language 
and input totality~\cite{ZakinthinosL97, FocardiG95} (aka receptiveness)
of the source, this forward simulation is flipped 
into a backward simulation (a simulation between target and source transition
system), as described by \citet{BeringerSDA14,Leroy09b}.
This flipping is useful because forward simulations are often much easier to
prove (by induction on the transitions of the source) than backward ones,
as it is the case here. 

We first give the main idea of the flipping proof, when the inputs are the same
in the source and the target~\cite{BeringerSDA14,Leroy09b}.
We only consider inputs, as it is the most interesting case, since
with determinacy, nondeterminism only occurs on inputs.
Given a forward simulation $\R$, and a target program $\WT$ that simulates 
a source program $\WS$, $\WT$ is able to perform an input iff  
so is $\WS$: otherwise, say for instance that $\WS$ performs an output, 
by forward simulation $\WT$ would also perform an output, which is 
impossible because of determinacy. By input totality of the source, 
$\WS$ must be able to perform the exact same input as $\WT$; 
 using forward simulation and determinacy, the resulting programs 
 must be related.

\centerline{
\xymatrix @R=3ex{
 \src{\WS}\ar@{-->}[dd]_{\src{i_1}}
 & = \ar@{-}[r]\ar@{-}[l] 
 & \WS \ar[dd]_{\src{i_2}} 
 & \ar@{-}[r]\ar@{-}[l]\R 
 & \WT \ar[dd]^{\trg{i_1}}
 \\
~~~~~ 
&
& ~~~~~ \ar@{=>}[ll]^{\txt{\tiny By input totality}}
&
& ~~~~~ \ar@{<=>}[ll]^{\txt{\tiny By contradiction,\\
                            \tiny using forward simulation\\
                            \tiny and determinacy}}
\\
\exists \src{{W_S}_1} \ar@/^-1.2pc/@{--}[rrrr]^\R_{\txt{\tiny By forward simulation and determinacy}} & &\cdot  & &
\trg{{W_T}_1}
}
}

However, our trace relation is not injective 
(both $\src{0}$ and \src{false} are
mapped to $\trg{0}$),  therefore these arguments do 
not apply: not all possible inputs of target programs are 
accounted for in the forward simulation. 
We thus have to strengthen the forward simulation assumption,
requiring the following additional property to hold, for any source program $\WS$
and target program $\WT$ related by the forward simulation $\R$. 


\begin{minipage}[t]{0.65\textwidth}
\centerline{
	\xymatrix{
		&\src {W_S} \ar[d]_{\src {{i_S}_1}}  \ar@{-->}[dl]_{\exists\src {{i_S}_2}}
			\ar@{-}[r] & \ar@{-}[r] \R 
				& \trg {W_T }\ar[d]^{\trg {{i_T}_1}} \ar[dr]^{\trg {{i_T}_2}}
		\\
		\exists \src{{W_S}_2} \ar@/^-1.0pc/@{--}[rrrr]_\R 
			&\src{{W_S}_1} 
				\ar@{-}[r] 
				&\ar@{-}[r] \R
					& \trg{{W_T}_1} 
						&  \trg{{W_T}_2}
	}}
\end{minipage}
\begin{minipage}[t]{0.25\textwidth}
\begin{align*}
\text{where } &\src{{i_S}_1}\sim\trg{{i_T}_1}\\
&\src{{i_S}_1}\sim\trg{{i_T}_2}\\
&\src{{i_S}_2}\sim\trg{{i_T}_2}
\end{align*}
\end{minipage}

We say that a forward simulation for which this property
holds is \emph{flippable}. 
For our example compiler, a flippable forward simulation works as follows:
 whenever a boolean input occurs in the source, the target 
program must perform every strictly positive input $\trg n$ (and not just $\trg 1$, 
as suggested by the compiler). 
Using this property, determinacy of the target, input totality of the 
source, as well as the fact that any target input has an inverse image  
through the relation, we can indeed show that the forward 
simulation can be turned into a backward one: starting from 
$\WS\RR\WT$ and an input $\trg{{i_T}_2}$, we show that 
there is $\src{{i_S}_1}$ and $\trg{{i_T}_2}$ as in the 
diagram above, using the same arguments as when the inputs are the same; 
because the simulation is flippable, we can close the diagram, and obtain
the existence of an adequate $\src{{i_S}_2}$. From this we obtain $\cctilde$.

In fact, we have proven a completely general `flipping theorem',
with this flippable hypothesis on the forward simulation 
(\coqhref{TypeRelationExampleInput.v}).
%
We have also shown that if the relation $\sim$ defines a bijection between the
inputs of the source and the target, then any forward simulation is flippable,
hence reobtaining the usual proof technique \cite{BeringerSDA14,Leroy09b} as a
special case.
This flipping theorem is further discussed in the online appendix.

%% file: instance-more-target-statements.tex

We now consider how to relate traces where a single source action is compiled to
multiple target ones.
To illustrate this, we take a pure, statically-typed source language that can
output (nested) pairs of arbitrary size, and a pure, \emph{untyped} target language
where sent values have a fixed size.
%
%
Concretely, the source is analogous to the language
of \autoref{sec:example-diff-values}, except that it does not have inputs or
booleans and it has an expression \sends{e},
which can emit a (nested) pair \src{e} of
values in a single action. That is, given that \src{e} reduces to a pair,
\EG \src{\pair{v1,\pair{v2,v3}}},
expression \sends{\src{\pair{v1,\pair{v2,v3}}}} emits
action \src{\pair{v1,\pair{v2,v3}}}.
%
%
That expression is compiled into a sequence of individual sends in the target
language \mbox{\sendt{v1} \trg{;}} \sendt{v2} \trg{;} \sendt{v3}, since
in the target, \sendt{e} sends the value that \trg{e} reduces to, but
the language has no pairs.
%

Due to space constraints we omit the full formalization of these simple languages
and of the homomorphic compiler ($\compm{\cdot} : \src{e}\to\trg{e}$).
The only interesting bit is the compilation of the \sends{\cdot} expression, which
relies on the \gensends{\cdot} function below.
That function takes a source expression of a given type and returns a sequence
of target \sendt{\cdot} instructions that send each element of the expression.
%
\begin{align*}
	\gensends{\vdash\src{e:\tau}} =
		\begin{cases}
			\trg{\sendt{{\compm{\vdash\src{{e}}:N }}}} 
			& \ift \src{\tau} = \src{N}
			\\
			\trg{
				\gensends{\vdash\src{\projone{e}:\tau'}};
				\gensends{\vdash\src{\projtwo{e}:\tau''}}
			}
			& \ift \src{\tau} = \src{\tau'\times\tau''}
		\end{cases}
\end{align*}

\myparagraph{Relating Traces}
We start with the trivial relation between numbers: $\src{n}\simnat\trg{n}$, \IE
numbers are related when they are the same.
%
%
We cannot build a relation between single actions since a single source action is
related to multiple target ones. Therefore, we define a relation between a source
action \src{M} and a target trace \trg{t} (a list of numbers),
inductively on the structure
of \src{M} (which is a pair of values, and values are natural numbers or pairs).
%
\begin{center}\small
	\typerule{Trace-Rel-N-N}{
		\src{n}\simnat\trg{n}
		&
		\src{n'}\simnat\trg{n'}
	}{
		\src{\pair{n,n'}} \simm\trg{n\listconcat n'}
	}{tr-rel-n}
	\typerule{Trace-Rel-N-M}{
		\src{n}\simnat\trg{n}
		&
		\src{M}\simm\trg{t}
	}{
		\src{\pair{n,M}}\simm\trg{n\listconcat t}
	}{tr-rel-n-m}
	\typerule{Trace-Rel-M-N}{
		\src{M}\simm\trg{t}
		&
		\src{n}\simnat\trg{n}
	}{
		\src{\pair{M,n}} \simm\trg{t\listconcat n}
	}{tr-rel-m-n}
	\typerule{Trace-Rel-M-M}{
		\src{M}\simm\trg{t}
		&
		\src{M'}\simm\trg{t'}
	}{
		\src{\pair{M,M'}} \simm\trg{t\listconcat t'}
	}{tr-rel-m-m}
\end{center}

A pair of naturals is related to the two actions that send each element of the
pair (\Cref{tr:tr-rel-n}). 
If a pair is made of sub-pairs, we require all such sub-pairs to be related
(\Cref{tr:tr-rel-n-m,tr:tr-rel-m-n,tr:tr-rel-m-m}).
We build on these rules to define the
\begin{wrapfigure}{r}{.25\textwidth}

		\typerule{Trace-Rel-Single}{
			\src{s}\simq\trg{t}
			&
			\src{M}\simm\trg{t'}
		}{
			\src{s\listconcat M}\simq \trg{t\listconcat t'}
		}{tr-rel-full}

		\phantom{a}
\end{wrapfigure}
$\src{s}\sim\trg{t}$ relation between source and target traces
for which the compiler is correct (\autoref{thm:inst-marco-cc}).
%
Trivially, traces are related when they are both empty. 
Alternatively, given related traces, we can concatenate
a source action and a second target trace provided that they are related
(\Cref{tr:tr-rel-full}).
%
%
%


\begin{theorem}[\compm{\cdot} is correct]\label{thm:inst-marco-cc}
\compm{\cdot} is \cctilde.
\end{theorem}

With our trace relation, the trace property mappings capture the following
intuitions:
\begin{itemize}[nosep]
	\item The target-to-source mapping states that a source property can
  reconstruct target action as it sees fit. For example, trace
  \trg{{4}\listconcat{6}\listconcat{5}\listconcat{7}} is related to
  \src{{\pair{4,6}}\listconcat{\pair{5,7}}} and
  \src{{\pair{\pair{4,\pair{6,\pair{5,7}}}}}} (and many more variations).
	This gives freedom to the source implementation of a target behavior, which
  follows from the non-injectivity of $\sim$.%
	\footnote{%
		Making $\sim$ injective is a matter of adding open and close parenthesis
    actions in target traces.
	}
      \item The source-to-target mapping ``forgets'' about the way pairs
        are nested, but is faithful w.r.t. the values $\src{v_i}$
        contained in a message.
%
        Notice that source safety properties are always mapped to
        target safety properties. For instance, if $\piS{\pi} \in \hS{Safety}$
        prescribes that some bad number is never sent, then
        $\tilde{\tau}(\piS{\pi})$ prescribes the same number
        is never sent in the target and $\tilde{\tau}(\piS{\pi}) \in \hT{Safety}$.
        Of course if
        $\piS{\pi} \in \hS{Safety}$ prescribes that a particular nested pairing
        like $\src{{\pair{4,\pair{6,\pair{5,7}}}}}$ never
        happens, then $\tilde{\tau}(\piS{\pi})$ is still a target safety
        property, but the trivial one, since
        $\tilde{\tau}(\piS{\pi}) = \trg{\top} \in \hT{Safety}$.
\end{itemize}

%% file: ANI.tex
%

When source and target observations are drawn from the same set, a
correct compiler (\ccequal) is enough to ensure the preservation of
all subset-closed
hyperproperties, in particular of
\emph{noninterference} (NI)~\cite{GoguenM82},
as also mentioned at the beginning of \autoref{sec:subset-closed}.
%
%
In the scenario where target observations are strictly more informative
than source observations, the best guarantee one may expect from a correct trace-relating compiler
(\cctilde) is a \emph{weakening} (or \emph{declassification}) of target
noninterference that matches the noninterference property satisfied in the source.
%
%
%
%
To formalize this reasoning, this section applies the trinitarian view of
trace-relating compilation to the general framework of abstract noninterference
(ANI)~\cite{giacobazzi2018abstract}.
%
%

We first define NI and explain the issue of preserving source NI via a
\cctilde compiler. We then introduce ANI, which allows
characterizations of various forms of noninterference, and formulate a
general theory of ANI preservation via \cctilde. We also study how to
deal with cases such as undefined behavior in the target. Finally, we
answer the dual question, \IE~which source NI should be satisfied to
guarantee that compiled programs are noninterfering with respect to
target observers.
%
%
\smallskip

Intuitively, NI requires that publicly observable outputs do not
reveal information about private inputs.  To define this formally, we
need a few additions to our setup.  We indicate the (disjoint)
\emph{input} and \emph{output} projections of a trace $t$ as $t\inp$
and $t\out$ respectively\footnote{Here we only require the projections
  to be disjoint. Depending on the scenario and the attacker model the
  projections might record information such as the ordering of
  events.}.
%
Denote with $\lowcls{t}$ the equivalence class of a trace $t$,
obtained using a standard low-equivalence relation that relates low
(public) events only if they are equal, and ingores any difference
between private events.
Then, NI for source traces can be defined as:
%
\begin{align*}
    \hS{NI} = \myset{\piS{\pi}}{\forall \tS{s_1} \tS{s_2} \in \piS{\pi}. ~
    \lowcls{\tS{s\inp_1}} = \lowcls{\tS{s\inp_2}}
    \Rightarrow
    \lowcls{\tS{s\out_1}} = \lowcls{\tS{s\out_2}} ~}.
\end{align*}
That is, source NI comprises the sets of traces that have equivalent low output projections as long as their low input projections are equivalent.


\myparagraph{Trace-Relating Compilation and Noninterference}
When additional observations are possible in the target, it is
unclear whether a noninterfering source program is compiled to a noninterfering
target program or not, and if so, whether the notion of NI in the
target is the expected or desired one.
We illustrate this issue considering a scenario where target traces
extend source ones by exposing the execution time.
While source noninterference $\hS{NI}$ requires
that private inputs do not affect public outputs, $\hT{NI}$
additionally requires that the execution time is not affected by
private inputs. 

To model the scenario described, let $\src{Trace_S}$ denote the set of
traces in the source, and
$\trg{Trace_T} = \src{Trace_S} \times
\trg{\mathbb{N}^{\trgb{\omega}}}$
be the set of target traces, where
$\trg{\mathbb{N}^{\trgb{\omega}}} \triangleq \trg{\mathbb{N} \cup \{
  \trgb{\omega} \}}$.
Target traces have two components: a source trace, and a natural
number that denotes the time spent to produce the trace
($\trgb{\omega}$ if infinite).
Notice that if two source traces $ \tS{s_1}, \tS{s_2} $, are
low-equivalent then $\{ \tS{s_1}, \tS{s_2} \} \in \hS{NI}$ and
$\{ (\tS{s_1}, \trg{42}), (\tS{s_1}, \trg{42}) \} \in \hT{NI}$, but
$\{ (\tS{s_1}, \trg{42}), (\tS{s_2}, \trg{43}) \} \not\in \hT{NI}$ and
$\{ (\tS{s_1}, \trg{42}), (\tS{s_2}, \trg{42}), (\tS{s_1}, \trg{43}), (\tS{s_2}, \trg{43}) \}
\not\in \hT{NI}$.

Consider the following straightforward trace relation, which relates a
source trace to any target trace whose first component is equal to it,
irrespective of execution time:
\[\tS{s} \sim \tT{t} \quad\equiv\quad \exists \trg{n}. ~\tT{t} = (\tS{s}, \trg{n}).\]
%
%
A compiler is \cctilde if any trace that can be exhibited in the
target can be simulated in the source in some amount of time.
%
%
%
%
%
For such a compiler \autoref{thm:ssch} says that if $\src{W}$
satisfies $\hS{NI}$, then $\cmp{W}$ satisfies
$\mathit{Cl_{\subseteq}} \circ \tilde{\tau}(\hS{NI})$, which however is
strictly weaker than \hT{NI}, as it contains, \EG
$\{ (\tS{s_1}, \trg{42}), (\tS{s_2}, \trg{42}), (\tS{s_1}, \trg{43}), (\tS{s_2}, \trg{43}) \}$, and one cannot conclude that $\cmp{W}$ is noninterfering in the target.
%
%
It is easy to prove that
\smallskip
\begin{align*}
  \mathit{Cl_{\subseteq}} \circ \tilde{\tau}(\hS{NI}) & = \ii{Cl_{\subseteq}} ~(\myset{ ~\piS{\pi} \times \trgb{\mathbb{N^{\omega}}}}{\piS{\pi} \in \hS{NI}}) 
                                                    = \myset{ ~\piS{\pi} \times \mathcal{I}} {\piS{\pi} \in \hS{NI} \wedge \mathcal{I} \subseteq \trg{\mathbb{N}^{\trgb{\omega}}}}, 
\end{align*}
the first equality coming from
$\tilde{\tau}(\piS{\pi}) = \piS{\pi} \times
\trgb{\mathbb{N^{\omega}}}$, and the second from $\hS{NI}$ being subset-closed.
%
%
As we will see, this hyperproperty {\em can} be characterized as a form of
NI, which one might call {\em timing-insensitive
noninterference}, and ensured only against attackers that cannot measure
execution time.
For this characterization, and to describe different forms of noninterference as well as formally analyze their preservation by a $\cctilde$ compiler, we rely on the general framework of {\em abstract noninterference}~\cite{giacobazzi2018abstract}.

\myparagraph{Abstract Noninterference}
%
ANI~\cite{giacobazzi2018abstract}
is a generalization of NI whose formulation relies on abstractions 
(in abstract interpretation sense~\cite{CousotCousot77-1}) in order
to encompass arbitrary variants of NI. ANI is parameterized by an {\em observer
abstraction} $\rho$, which denotes the distinguishing power of the attacker, and
a {\em selection abstraction} $\phi$, which specifies when to check NI, and therefore captures a form of
declassification~\cite{sabelfeld05:_dimensions_declass}.%
\footnote{ANI includes a third parameter $\eta$, which describes the maximal input variation that the
attacker may control. Here we omit $\eta$ (\IE take it to be the identity) in
order to simplify the presentation.} Formally:
%
  \begin{equation*}
   \ani{\phi}{\rho} = \myset{\pi}{\forall t_1 t_2 \in \pi. ~ \phi(t\inp_1) = \phi(t\inp_2) \Rightarrow \rho(t\out_1) = \rho(t\out_2) }.
  \end{equation*}
%
By picking $\phi = \rho = \lowcls{\cdot}$, we recover the standard noninterference
defined above, where NI must hold for all low inputs (\IE no
declassification of private inputs), and the observational power of the attacker
is limited to distinguishing low outputs.

The observational power of the attacker can be weakened by
choosing a more liberal relation for $\rho$. For instance, one may limit the
attacker to observe the {\em parity} of output integer values. Another way to
weaken ANI is to use $\phi$ to specify that noninterference is only required to
hold for a subset of low inputs.


To be formally precise, $\phi$ and $\rho$ are defined over sets of
(input and output projections of) traces, so when we write $\phi(t)$ above, this
should be understood as a convenience notation for $\phi(\{t\})$. Likewise,
$\phi = \lowcls{\cdot}$ should be understood as $\phi
= \lambda \pi. \bigcup_{t \in \pi} \lowcls{t}$, \IE~the powerset lifting of
$\lowcls{\cdot}$.
%
%
Additionally, $\phi$ and $\rho$ are required to be upper-closed operators
(\ii{uco})---\IE monotonic, idempotent and extensive%
%
%
---on the poset that is the powerset of (input and output projections of) traces ordered by
inclusion~\cite{giacobazzi2018abstract}.
\myparagraph{Trace-Relating Compilation and ANI for Timing}
We can now reformulate our example with observable execution times in the target
in terms of ANI. We have $\hS{NI} = \ani{\src{\rho_S}}{\src{\phi_S}}$ with
$\src{\phi_S} = \src{\rho_S} = \lowcls{\cdot}$. In this case, we can formally
describe the hyperproperty that a compiled program $\cmp{W}$ satisfies whenever
$\src{W}$ satisfies $\hS{NI}$ as an instance of ANI:
\begin{align*}
&\mathit{Cl_{\subseteq}} \circ \tilde{\tau}(\hS{NI}) = \ani{\trgb{\phi}_\trg{T}}{\trgb{\rho}_\trg{T}},
\\
 \text{ for } \trgb{\phi}_\trg{T} =&\ \src{\phi_S} \text{ and }
   \trgb{\rho}_\trg{T} (\piT{\pi}) = \myset{\trg{(\tS{s}, n)}}{
   \exists 
  \trg{(\tS{s_1}, n_1)} \in \piT{\pi} .\; \lowcls{\src{s\out}} = \lowcls{\src{s\out_1}}}.
\end{align*}
%
%
%
%
%
%
%
%
%
The definition of $\trgb{\phi}_\trg{T}$ tells us that the trace relation does not affect
the selection abstraction. The definition of $\trgb{\rho}_\trg{T}$ characterizes
an observer that cannot distinguish execution times for noninterfering traces
(notice that $\trg{n_1}$ in the definition of $\trgb{\rho}_\trg{T}$ is discarded). For
instance,
$\trgb{\rho}_\trg{T}(\{ \trg{(\tS{s}, n_1)} \}) = \trgb{\rho}_\trg{T}(\{ \trg{(\tS{s}, n_2)} \})$,
for any $\tS{s}$, $\trg{n_1}$, $\trg{n_2}$. Therefore, in this setting,
we know explicitly through $\trgb{\rho}_\trg{T}$
that a $\cctilde$ compiler degrades source noninterference to target
{\em timing-insensitive} noninterference.



\myparagraph{Trace-Relating Compilation and ANI in General}
While the particular $\trgb{\phi}_\trg{T}$ and $\trgb{\rho}_\trg{T}$ above can be
discovered by intuition, we want to know whether there is a systematic
way of obtaining them in general.
%
%
In other words, for {\em any} trace relation $\sim$ and {\em any} notion of
source NI, what property is guaranteed on noninterfering source programs by any
$\cctilde$ compiler?

We can now answer this question generally (\autoref{thm:compilingANI}): any source notion of
noninterference expressible as an instance of ANI is mapped to a corresponding
instance of ANI in the target, whenever source traces are an
abstraction of target ones (\IE when $\sim$ is a total and surjective map). 
For this result we consider trace relations that can be split into input and
output trace relations (denoted as 
\mbox{$\sim\; \triangleq \langle \simin, \simout \rangle$}) such that
%
  $\tS{s} \sim \tT{t} \iff \tS{s\inp} \simin \tT{t\inp} \wedge \tS{s\out}
     \simout \tT{t\out}$.
  %
  The trace relation $\sim$ corresponds to a Galois connection between the
  sets of trace properties
  $\tilde{\tau} \leftrightarrows \tilde{\sigma}$ as described in
  \autoref{sec:trinity}. Similarly, the pair $\simin$ and $\simout$ corresponds to
  a pair of Galois connections,
  $\tilde{\tau}\inp \leftrightarrows \tilde{\sigma}\inp$ and
  $\tilde{\tau}\out \leftrightarrows \tilde{\sigma}\out$, between the sets
  of input and output properties. \ca{In general is not true that $\tilde{\tau}$
  is (reconducible to) $\tilde{\tau\inp} \cap \tilde{\tau\out}$}%
In the timing example, time is an output so
we have $\sim\; \triangleq \langle = , \simout \rangle$ and $\simout$ is defined as 
$\tS{s\out} \simout \tT{t\out} \equiv \exists \trg{n}. ~\tT{t\out} = \trg{(\tS{s\out}, n)}$.
\begin{theorem}[Compiling ANI] \label{thm:compilingANI} Assume traces
  of source and target languages are related via
  $\sim\; \subseteq {\src{Trace_S}} \times {\trg{Trace_T}}$,
  $\sim\;\triangleq \langle \simin , \simout \rangle$ such that
  $\simin$ and $\simout$ are both total maps from target to source
  traces, and $\simin$ is surjective.
  Assume $\downarrow$ is a $\cctilde$ compiler, and
  $\src{\phi_S} \in \uco{2^{\piS{Trace\inp}}}, ~\src{\rho_S} \in \uco{2^{\piS{Trace\out}}}$. \\
  If $\src{W}$ satisfies \ani{\src{\phi_S}}{\src{\rho_S}}, then $\cmp{W}$
  satisfies \ani{\trgb{\phi^{\#}_\trg{T}}}{\trgb{\rho^{\#}_\trg{T}}}, where
  $\trgb{\phi^{\#}_\trg{T}}$ and $\trgb{\rho^{\#}_\trg{T}}$ are defined as:
  \begin{align*}
    \trgb{\phi^{\#}_\trg{T}} &= g\inp \circ \src{\phi_S} \circ f\inp;
    &
    \trgb{\rho^{\#}_\trg{T}} &= g\out \circ \src{\rho_S} \circ f\out \quad\text{and}
    \\
    f\inp (\piT{\pi\inp}) &= \myset{\tS{s\inp}}{\exists \tT{t\inp} \in \piT{\pi\inp}. ~ \tS{s\inp} \simin \tT{t\inp}};
    &
    g\inp (\piS{\pi\inp}) &= \myset{\tT{t\inp}}{\forall \tS{s\inp}. ~ \tS{s\inp} \simin \tT{t\inp} \Rightarrow \tS{s\inp} \in \piS{\pi\inp}}
  \end{align*}
(and both $f\out$ and $g\out$ are defined analogously).
\end{theorem}
For the example above we recover the definitions we
justified intuitively, \IE  
$\trgb{\phi^{\#}_\trg{T}} = g\inp \circ \src{\phi_S} \circ f\inp = \trgb{\phi}_\trg{T}$ and
$\trgb{\rho^{\#}_\trg{T}} = g\out \circ \src{\rho_S} \circ f\out = \trgb{\rho}_\trg{T}$.
%
Moreover, we can prove that if $\simout$ also is surjective,
$\ani{\trgb{\phi^{\#}_\trg{T}}}{\trgb{\rho^{\#}_\trg{T}}} \subseteq
\ii{Cl}_{\subseteq} \circ \tilde{\tau}(\ani{\src{\phi_S}}{\src{\rho_S}})$.
Therefore, the derived guarantee
$\ani{\trgb{\phi^{\#}_\trg{T}}}{\trgb{\rho^{\#}_\trg{T}}}$ is at least as strong as
the one that follows by just knowing that the compiler $\downarrow$ is
\cctilde.

%
%
%
%
%
\myparagraph{Noninterference and Undefined Behavior} As stated above,
\Cref{thm:compilingANI} does not apply to several scenarios from
\autoref{sec:instances} such as undefined behavior
(\autoref{sec:example-undef}), as in those cases the relation
$\simout$ is not a total map. Nevertheless, we can still exploit our
framework to reason about the impact of compilation on
noninterference.

Let us consider $\sim\; \triangleq \langle \simin , \simout \rangle$
where $\simin$ is any total and surjective map from target to source
inputs (\EG~equality) and $\simout$ is defined as
$\tS{s\out} \simout \tT{t\out} \equiv \tS{s\out} = \tT{t\out} \vee
\exists m\out \leq \tT{t\out}.~ \tS{s\out} =
{m\out}\cdot\mathit{\eundef}$.
Intuitively, a \cctilde compiler guarantees that no interference
can be observed by a target attacker that cannot exploit undefined
behavior to learn private information.
\ca{Writing down a concrete
  $\trgb{\rho^{\#}_\trg{T}}$ for CompCert ub would take a lot of space and
  produce a lot of unreadable symbols...}%
This intuition can be made formal by the following theorem.
%
%
\begin{theorem}[Relaxed Compiling ANI]\label{thm:genANI}
Relax the assumptions of \Cref{thm:compilingANI} by allowing $\simout$ to be
{\em any} output trace relation.
  If $\src{W}$ satisfies
  \ani{\src{\phi_S}}{\src{\rho_S}}, then $\cmp{W}$ satisfies
  \ani{\trgb{\phi^{\#}_\trg{T}}}{\trgb{\rho^{\#}_\trg{T}}} where
  $\trgb{\phi^{\#}_\trg{T}}$ is defined as in \Cref{thm:compilingANI}, and
  $\trgb{\rho^{\#}_\trg{T}}$ is such that:
  \begin{equation*}
    \forall \tS{s} ~\tT{t}. ~ \tS{s\out} \simout \tT{t\out} \Rightarrow ~
    \trgb{\rho^{\#}_\trg{T}}(\trg{\tT{t\out}}) = \trgb{\rho^{\#}_\trg{T}}(\tilde{\tau}\out(\src{\rho_S}(\tS{s\out)})).
  \end{equation*}
\end{theorem}
Technically, instead of giving us a {\em definition} of $\trgb{\rho^{\#}_\trg{T}}$, the
theorem gives a {\em property} of it. The property states that, given a target
output trace $\tT{t\out}$, the attacker cannot distinguish it from any other
target output traces produced by other possible compilations
($\tilde{\tau}\out$) of the source trace \tS{s} it relates to, up to the
observational power of the source level attacker~$\src{\rho_S}$.
Therefore, given a source attacker \src{\rho_S},
the theorem characterizes a {\em family} of attackers that cannot
observe any interference for a correctly compiled noninterfering
program.
%
%
%
\MP{And the intuition is?}%
Notice that the target attacker
$\trgb{\rho^{\#}_\trg{T}} = \lambda \_. ~ \trgb{\top}$ satisfies the premise
of the theorem, but defines a trivial hyperproperty, so that we cannot
prove in general that
$\ani{\trgb{\phi^{\#}_\trg{T}}}{\trgb{\rho^{\#}_\trg{T}}} \subseteq
\ii{Cl}_{\subseteq} \circ \tilde{\tau}(\ani{\src{\phi_S}}{\src{\rho_S}})$.
The same $\trgb{\rho^{\#}_\trg{T}} = \lambda \_. ~ \trgb{\top}$ shows
that the family of attackers described in \Cref{thm:genANI} is
nonempty, and this ensures the existence of a most powerful
attacker among them~\cite{giacobazzi2018abstract}, whose explicit characterization we
leave for future work.

\myparagraph{From Target NI to Source NI} 
We now explore the dual question: under what hypotheses does trace-relating compiler
correctness alone allow
target noninterference to be reduced to source noninterference?
%
%
%
%
This is of practical interest, as one would be able to
protect from target attackers by ensuring noninterference in the
source.  This task can be made easier if the source language has some
static enforcement mechanism~\cite{AbadiCCD, mastroeni2018verifying}.

%
%
%

Let us consider the languages from \autoref{sec:instance-marco} extended
with inputting of (pairs of) values.
It is easy to show that the compiler described in \autoref{sec:instance-marco} is still \cctilde.
\ca{the reasons should be obvious. To simulate a
  target trace $\trgb{t}$ you give to the source program the same
  inputs, \IE $\src{s\inp} = \trgb{t\inp}$, the compiler in
  \autoref{sec:instance-marco} ensures you can simulate the same
  outputs in the src, \IE $\src{s\out} \simout \trgb{t\out}$.}%
Assume that we want to satisfy a given notion of target noninterference after
compilation, \IE~$\cmp{W} {\trgb{\models}} \ani{\trgb{\phi}_\trg{T}}{\trgb{\rho}_\trg{T}}$. 
Recall that the observational power of the target attacker,
$\trgb{\rho}_\trg{T}$, is expressed as a property of sequences of values.  To
express the same property (or attacker) in the source, we have to
abstract the way pairs of values are nested. For instance, the source
attacker should not distinguish
$\src{\langle v_1, \langle v_2, v_3 \rangle \rangle}$ and
$\src{\langle \langle v_1, v_2 \rangle, v_3 \rangle}$. 
In general (\IE~ when $\simin$ is not the identity), this argument is
valid only when $\trgb{\phi}_\trg{T}$ can be represented in the source. More
precisely, $\trgb{\phi}_\trg{T}$ must consider as equivalent all target inputs
that are related to the same source one, because in the source it
is not possible to have a finer distinction of inputs.
%
This intuitive correspondence can be formalized as follows:
%
%
\begin{theorem}[Target ANI by source ANI] \label{thm:srcANI} Let
  $\trgb{\phi}_\trg{T} \in \uco{2^{\hT{Trace\inp}}}$,
  $\trgb{\rho}_\trg{T} \in \uco{2^{\hT{Trace\out}}}$ and $\simout$ a total and
  surjective map from source outputs to target ones and assume that
  \begin{equation*}
    \forall \tS{s} ~\tT{t}. ~ \tS{s\inp} \simin \tT{t\inp} \Rightarrow \trgb{\phi}_\trg{T}(\tT{t\inp}) = \trgb{\phi}_\trg{T}(\tilde{\tau}\inp(\tS{s\inp})).
  \end{equation*}
  If $\cmp{\cdot}$ is a \cctilde compiler and $\src{W}$
  satisfies $\ani{\src{\phi^{\#}_S}}{\src{\rho^{\#}_S}}$, then $\cmp{W}$
  satisfies \ani{\trgb{\phi}_\trg{T}}{\trgb{\rho}_\trg{T}} for
  \begin{align*}
   \src{\phi^{\#}_S} =&\ \tilde{\sigma}\inp \circ \trgb{\phi}_\trg{T} \circ \tilde{\tau}\inp;
   &  
   \src{\rho^{\#}_S} =&\ \tilde{\sigma}\out \circ \trgb{\rho}_\trg{T} \circ \tilde{\tau}\out.
  \end{align*}

\end{theorem}

To wrap up the discussion about noninterference, the results presented in this section formalize and generalize some intuitive facts about compiler correctness and noninterference. Of course, they all place some restrictions on the shape of the noninterference instances that can
be considered, because compiler correctness alone is in general not a strong enough criterion for dealing with many security properties~\cite{BartheGL18, DSilvaPS15}.



%% file: secure-compilation-motivation.tex
%
So far we have studied compiler correctness criteria 
for whole, standalone programs.
However, in practice, programs do not exist in isolation, but in a
context where they interact with other programs, libraries, \ETC
%
In many cases, this context cannot be assumed to be benign and
could instead behave maliciously to try to disrupt a compiled program.

Hence, in this section we consider the following {\em secure compilation}
scenario: a source program is compiled and linked with an arbitrary
target-level context, \IE one that may not be expressible as the compilation of
a source context.
Compiler correctness does not address this case, as it does not consider
arbitrary target contexts,
%
looking instead at whole programs (empty context~\cite{Leroy09})
or
well-behaved target contexts that behave like source ones (as in 
compositional compiler
correctness~\cite{NeisHKMDV15,StewartBCA15,HurD11,KangKHDV15}).
%

To account for this scenario, \citet{AbateBGHPT19} describe several secure
compilation criteria based on the preservation of classes of (hyper)properties (e.g., trace properties, safety, hypersafety, hyperproperties, etc.) against
arbitrary target contexts.
For each of these criteria, they give an equivalent ``property-free'' criterion,
analogous to the equivalence between \(\traceP\) and \(\ccbasic^=\).
For instance, their {\em robust} trace property preservation criterion (\rtp)
states that, for any trace property \(\pi\), if a source \emph{partial} program \(\src{P}\) plugged into any
context \(\src{C_S}\) satisfies $\pi$, then the compiled
program \(\cmp{P}\) plugged into any target context \(\trg{C_T}\) satisfies
$\pi$.
Their equivalent criterion to \rtp is \pf{\rtp}, which states that for any trace
produced by the compiled program, when linked with any target context, there is a
source context that produces the same trace. Formally (writing \(C\hole{P}\) to
mean the whole program that results from linking partial program \(P\) with
context \(C\)) they define:
\begin{align*}
  \rtp\equiv&\ \forall \src{P}.~ \forall \pi.~ (\forall \src{C_S}. ~\forall t . \src{C_S\hole{P}} \src{\sem} t \Rightarrow t\in\pi) \Rightarrow
                       (\forall \trg{C_T}. ~\forall \trg{t}.~ \trg{C_T\hole{\cmp{P}}} \trg{\sem} t \Rightarrow t\in\pi);
  \\
  \pf{\rtp} \equiv&\ \forall \src{P} . ~\forall \trg{C_T} . \forall t . \trg{C_T\hole{\cmp{P}}} \trg{\sem} t \Rightarrow
  \exists \src{C_S} .  ~\src{C_S\hole{P}} \src{\sem} t.
\end{align*}
In the following we adopt the notation $P \models_R \pi$ to mean
``$P$ \emph{robustly} satisfies $\pi$,''
\IE $P$ satisfies $\pi$ irrespective of the contexts it is linked with.
Thus, we write more compactly:
\begin{align*}
\rtp \equiv \forall \pi. ~\forall \src{P}. ~ \src{P}~\src{\models_R} \pi \Rightarrow \cmp{P} ~\trg{\trgb{\models}_R} \pi.
\end{align*}

%
All the criteria of \citet{AbateBGHPT19}
share this flavor of stating the existence of some source context that
simulates the behavior of any given target context, with some variations
depending on the class of (hyper)properties under consideration.
All these criteria are
stated in a setting where source and target traces are the same.
In this section, we extend their result to our trace-relating
setting, obtaining 
trintarian views for secure compilation. Despite the similarities with
\autoref{sec:compiler-correctness}, more challenges show up, in
particular when considering the robust preservation of proper
sub-classes of trace properties. For example, after 
application of $\tilde{\sigma}$ or $\tilde{\tau}$, a property may not be
safety anymore, a crucial point for the
equivalence with the property-free criterion for safety properties by
\citet{AbateBGHPT19}.
We solve this by interpreting the class of safety
properties as an \emph{abstraction} of the class of all trace
properties induced by a closure operator (\autoref{sec:sec-comp-trini}).
%
%
The remaining subsections provide example compilation chains satisfying our
trace-relating secure compilation criteria for trace properties
(\autoref{sec:sec-comp-traces}) and for safety properties
hypersafety~(\autoref{sec:comp-summ}).

%% file: sec-comp-explain.tex

In this subsection we generalize many of the criteria of \citet{AbateBGHPT19} using
the ideas of \autoref{sec:compiler-correctness}.
%
%
Before discussing how we solve the challenges for classes such as
safety and hypersafety, we show the simple generalization of
$\pf{\rtp}$ to the trace-relating setting (\rtctilde) and its corresponding
trinitarian view (\Cref{thm:rtc-trinity}):
%
%


 


\begin{theorem}[Trinity for Robust Trace Properties \coqhref{RobustTraceCriterion.v}] \label{thm:rtc-trinity}
For any trace relation $\sim$ and induced property mappings
 $\tilde{\tau}$ and $\tilde{\sigma}$, we have: 
    $\rtptautilde \iff \rtctilde \iff \rtpsigmatilde$, where
   \begin{align*}
     \rtctilde &\equiv \forall \src{P} ~\forall\trg{C_T} ~\forall\tT{t}. ~
   \trg{C_T\hole{\cmp{P}}} \trg{\sem} \tT{t} \Rightarrow 
   \exists \src{C_S} ~\exists \tS{s} \sim \tT{t}.~
      \src{C_S\hole{P}} \src{\sem} \tS{s};\\
     \rtptautilde & \equiv \forall \src{P}~ \forall \piS{\pi} \in \propS.~ \src{P} ~\src{\models_{R}}~ \piS{\pi} \Rightarrow \cmp{P} ~\trg{\trgb{\models}_R} ~ \tilde{\tau}(\piS{\pi});\\
    \rtpsigmatilde & \equiv \forall \src{P}~ \forall \piT{\pi} \in \propT.~
     \src{P} ~\src{\models_R}~ \tilde{\sigma}(\piT{\pi}) \Rightarrow \cmp{P} ~\trg{\trgb{\models}_R}~ \piT{\pi}.
  \end{align*} 
\end{theorem}

%% file: expand-sec-diagram.tex
\citet{AbateBGHPT19} propose many more equivalent pairs of criteria,
each preserving different classes of (hyper)properties, which we
briefly recap now.  For trace properties, they also have criteria
that preserve
safety properties plus their version of liveness properties.  For
hyperproperties, they have criteria that preserve hypersafety
properties, subset-closed hyperproperties, and arbitrary
hyperproperties.  Finally, they define \emph{relational}
hyperproperties, which are relations between the behaviors
of multiple programs for
expressing, \EG that a program always runs faster than another.
For relational hyperproperties, they have criteria that preserve arbitrary
relational properties, relational safety properties, relational hyperproperties
and relational subset-closed hyperproperties.  Roughly speaking, the
security guarantees due to robust preservation of trace properties
regard only protecting the integrity of the program from the context,
the guarantees of hyperproperties also regard data confidentiality,
and the guarantees of relational hyperproperties
even regard code confidentiality.
Naturally, these stronger guarantees are increasingly harder to enforce and prove.

%

While we have lifted the most significant criteria from \citet{AbateBGHPT19}
to our trinitarian view, due to space constraints we provide the
formal definitions only for the two most interesting criteria.
We summarize the generalizations of many other criteria in
\autoref{fig:diagram-preserve-props-sec}, described at the end.
Omitted definitions are available in the online appendix.

%
%
%
\myparagraph{Beyond Trace Properties: Robust Safety and Hyperproperty Preservation}
We detail robust preservation of safety properties and of arbitrary
hyperproperties since they are both relevant from a security point of
view and their generalization is interesting.
\begin{theorem}[Trinity for Robust Safety Properties \coqhref{RobustSafetyCriterion.v}] \label{thm:rsp-sec-trinity}
For any trace relation $\sim$ and for the induced property mappings $\tilde{\tau}$ and $\tilde{\sigma}$, we have: 
\begin{align*}
  &\rsptautilde \iff \rsctilde \iff \rspsigmatilde,  &\text{ where }
\end{align*}
\begin{align*}
   \rsctilde &\equiv \forall \src{P} ~\forall\trg{C_T} ~\forall\tT{t} ~\forall \tT{m} \leq \tT{t}. 
   \trg{C_T\hole{\cmp{P}}} \trg{\sem} \tT{t} \Rightarrow 
   \exists \src{C_S} ~\exists \tT{t'} \geq \tT{m} ~\exists  \tS{s} \sim \tT{t'}. ~ \src{C_S\hole{P}} \src{\sem} \tS{s};
  \\
  \rsptautilde & \equiv \forall \src{P} \forall \piS{\pi} \in \propS. 
    \src{P} ~\src{\models_{R}}~ \piS{\pi} \Rightarrow \cmp{P} ~\trg{\trgb{\models}_R} ~ (\mi{Safe}\circ\tilde{\tau})(\piS{\pi});
  \\
  \rspsigmatilde & \equiv \forall \src{P} \forall \piT{\pi} \in \hT{Safety}.
    \src{P} ~\src{\models_R}~ \tilde{\sigma}(\piT{\pi}) \Rightarrow \cmp{P} ~\trg{\trgb{\models}_R}~ \piT{\pi}.
\end{align*} 
\end{theorem}
There is an interesting asymmetry between the last two characterizations above,
which we explain now in more detail.
$\rspsigmatilde$ quantifies over target safety properties, while
$\rsptautilde$ quantifies over {\em arbitrary} source properties,
but 
imposes the composition of $\tilde{\tau}$ with $\ii{Safe}$, which maps
an arbitrary target property \piT{\pi} to the target safety property
that best over-approximates \piT{\pi}%
\footnote{$\ii{Safe}(\piT{\pi}) = \cap \myset{\piT{S}}{\piT{\pi}
    \subseteq \piT{S} \wedge \piT{S} \in \hT{Safety}}$
  is the topological closure in the topology of \citet{ClarksonS10},
  where safety properties coincide with the closed sets.} 
(an analogous \emph{closure} was needed for subset-closed
hyperproperties in \autoref{thm:ssch}).  More precisely, $\ii{Safe}$ is
a closure operator on target properties, with
$\hT{Safety} = \myset{\ii{Safe}(\piT{\pi})}{\piT{\pi} \in \propT}$.
%
The mappings
\begin{equation*}
 \ii{Safe} \circ \tilde{\tau} : \propS \leftrightarrows \hT{Safety} : \tilde{\sigma}  
\end{equation*}
determine a Galois connection between source trace properties and
target safety properties, and ensure the equivalence
$\rsptautilde \iff \rspsigmatilde$ (\coqhref{RobustSafetyPreservation.v}).
This argument generalizes to arbitrary closure operators on target
properties (\coqhref{UcoRobustPreservation.v}) and on hyperproperties,
as long as the corresponding class is a sub-class of subset-closed
hyperproperties, and explains all but one of the asymmetries in
\autoref{fig:diagram-preserve-props-sec}, the one that concerns the robust
preservation of arbitrary hyperproperties:

\begin{theorem}[Weak Trinity for Robust
  Hyperproperties \coqhref{RobustHyperCriterion.v}] \label{thm:rhp-sec-trinity} For a trace relation
  ${\sim} \subseteq \piS{Trace} \times \trg{Trace_T}$ and induced
  property mappings $\tilde{\sigma}$ and $\tilde{\tau}$, $\rhctilde$
  is equivalent to \rhptautilde;  moreover, if
  $\tilde{\tau} \leftrightarrows \tilde{\sigma}$ is a Galois insertion
  (\IE $\tilde{\tau} \circ \tilde{\sigma} = id$), $\rhctilde$ implies
  \rhpsigmatilde, while if
  $\tilde{\sigma} \leftrightarrows \tilde{\tau}$ is a Galois
  reflection (\IE $\tilde{\sigma} \circ \tilde{\tau} = id$),
  \rhpsigmatilde implies $\rhctilde$,
\begin{align*}
  \mbox{where }\rhctilde & \equiv\ 
    \forall \src{P} ~\forall\trg{C_T} ~\exists \src{C_S} ~\forall\tT{t}. 
        ~\trg{C_T\hole{\cmp{P}}} \trg{\sem} \tT{t} \iff 
           (\exists \tS{s} \sim \tT{t}. ~\src{C_S\hole{P}} \src{\sem} \tS{s});
                     \\ 
  \rhptautilde & \equiv \forall \src{P}~ \forall \hS{H}.~
    \src{P} ~\src{\models_{R}}~ \hS{H} \Rightarrow \cmp{P} ~\trg{\trgb{\models}_R} ~ \tilde{\tau}(\hS{H});
  \\
  \rhpsigmatilde & \equiv \forall \src{P}~ \forall \hT{H}.~
    \src{P} ~\src{\models_R}~ \tilde{\sigma}(\hT{H}) \Rightarrow \cmp{P} ~\trg{\trgb{\models}_R}~ \hT{H}.
\end{align*} 
\end{theorem}

This trinity is \emph{weak} since extra hypotheses are needed to prove
some implications. While the equivalence $\rhctilde \iff \rhptautilde$
holds unconditionally, the other two implications hold only under
distinct, stronger assumptions.
%
For $\rhpsigmatilde$ it is still possible and correct to deduce a source
obligation for a given target hyperproperty $\hT{H}$ when no
information is lost in the the composition
$\tilde{\tau} \circ \tilde{\sigma}$ (\IE the two maps are a Galois
\emph{insertion}).
On the other hand, $\rhptautilde$ is a consequence of $\rhpsigmatilde$
when no information is lost in composing in the other direction,
$\tilde{\sigma} \circ \tilde{\tau}$ (\IE the two maps are a Galois
\emph{reflection}). \ca{maybe it is possible to improve a bit this
  intuition, it is not clear why we don't have to lose information!}
\ch{Where are the missing implications?}

\myparagraph{Navigating the Diagram}
For a given trace relation $\sim$,
\autoref{fig:diagram-preserve-props-sec} orders the generalized
criteria according to their relative strength.  If a trinity implies
another (denoted by $\Rightarrow$), then the former provides stronger
security for a compilation chain than the latter.

As mentioned, some property-full criteria regarding proper subclasses
(i.e., subset-closed hyperproperties, safety, hypersafety,
2-relational safety and 2-relational hyperproperties) quantify over
arbitrary (relational) (hyper)properties and compose $\tilde{\tau}$
with an additional operator.
We have already presented the \mi{Safe} operator; other operators are
\mi{Cl_\subseteq}, \mi{HSafe}, and \mi{2rSafe}, which approximate the
image of $\tilde{\tau}$ with a subset-closed hyperproperty, a
hypersafety and 2-relational safety respectively.

%
As a reading aid, when quantifying over arbitrary trace properties we
use the shaded blue as background color, we use the red when
quantifying over arbitrary subset-closed hyperproperties and green 
for arbitrary 2-relational properties.

We now describe how to interpret the acronyms in
\Cref{fig:diagram-preserve-props-sec}. All criteria start with \ms{R}
meaning they refer to robust preservation.  Criteria for relational
hyperproperties---here only arity 2 is shown
---contain \ms{2r}.
Next, criteria names spell the class of hyperproperties they preserve:
\ms{H} for hyperproperties, \ms{SCH} for subset-closed
hyperproperties, \ms{HS} for hypersafety, \ms{T} for trace
properties, and \ms{S} for safety properties.  Finally,
property-free criteria end with a $\ms{C}$ while property-full
ones involving $\tilde{\sigma}$ and $\tilde{\tau}$ end with
$\ms{P}$.  Thus, \emph{robust (\ms{R}) subset-closed
  hyperproperty-preserving (\ms{SCH}) compilation (\ms{C})} is
\rschptilde, \emph{robust (\ms{R}) two-relational (\ms{2r})
  safety-preserving (\ms{S}) compilation (\ms{C})} is \rtworelsctilde,
etc.

\input{diagram-prop-pres-comp-security}

%% file: diagram-prop-pres-comp-security.tex

\begin{figure}[!t]
  \begin{center}
    \begin{tikzpicture}[every node/.style={align=center,font=\footnotesize}]
      \node[](rtp){
        \rtctilde
      };
      \node[below right=.2 and .01 of rtp](rtptau){
        \rtptautilde
      };
      \node[below left=.2 and .01 of rtp](rtpsigma){
        \rtpsigmatilde
      };

      \draw[myiff] (rtp.south east) to (rtptau.90);
      \draw[myiff] (rtp.south west) to (rtpsigma.north);
      \draw[myiff] (rtptau.180) to (rtpsigma.0);
      

       \node[below right =.7 and 3 of rtp](rsc){
         \rsctilde
      };
      \node[below right=.2 and .01 of rsc,](rsctau){
        \rsptautilde
      };
      \node[below left=.2 and .01 of rsc](rscsigma){
        \rspsigmatilde
      };

      \draw[myiff] (rsc.south east) to (rsctau.90);
      \draw[myiff] (rsc.south west) to (rscsigma.north);
      \draw[myiff] (rsctau.180) to (rscsigma.0);


      \node[above = 1 of rtp](rschp){
        \rschptilde
      };
      \node[below right=.2 and -.5 of rschp](rschptau){
        \rschptautilde
      };
      \node[below left=.2 and -.5 of rschp](rschpsigma){
        \rschpsigmatilde
      };

      \draw[myiff] (rschp.south east) to (rschptau.90);
      \draw[myiff]
      (rschp.south west) to (rschpsigma.north);
      \draw[myiff] 
      (rschptau.180) to (rschpsigma.0);


      \node[above = 1 of rschp](rhp){
        \rhctilde
      };
      \node[below right=.2 and .01 of rhp](rhptau){
        \rhptautilde
      };
      \node[below left=.2 and .01 of rhp](rhpsigma){
        \rhpsigmatilde
      };

      \draw[myiff] (rhp.south east) to (rhptau.90);
      \draw[myimpl] (rhp.south west)
      to node[midway,sloped,above] (hpi) {\scalebox{.8}{Ins.}} (rhpsigma.north);
      \draw[myimpl] (rhpsigma.0) 
      to node[midway,sloped,above] (hpi) {\scalebox{.8}{Refl.}} (rhptau.180);


      \node[above = 1 of rsc](rhsp){
        \rhsctilde
      };
      \node[below right=.2 and -.3 of rhsp](rhsptau){
        \rhsptautilde
      };
      \node[below left=.2 and -.3 of rhsp](rhspsigma){
        \rhspsigmatilde
      };

      \draw[myiff] (rhsp.south east) to (rhsptau.90);
      \draw[myiff] (rhsp.south west) to (rhspsigma.north);
      \draw[myiff] (rhsptau.180) to (rhspsigma.0);


      \node[above = 1.3 of rhsp](rrtp){
        \rtworeltctilde
      };
      \node[below right=.2 and -.3 of rrtp](rrtptau){
        \rtworeltptautilde
      };
      \node[below left=.2 and -.3 of rrtp](rrtpsigma){
        \rtworeltpsigmatilde
      };

      \draw[myiff] (rrtp.south east) to (rrtptau.90);
      \draw[myiff] (rrtp.south west) to (rrtpsigma.north);
      \draw[myiff] (rrtptau.180) to (rrtpsigma.0);


      \node[above right= .8 and 1 of rrtp](r2scp){
        \rtworelschpctilde
      };
      \node[below right=.2 and -.5 of r2scp](r2scptau){
        \rtworelschptautilde
      };
      \node[below left=.2 and -.5 of r2scp](r2scpsigma){
        \rtworelschpsigmatilde
      };

      \draw[myiff] (r2scp.south east) to (r2scptau.90);
      \draw[myiff] (r2scp.south west) to (r2scpsigma.north);
      \draw[myiff] (r2scptau.180) to (r2scpsigma.0);


      \node[above right= .5 and 1.7 of rhsp](rrsp){
        \rtworelsctilde
      };
      \node[below right=.2 and -.3 of rrsp](rrsptau){
        \rtworelsptautilde
      };
      \node[below left=.2 and -.3 of rrsp](rrspsigma){
        \rtworelspsigmatilde
      };

      \draw[myiff] (rrsp.south east) to (rrsptau.90);
      \draw[myiff] (rrsp.south west) to (rrspsigma.north);
      \draw[myiff] (rrsptau.180) to (rrspsigma.0);


      \draw[rounded corners, thick, opacity=0.2, dotted]
      (rsc.north east) -- (rsctau.north east) |- (rsctau.south) -- (rscsigma.south) -| (rscsigma.north west) -- (rsc.north west) -- (rsc.north east);

      \draw[rounded corners, thick, opacity=0.2, dotted, fill=\tcol!20!white]
      (rtp.north east) -- (rtptau.north east) |- (rtptau.south) -- (rtpsigma.south) -| (rtpsigma.north west) -- (rtp.north west) -- (rtp.north east);

      \draw[rounded corners, thick, opacity=0.2, dotted, fill=\sccol!20!white]
      (rschp.north east) -- (rschptau.north east) |- (rschptau.south) -- node[midway](p1){} (rschpsigma.south) -| (rschpsigma.north west) -- (rschp.north west) -- (rschp.north east);

      \draw[rounded corners, thick, opacity=0.2, dotted]
      (rhp.north east) -- (rhptau.north east) |- (rhptau.south) -- node[midway](p2){} (rhpsigma.south) -| (rhpsigma.north west) -- (rhp.north west) -- (rhp.north east);

      \draw[rounded corners, thick, opacity=0.2, dotted]
      (rhsp.north east) -- (rhsptau.north east) |- (rhsptau.south) -- node[midway](p3){} (rhspsigma.south) -| (rhspsigma.north west) -- (rhsp.north west) -- (rhsp.north east);

      \draw[rounded corners, thick, opacity=0.2, dotted, fill=\trtcol!20!white]
      (rrtp.north east) -- node[pos=.3](p6){} (rrtptau.north east) |- (rrtptau.south) -- node[midway](p4){} (rrtpsigma.south) -| (rrtpsigma.north west) -- (rrtp.north west) -- (rrtp.north east);

      \draw[rounded corners, thick, opacity=0.2, dotted]
      (r2scp.north east) -- (r2scptau.north east) |- (r2scptau.south) -- node[pos=.6](p5){} (r2scpsigma.south) -| (r2scpsigma.north west) -- (r2scp.north west) -- (r2scp.north east);

      \draw[rounded corners, thick, opacity=0.2, dotted]
      (rrsp.north east) -- (rrsptau.north east) |- (rrsptau.south) -- (rrspsigma.south) -| (rrspsigma.north west) -- (rrsp.north west) -- (rrsp.north east);


      \draw[myimpl] (rtptau.south east) to (rscsigma.north west);
      \draw[myimpl] (rschptau.south) to (rhspsigma.north west);

      \draw[myimpl] (p1.center) to (rtp.north) ;
      \draw[myimpl] (p2.center) to (rschp.north) ;
      \draw[myimpl] (rsc.north |- p3.center) to (rsc.north) ;
      \draw[myimpl] (rhsp.north |- p4.center) to (rhsp.north) ;

      \draw[myimpl] (p5.center) to (p6.east) ;

      \draw[myimpl] (rrtptau.south) to (rrspsigma.north west) ;
      \draw[myimpl] (rrspsigma.south east) to (rhsptau.north east) ;


      \draw[rounded corners, thick, opacity=0.2, dotted, fill=\tcol!20!white]
      ([xshift=.1em,yshift=.2em]rsctau.south west) -| ([xshift=-2.3em,yshift=-.2em]rsctau.north east) -| ([xshift=.1em,yshift=.2em]rsctau.south west);
      
      \draw[rounded corners, thick, opacity=0.2, dotted, fill=\sccol!20!white]
      ([xshift=.1em,yshift=.2em]rhsptau.south west) -| ([xshift=-2.7em,yshift=-.2em]rhsptau.north east) -| ([xshift=.1em,yshift=.2em]rhsptau.south west);
      
      \draw[rounded corners, thick, opacity=0.2, dotted, fill=\trtcol!20!white]
      ([xshift=.1em,yshift=.2em]rrsptau.south west) -| ([xshift=-2.7em,yshift=-.2em]rrsptau.north east) -| ([xshift=.1em,yshift=.2em]rrsptau.south west);

      \node[]at([xshift=-.5em]rhp -|rhpsigma){\scalebox{1.2}{\coqhref{RobustHyperCriterion.v}}};
      \node[]at([xshift=-.5em]rschp -|rschpsigma){\scalebox{1.2}{\coqhref{RobustSSCHCriterion.v}}};
      \node[]at([xshift=-.5em]rtp -|rtpsigma){\scalebox{1.2}{\coqhref{RobustTraceCriterion.v}}};
      \node[]at([xshift=-.5em]rsc -|rscsigma){\scalebox{1.2}{\coqhref{RobustSafetyCriterion.v}}};
      \node[]at([xshift=-.5em]rhsp -|rhspsigma){\scalebox{1.2}{\coqhref{RobustHyperSafetyCriterion.v}}};
      \node[]at([xshift=-.5em]r2scp -|r2scpsigma){\scalebox{1.2}{\coqhref{Robust2relSSCHCriterion.v}}};
      \node[]at([xshift=-.5em]rrtp -|rrtpsigma){\scalebox{1.2}{\coqhref{Robust2relTraceCriterion.v}}};
    \end{tikzpicture}
{\small
\begin{align*}
  \ms{R} & \text{ robust}
  &
  \ms{2r} & \text{ 2-relational }
  \\
  \hline
  \ms{H} & \text{ hyperproperties}
  & 
  \ms{SCH} & \text{ subset-closed hyperproperties}
  &
  \ms{HS} & \text{ hypersafety}
  \\
  \ms{T} & \text{ trace properties}
  &
  \ms{S} & \text{ safety properties}
  \\
  \hline
  \ms{P} & \text{ property-full criterion}
  &
  \ms{C} & \text{ property-free criterion based on $\sigma$ and $\tau$} 
\end{align*}
\MP{
  last line seems flipped, right?
}
}
  \end{center}
  \caption{Hierarchy of trinitarian views of secure compilation
    criteria preserving classes of hyperproperties and the key to read
    each acronym.  Shorthands `Ins.' and `Refl.' stand for Galois
    Insertion and Reflection. The \CoqSymbol\ symbol denotes trinities
    proven in Coq.\vspace{1em}}
\label{fig:diagram-preserve-props-sec}
\end{figure}

%% file: robust-trace.tex


This subsection illustrates trace-relating secure compilation when the target
language has strictly more events than the source that target contexts can
exploit to break security.

\myparagraph{Source and Target Languages}
The source and target languages used here
are nearly identical expression languages, borrowing from the
syntax of the source language of \autoref{sec:example-diff-values}. Both
languages add \emph{sequencing} of expressions
, two kinds of \emph{output
events}, and the expressions that generate them: $\SrOutL\ \src{n}$ and
$\TrOutL\ \trg{n}$ usable in source and target, respectively,
and $\TrOutH\ \trg{n}$
usable only in the target,
which is the only difference between source and target.
%
The extra events in the target model the fact that the target language has
an increased ability to perform certain operations, some of them potentially
dangerous (such as writing to the hard drive), which cannot be performed by the
source language, and against which source-level reasoning can therefore offer no
protection.



Both languages and compilation chains now deal with partial programs, contexts
and linking of those two to produce whole programs.
In this setting, a whole program is the combination of a \emph{main expression}
to be evaluated and a set of \emph{function definitions} (with distinct names) that can refer to their
argument symbolically and can be called by the main expression and by
other functions. The set of functions of a whole program is the union of the functions of a partial
program and a context; the latter also contains the main expression.
\rb{Write about the union of function sets explicitly. Maybe a bit clearer, but
    does it mess up spacing?}%
%
%
\MP{
  dropped some standard assumptions, which don't really add to the description here
  they're in commented latex below
}%
%
\rb{Citations? Later when discussing compartments?}%
\MP{
  the disjointedness conditions on function names do not really belong to the grammar, assume them outside
}%
%
The extensions of the typing rules and the
operational semantics for whole programs 
are unsurprising and therefore elided.
The trace model also follows closely that of \autoref{sec:example-diff-values}: 
it consists of a list of \emph{regular events}
(including the new outputs) terminated by a \emph{result event}.
Finally, a partial program and a context can be linked into a whole program
when their functions satisfy the requirements mentioned above.

%

\myparagraph{Relating Traces}
In the present model, source and target traces differ only in the fact that the
target draws (regular) events from a strictly larger set than the source, \IE
$\trg{\Sigma_{T}} \supset \src{\Sigma_S}$.
A natural relation between source and target traces
essentially maps to a given target trace $\tT{t}$ the source trace that erases
from $\tT{t}$ those events that exist only at the target level.
Let $\trg{t}\myfilter{\src{\Sigma_S}}$ indicate trace \trg{t} filtered to retain only those elements included in alphabet \src{\Sigma_S}.
We define the trace relation as:
%
\MP{ inline for space ?}
\begin{equation*}
  \tS{s} \sim \tT{t} \quad \equiv \quad \tS{s} = \tT{t}\myfilter{\src{\Sigma_S}}.
\end{equation*}
\ca{We are in the hypothesis of \Cref{thm:compilingANI} :) }%
In the opposite direction, a source trace $\tS{s}$ is related to many target
ones, as any 
target-only events can be inserted at any point in
$\tS{s}$. The induced mappings for \iffull\else $\sim$ \fi are:
%
%
\begin{align*}
  \tilde{\tau}(\piS{\pi})   & = \myset{\tT{t}}{\exists \tS{s} \ldotp
                            \tS{s} = \tT{t}\myfilter{\src{\Sigma_S}}
                            \wedge
                            \tS{s} \in \piS{\pi}};
  &
  \tilde{\sigma}(\piT{\pi}) & = \myset{\tS{s}}{\forall \tT{t} \ldotp
                            \tS{s} = \tT{t}\myfilter{\src{\Sigma_S}}
                            \Rightarrow
                            \tT{t} \in \piT{\pi}}.
\end{align*}
\jt{Can we simplify the expressions instead of keeping useless universal
and existential quantifiers?}

That is, the target guarantee of a source property is that the target has the
same source-level behavior, sprinkled with arbitrary target-level behavior.
Conversely, the source-level obligation of a target property is the aggregate of
those source traces all of whose target-level enrichments are in the target
property.
\rb{Are these intuitions clear enough? The obligation seems a little involved.}

Since $\Sr$ and $\Tr$ are very similar, it is simple  
to prove that the identity compiler (\cmpr{\cdot}) from $\Sr$ to $\Tr$ is secure according to the trace relation $\sim$ defined above.

\begin{theorem}[\cmpr{\cdot} is Secure \coqhref{MoreTargetEventsExample.v}]\label{thm:rob}
\cmpr{\cdot} is \rtctilde.
%
\end{theorem}

%% file: rw-sec-crits-short.tex
To provide examples of cross-language trace-relations that preserve safety and hypersafety properties, we show how existing secure compilation results can be interpreted in our framework.
This indicates how the more general theory developed here can already be instantiated to encompass existing results, and that existing proof techniques can be used in order to achieve the secure compilation criteria we define.

For the preservation of safety, \citet{PatrignaniG18} study a compiler from a typed, concurrent \textsc{while} language to an untyped, concurrent \textsc{while} language with support for memory capabilities.
As in \autoref{sec:example-diff-values}, their source has \src{bool}s and \src{nat}s while their target only has \trg{nat}s.
Additionally, their source has an ML-like memory (where the domain is locations \src{\ell}) while their target has an assembly-like memory (where the domain is natural numbers \trg{n}).
Their traces consider context-program interactions and as such they are concatenations of call and return actions with parameters, which can include booleans as well as locations.
Because of the aforementioned differences, they need a cross-language relation to relate source and target actions.

Besides defining a relation on traces (i.e., an instance of $\sim$), they also define a relation between source and target safety properties.
They provide an instantiation of $\tau$ that maps all safe source traces to the related target ones.
This ensures that no additional target trace is introduced in the target property, and source safety properties are mapped to target safety ones by $\tau$.
Their compiler is then proven to generate code that respects $\tau$, so they achieve a variation of $\rsptautilde$. 

\smallskip

Concerning the preservation of hypersafety, \citet{PatrignaniG17} consider compilers in a reactive setting where traces are sequences of input ($\alpha?$) and output ($\alpha!$) actions.
In their setting, traces are different between source and target, so they define a cross-language relation on actions that is total on the source actions and injective.
Additionally, their set of target output actions is strictly larger than the source one, as it includes a special action \trgb{\tick}, which is how compiled code must respond to invalid target inputs (i.e., receiving a \trg{bool} when a \trg{nat} was expected).
Starting from the relation on actions, they define \trg{TPC}, which is an instance of what we call $\tau$.
Informally, given a set of source traces, \trg{TPC} generates all target traces that are related (pointwise) to a source trace.
Additionally, it generates all traces with interleavings of undesired inputs \trgb{\alpha?} followed by \trgb{\tick} as long as removing $\trgb{\alpha?\tick}$ leaves a trace that relates to the source trace.
\trg{TPC} preserves hypersafety across languages, \IE it is an instance of \rhsptautilde mapping source hypersafety to target hypersafety (and safety to safety).

%% file: appendixintro.tex
\section*{Appendix}
The appendix contains the following addenda.
First of all, it presents the proof index (\Cref{sec:pi}).
Secondly, it discusses extension of the notion of compiler
correctness to other (hyper)property classes: safety properties
and non-subset closed hyperproperties (\Cref{sec:other-classes}).
Then, it discusses the relation with the work of \citet{Melton:1986} and
\citet{sabry1997reflection} in more details (\Cref{sec:relatedwork}).
Then it contains missing formalization bits from \Cref{sec:instances}
(\Cref{sec:elided}).
Then it contains a further discussion on the composition of relations and how
that affects our criteria (\Cref{sec:composition}).
Finally, it presents a more in-depth explanation of how to scale our results to
the setting of secure compilation (\Cref{sec:in-depth-trsec}).

%% file: proofindex.tex

\section{Proof Index}\label{sec:pi}

\begin{proof}[Of \autoref{thm:galois} (\CoqSymbol)] See Corollary Adj\_{$\sigma$}TP\_iff\_{$\tau$}TP in \texttt{Def.v}.
  In general,
  if a program satisfies a property $\pi$, then it satisfies every extension
  $\pi' \supseteq \pi$. Using this, the theorem follows by:\\
  ($\Rightarrow$)\quad Assume \tptau and that $\src{W}$ satisfies $\sigma(\piT{\pi})$. Apply
  \tptau to $\src{W}$ and $\sigma(\piT{\pi})$ and deduce that $\cmp{W}$ satisfies
  $\tau(\sigma (\piT{\pi})) \subseteq \piT{\pi}$.\\
  ($\Leftarrow$)\quad Assume \tpsigma and that $\src{W}$
  satisfies $\src{\pi_S} \subseteq \sigma( \tau (\src{\pi_S} ) )$. Apply \tpsigma to $\src{W}$
  and $\sigma( \tau (\src{\pi_S} ) )$ deducing $\cmp{W}$ satisfies $\tau (\src{\pi_S} )$.
\end{proof} \hrule


\begin{proof}[Of \autoref{thm:trinitarianview} (\CoqSymbol)] See Theorems rel\_TC\_{$\tau$}TP and rel\_TC\_{$\sigma$}TP
  in \texttt{TraceCriterion.v}
, where the $\tptautilde \iff \tpsigmatilde$ part follows directly from \autoref{thm:galois}.
\end{proof} \hrule

\begin{lemma}[Special relations and consequences on the adjoints] \label{lem:surj_rel}
  Let $X, Y$ be two arbitrary sets and $\sim \subseteq X \times
  Y$.
  Assume $\sim$ is a total and surjective map from $Y$ to $X$. Let
  $\alpha \leftrightarrows \gamma$ be its existential and universal image, i.e.
  \begin{align*}
   \tilde{\alpha} &= \lambda ~ \pi_X. ~\myset{y}{\exists x \in \pi_X. ~ x \sim y} \\
   \tilde{\gamma} &= \lambda ~ \pi_Y. ~\myset{x}{\forall y. ~ x \sim y \Rightarrow y \in \pi_Y} 
  \end{align*}
  Then $\tilde{\gamma} = \lambda ~ \pi_Y. ~\myset{x}{\exists y \in \pi_Y. ~ x \sim y}$, and $\tilde{\gamma}$ is injective.
\end{lemma}

\begin{proof}[Of \autoref{lem:surj_rel}] See Lemma
  rel\_total\_surjective and rel\_total\_surjective\_up\_inj in Galois.v
\end{proof} \hrule

\begin{proof}[Of \autoref{thm:ssch} (\CoqSymbol)]
 rel\_TC\_sCl${\tau}$SCHP, rel\_TC\_sCl\_${\sigma}$RSCHP
 in \texttt{SSCHCriterion.v}.
\end{proof} \hrule

\begin{proof}[Of \autoref{thm:inst-dv-cc} (\CoqSymbol)]
        See Theorem correctness in \texttt{TypeRelationExampleInput.v}.
\end{proof} \hrule

\begin{proof}[Of \autoref{thm:inst-marco-cc}]
  See in \autoref{sec:proofs-marco}.
\end{proof}
\hrule

\begin{proof}[Of \autoref{thm:compilingANI}] 
  Assume $\src{W {\models}} \ani{\src{\phi}}{\src{\rho}}$ and
  $\cmp{W} \trg{\rightsquigarrow} \tT{t_1}, \tT{t_2}$ with 
  $\trgb{\phi^{\#}}(\tT{t_1 \inp}) = \trgb{\phi^{\#}}(\tT{t_2 \inp})$.
  We have to show that
  $\trgb{\rho^{\#}}(\tT{t_1\out}) = \trgb{\rho^{\#}}(\tT{t_1\out})$. \\ 
  By \cctilde there exists
  $\tS{s_1} \sim \tT{t_1}$ and $\tS{s_2} \sim \tT{t_2}$ such that
  $\src{W \rightsquigarrow} \tS{s_1}, \tS{s_2}$.
  As a preliminary, apply \Cref{lem:surj_rel} to the relations
  $\simin \circ\; \ii{swap}$ and deduce $g\inp$ is injective. 
  Notice also that by functionality and totality, of $\simin$ and of
  $\simout$, $f\inp(\tT{t_1\inp}) = \{ \tS{s_1\inp} \}$ and
  $f\out(\tT{t_1\out}) = \{ \tS{s_1\out} \}$ and a similar fact holds
  for $\tS{s_2}$ and $\tT{t_2}$.
  \begin{align*}
    & \trgb{\phi^{\#}}(\tT{t_1 \inp}) = \trgb{\phi^{\#}}(\tT{t_2 \inp}) \Rightarrow &[~\text{definition of} ~\trgb{\phi^{\#}}] \\
    & g\inp \circ \src{\phi} \circ f\inp  (\tT{t_1\inp}) = g\inp \circ \src{\phi} \circ f\inp (\tT{t_2\inp}) \Rightarrow &[g\inp ~\text{injective}] \\
    & \src{\phi} \circ f\inp  (\tT{t_1\inp}) = \src{\phi} \circ f\inp (\tT{t_2\inp}) \Rightarrow &[f\inp(\tT{t_i\inp}) = \tS{s_i\inp} ~ i=1,2] \\ 
    & \src{\phi} (\tS{s_1\inp}) = \src{\phi} (\tS{s_2\inp}) \Rightarrow &[\src{W {\models}} \ani{\src{\phi}}{\src{\rho}}] \\
    & \src{\rho} (\tS{s_1\out}) = \src{\rho} (\tS{s_2\out}) \Rightarrow &[\tS{s_i\out} = f\out(\tT{t_i\out})  ~ i=1,2] \\ 
    & \src{\rho} \circ f\out  (\tT{t_1\out}) = \src{\rho} \circ f\out (\tT{t_2\out}) \Rightarrow &[~\text{functionality of} ~g\out] \\ 
    & g\out \circ \src{\rho} \circ f\out  (\tT{t_1\out}) = g\out \src{\rho} \circ f\out (\tT{t_2\out}) \Rightarrow 
                                                                                                       & [~\text{definition of} ~\trgb{\rho^{\#}}] \\ 
    & \trgb{\rho^{\#}}(\tT{t_1\out}) = \trgb{\rho^{\#}}(\tT{t_2\out}),
  \end{align*}
  so that $\cmp{W} {\trg{\models}} \aniT$. \\
  We now show that if $\simout$ is surjective, \IE~$g\out$ injective,
  $\aniT \subseteq \ii{Cl}_{\subseteq} \circ \tilde{\tau}(\aniS)$. \\
  Let $\piT{\pi} \in \aniT$, we show that
  $\piT{\pi} \subseteq \tilde{\tau}(\piS{\pi})$ for some
  $\piS{\pi} \in \aniS$. \\
  %
  The source property
  $\piS{\pi} = \myset{\tS{s}}{\exists \tT{t} \in \piT{\pi}. ~\tS{s} \sim \tT{t}} = f(\piT{\pi})$
  is such that $\piT{\pi} \subseteq \tilde{\tau}(\piS{\pi})$. We only
  need to show $\piS{\pi} \in \aniS$. Let
  $\tS{s_1}, \tS{s_2} \in \piS{\pi}$,
  \begin{align*}
   &\src{\phi}(\tS{s_1}\inp) = \src{\phi}(\tS{s_2}\inp) \Rightarrow & [\text{by} ~ f\inp(\tT{t\inp}) = \tS{s\inp} ~ \text{for some} ~ \tT{t} \in \piT{\pi}] \\ 
   &\src{\phi}(f\inp(\tT{t_1\inp})) = \src{\phi}(f\inp(\tT{t_2\inp})) \Rightarrow & [g\inp ~\text{is a function}]\\
   &g\inp(\src{\phi}(f\inp(\tT{t_1\inp}))) = g\inp(\src{\phi}(f\inp(\tT{t_2\inp}))) \Rightarrow & [\text{by definition of} ~ \trg{\phi^{\#}}] \\ 
   &\trg{\phi^{\#}}(\tT{t_1\inp}) = \trg{\phi^{\#}}(\tT{t_2\inp}) \Rightarrow & [\piT{\pi}\in \aniT] \\
   &\trg{\rho^{\#}}(\tT{t_1\out}) = \trg{\rho^{\#}}(\tT{t_2\out}) \Rightarrow & [\text{definition of} ~ \trg{\rho^{\#}}] \\ 
   &g\out(\src{\rho}(f\out(\tT{t_1\out}))) = g\out(\src{\rho}(f\out(\tT{t_2\out}))) \Rightarrow &[\text{by injectivity of} ~g\out] \\ 
   &\src{\rho}(f\out(\tT{t_1\out})) = \src{\rho}(f\out(\tT{t_2\out}))   \Rightarrow &[f\out(\tT{t_i}) = \tS{s_i\out}, i=1,2] \\ 
   &\src{\rho}(\tS{s_1\out}) = \src{\rho}(\tS{s_2\out}), 
  \end{align*}
  that shows $\piS{\pi} \in \aniS$ and concludes the proof.

\end{proof} \hrule

\begin{proof}[Of \autoref{thm:genANI}] 
  Assume $\src{W {\models}} \ani{\src{\phi}}{\src{\rho}}$ and
  $\cmp{W} \trg{\rightsquigarrow} \tT{t_1}, \tT{t_2}$ with 
  $\trgb{\phi^{\#}}(\tT{t_1 \inp}) = \trgb{\phi^{\#}}(\tT{t_2 \inp})$.
  We have to show that
  $\trgb{\rho^{\#}}(\tT{t_1\out}) = \trgb{\rho^{\#}}(\tT{t_1\out})$,
  for an arbitrary $\trgb{\rho^{\#}}$ that satisfies the condition 
  $ H \equiv \forall \tS{s} ~\tT{t}. ~ \tS{s\out} \simout \tT{t\out} \Rightarrow ~\trgb{\rho^{\#}}(\tilde{\tau}\out(\src{\rho}(\tS{s\out)})) = \trgb{\rho^{\#}}(\trg{\tT{t\out}})$.\\
  By \cctilde there exists
  $\tS{s_1} \sim \tT{t_1}$ and $\tS{s_2} \sim \tT{t_2}$ such that
  $\src{W \rightsquigarrow} \tS{s_1}, \tS{s_2}$.
  As a preliminary, recall that \Cref{lem:surj_rel} ensures $g\inp$ is
  injective. Morevoer notice that by functionality and totality, of
  $\simin$, $f\inp(\tT{t_1\inp}) = \{ \tS{s_1\inp} \}$ and
  $f\inp(\tT{t_2\inp}) = \{ \tS{s_2\inp} \}$. \\
  \begin{align*}
    & \trgb{\phi^{\#}}(\tT{t_1 \inp}) = \trgb{\phi^{\#}}(\tT{t_2 \inp}) \Rightarrow &[~\text{definition of} ~\trgb{\phi^{\#}}] \\
    & g\inp \circ \src{\phi} \circ f\inp  (\tT{t_1\inp}) = g\inp \circ \src{\phi} \circ f\inp (\tT{t_2\inp}) \Rightarrow &[g\inp ~\text{injective}] \\
    & \src{\phi} \circ f\inp  (\tT{t_1\inp}) = \src{\phi} \circ f\inp (\tT{t_2\inp}) \Rightarrow &[f\inp(\tT{t_i\inp}) = \tS{s_i\inp} ~ i=1,2] \\ 
    & \src{\phi} (\tS{s_1\inp}) = \src{\phi} (\tS{s_2\inp}) \Rightarrow &[\src{W {\models}} \ani{\src{\phi}}{\src{\rho}}] \\
    & \src{\rho} (\tS{s_1\out}) = \src{\rho} (\tS{s_2\out}) \Rightarrow &[~\text{functionality of} ~\tilde{\tau}\out] \\ 
    & \tilde{\tau}\out(\src{\rho} (\tS{s_1\out})) = \tilde{\tau}\out(\src{\rho} (\tS{s_2\out})) \Rightarrow &[~\text{by functionality of} ~\trgb{\rho^{\#}}] \\ 
    & \trgb{\rho^{\#}}(\tilde{\tau}\out(\src{\rho} (\tS{s_1\out}))) =
      \trgb{\rho^{\#}}(\tilde{\tau}\out(\src{\rho} (\tS{s_2\out}))) \Rightarrow 
       &[~\text{by condition} ~H] \\ 
    & \trgb{\rho^{\#}}(\tT{t_1 \out}) = \trgb{\rho^{\#}} (\tT{t_2\out})
  \end{align*}
  so that $\cmp{W} {\trg{\models}} \aniT$. \\
\end{proof} \hrule


\begin{proof}[Of \autoref{thm:srcANI}]
  Assume $\src{W {\models}} \ani{\src{\phi^{\#}}}{\src{\rho^{\#}}}$
  and $\cmp{W} \trg{\rightsquigarrow} \tT{t_1}, \tT{t_2}$ with
  $\trgb{\phi}(\tT{t_1 \inp}) = \trgb{\phi}(\tT{t_2 \inp})$ and
  $\trgb{\phi}$ satisfying the condition 
  $ H \equiv \forall \tS{s} ~\tT{t}. ~ \tS{s\inp} \simin \tT{t\inp} \Rightarrow ~\trgb{\phi}(\tT{t\inp}) = \trgb{\phi}(\tilde{\tau}\inp(\tS{s\inp}))$.
  We have to show that
  $\trgb{\rho}(\tT{t_1\out}) = \trgb{\rho}(\tT{t_2\out})$.  By
  \cctilde there exists $\tS{s_1} \sim \tT{t_1}$ and
  $\tS{s_2} \sim \tT{t_2}$ such that
  $\src{W \rightsquigarrow} \tS{s_1}, \tS{s_2}$.
  As a preliminary, recall that \Cref{lem:surj_rel} ensures
  $\tilde{\sigma}\out$ is injective. Morevoer notice that by
  functionality and totality, of $\simout$,
  $\tilde{\tau}\out(\tS{s_1\out}) = \{ \tT{t_1\out} \}$ and
  $\tilde{\tau}\out(\tS{s_2\out}) = \{ \tT{t_2\out} \}$. \\
  \begin{align*}
    &\trgb{\phi}(\tT{t_1 \inp}) = \trgb{\phi}(\tT{t_2 \inp}) \Rightarrow &[~\text{by} ~H]\\
    &\trgb{\phi}(\tilde{\tau}\inp(\tS{s_1 \inp})) = \trgb{\phi}(\tilde{\tau}\inp(\tS{s_2 \inp}))  \Rightarrow 
                                                                        & [~\text{functionality of} ~\tilde{\sigma}\inp] \\
    & \tilde{\sigma}\inp (\trgb{\phi}(\tilde{\tau}\inp(\tS{s_1 \inp}))) = \tilde{\sigma}\inp (\trgb{\phi}(\tilde{\tau}\inp(\tS{s_1 \inp})))
                                                            \Rightarrow & [~\text{definition of} ~\src{\phi^{\#}}] \\ 
    & \src{\phi^{\#}}(\tS{s_1 \inp}) = \src{\phi^{\#}}(\tS{s_2 \inp}) \Rightarrow 
                                                                        & [\src{W \models} \ani{\src{\phi^{\#}}}{\src{\phi^{\#}}}] \\  
    & \src{\rho^{\#}}(\tS{s_1 \out}) = \src{\rho^{\#}}(\tS{s_2 \out}) \Rightarrow 
                                                                        & [~\text{by definition of} ~\src{\rho^{\#}}] \\
    & \tilde{\sigma}\out (\trgb{\rho}(\tilde{\tau}\out(\tS{s_1 \out}))) = \tilde{\sigma}\out (\trgb{\rho}(\tilde{\tau}\out(\tS{s_2 \out})))
                                                            \Rightarrow & [~\text{injectivity of} ~\tilde{\sigma}\out] \\ 
    & \trgb{\rho}(\tilde{\tau}\out(\tS{s_1 \out})) = \trgb{\rho}(\tilde{\tau}\out(\tS{s_2 \out}))  \Rightarrow
                                                                        & [ \tilde{\tau}\out(\tS{s_i\out}) = \{ \tT{t_i\out} \} ~i=1,2] \\
    & \trgb{\rho}(\tT{t_1 \out}) = \trgb{\rho}(\tT{t_2 \out})
  \end{align*}
  so that $\cmp{W} \trgb{\models} \ani{\trgb{\phi}}{\trgb{\rho}}$. 
  
\end{proof} \hrule

\begin{proof}[Of \autoref{thm:rtc-trinity} (\CoqSymbol)]
  Theorems rel\_RTC\_$\tau$RTP and rel\_RTC\_$\sigma$RTP in \texttt{RobustTraceCriterion.v}. 
\end{proof} \hrule

\begin{proof}[Of \autoref{thm:rsp-sec-trinity} (\CoqSymbol)]
 Theorems tilde\_RSC\_$\sigma$RSP and tilde\_RSC\_Cl\_$\tau$RTP in \texttt{RobustSafetyCriterion.v}.
\end{proof} \hrule

\begin{proof}[Of \autoref{thm:rhp-sec-trinity} (\CoqSymbol)]
  Lemmas $\sigma$RHP\_rel\_RHC and rel\_RHC\_$\sigma$RHP and Theorem rel\_RHC\_$\tau$RHP in \texttt{RobustHyperCriterion.v}.
\end{proof} \hrule

\begin{proof}[Of \autoref{thm:rob} (\CoqSymbol)] (See theorem extra\_target\_RTCt in \texttt{MoreTargetEventsExample.v}, mechanizing a slightly simplified model.)
By definition of \rtctilde we need to find a source context and source trace given a source program, target context and target trace related by compilation and program semantics:
This instantiation is simple since the trace relation is a \emph{function} from
target traces to source traces, and it is easy to clean target contexts to
produce equivalent source context without target-only events.
The proof is a trivial instance of \emph{precise, context-based backtranslation}
\cite{AbateBGHPT19,MarcosSurvey,Ahmed:2016:ICFP:UnivEmb,skorstengaard-stktokens:2019},
aided by a few straightforward lemmas and where the case of function calls is
guaranteed to terminate by the language.
%
%
\end{proof} \hrule

\begin{proof}[Of \autoref{thm:trini-safe} (\CoqSymbol)]
  Theorems
  tilde\_SC\_$\sigma$SP and tilde\_SC\_Cl\_$\tau$TP in \texttt{SafetyCriterion.v}. 
\end{proof} \hrule

\begin{proof}[Of \autoref{thm:eq-hp}]
  For the implication from left to right, assume $\src{W} \models
  H$. By \ccequal have $\trg{beh(\cmp{W})} = \src{beh(W)}$,
  so that $\cmp{W} \models H$ as well. \\
  For the implication from right to left, instantiate $\hyperP$ with
  the hyperproperty $\{ \src{beh(W)} \}$, for a given $\src{W}$, and
  deduce that $\cmp{W} \models \{ \src{beh(W)} \}$ \IE~
  $\trg{beh(\cmp{W})} = \src{beh(W)}$.
\end{proof} \hrule

\begin{proof}[Of \autoref{thm:wktr} (\CoqSymbol)]
  Theorems rel\_HC\_$\tau$HP, rel\_HC\_$\sigma$HP and  $\sigma$HP\_rel\_HC in \texttt{HyperCriterion.v}. 
\end{proof} \hrule

%% file: appendix-other.tex
\section{Preserving Other (Hyper)Property Classes}
\label{sec:other-classes}
\ca{ 
  \begin{itemize}
  \item  why it interesting/useful to look at specific classes of (hyper)properties 
  \item  in the equal trace case preservation of trace properties and subset-closed coincide. 
         Something similar happens for different traces. 
  \item  Explain general story about closure operators and $\tau, \sigma$ preservation  
         (then (hyper)safety, subset-closed and even liveness will be just an instance of this) 
  \end{itemize}
}

In this section we investigate how to preserve two classes of (hyper)properties
beyond trace properties: safety properties (\autoref{sec:safety-preservation})
and arbitrary hyperproperties that are not just subset-closed
(\autoref{sec:correct-hyper}).
For each of these classes, we start by giving an intuition of what it means to
preserve such a class in the equal-trace setting, then we study preservation of
that class in the trace-relating setting.
In both cases, we obtain a somewhat weaker preservation result that
reflects the loss of precision incurred by the properties.
\rb{Shoutout to intro.}

\subsection{Preserving Safety Properties}
\label{sec:safety-preservation}



\emph{Safety} collects all trace properties prescribing that {\it ``something
  bad never happens,''} so that their violation can be monitored and, when
observed, no longer restored \cite{ClarksonS10}.
%
In order to define this class, we assume traces are built in a standard manner
as lists or streams of events, so that we can consider \emph{finite prefixes}
\(m\) of a trace \(t\), written \(m \leq t\).
%
%
\begin{definition}[Safety Properties]
  A property $\pi$ is a \emph{safety property} if
    $\forall t \notin \pi. ~\exists m \leq t. ~\forall t'. ~ m\leq t' \Rightarrow t' \notin \pi $.
\end{definition}
\jt{should it be described in words or is the above intuition enough?}
A proof of the preservation of all trace properties, \IE of $\ccequal$,
may require to show that a source program can produce an infinite stream of events.
In order to prove the same for safety properties, one only has to
show the following weaker variant of $\ccequal$: \MP{ inline for space ?}
\rb{Shoutout to weakening, closure on safety?}
\begin{equation*}
  \scequal \equiv \forall \src{W},m \ldotp \wtmkt{W}{\com{m}} \Rightarrow \wsmkt{W}{\com{m}}
\end{equation*}
where $W {\sem} m$ means that \(W\) produces the finite prefix $m$, \IE that \(m\)
is a prefix of a trace produced by \(W\), that is $\exists t\ldotp m\leq t \wedge
W {\sem} t$.

Recalling the intention of safety properties to prescribe {\it ``something bad
never happens,''} $W {\sem} m$ can be interpreted as a violation of a
safety property, with \(m\) being a bad prefix.
%
The statement of $\scequal$ can therefore be read as {\it ``whenever $\cmp{W}$
violates a safety property, then $\src{W}$ does.'' }
By contraposition, this was proven equivalent to the preservation of arbitrary
safety properties~\cite{AbateBGHPT19}: \jt{Should we really add a citation for
this? The paper really only do the proof for the ``robust'' part}
\begin{equation*}
  \scequal \iff \forall \pi \in \ii{Safety}.~\src{W} ~\satS~ \pi \Rightarrow \cmp{W} ~\satT~ \pi
\end{equation*}
In the trace-relating setting, we generalize $\scequal$ by requiring that if $\cmp{W}$ violates 
a target safety property $\piT{\pi} \in \hT{Safety}$, then $\src{W}$ violates its source interpretation, $\tilde{\sigma}(\piT{\pi})$. 

\begin{definition}[$\sctilde$]
 Given a trace relation $\sim \subseteq \piS{Trace} \times \trg{Trace_T}$, a compilation chain $\downarrow$ is $\sctilde$ if and only if the following holds:
  \MP{ inline for space ?}
  \begin{align*}
      \sctilde \equiv \forall \src{W}\ldotp \forall & \tT{m}\ldotp \wtmkt{W}{m} \Rightarrow 
                                            \exists \tT{t}\ldotp \exists \tS{s}\ldotp \tT{m} \leq \tT{t}
                                            \wedge \tS{s} \sim \tT{t} \wedge \wsmkt{W}{s} 
  \end{align*}

\end{definition}
  
Again, we propose two generalizations of the preservation of safety properties
that are equivalent to $\sctilde$.
Similarly to what we had to do in \autoref{sec:subset-closed}, we have to take
into account the fact that for $\piS{\pi} \in \hS{Safety}$,
$\tilde{\tau}(\piS{\pi})$ is not necessarily in $\hT{Safety}$.
Hence, we have to close $\tilde{\tau}(\piS{\pi})$ in the class of target safety
properties.
\MP{
  footnote elided, did not have an intuition
}

\begin{theorem}[Trinitarian view for Safety] \label{thm:trini-safe}
  For a trace relation $\sim\ \subseteq \piS{Trace} \times \trg{Trace_T}$ and its induced property mappings $\tilde{\sigma}$ and $\tilde{\tau}$, 
  $\sctilde$ is equivalent to these criteria:

  \hspace{-1em}\begin{minipage}{0.7\textwidth}\small
    \begin{align*}
      \spsigmatilde & \equiv \forall \src{W} ~\forall \piT{\pi} \in \hT{Safety}. ~ \src{W} ~\satS~ \tilde{\sigma}(\piT{\pi}) \Rightarrow 
                                                                                                       \cmp{W} ~\satT~ \piT{\pi} \\
      \sptautilde & \equiv \forall \src{W} ~\forall \piS{\pi} \in \propS. ~ \src{W} ~\satS~ (\piS{\pi}) \Rightarrow 
                                                                                                       \cmp{W} ~\satT~ (\ii{Safe}\circ\tilde{\tau})(\piS{\pi})
    \end{align*}  
  \end{minipage}
  \begin{minipage}{0.3\textwidth}
    \input{diagram-sc-only} 
  \end{minipage}
  where $\ii{Safe}(\piT{\pi}) = \bigcap \myset{\hT{S}}{\hT{S} \in \hT{Safety} \wedge \piT{\pi} \subseteq \hT{S}}$ is the smallest target safety property that 
  contains $\piT{\pi}$.
\end{theorem}

Notice that in $\sptautilde$, we quantify over arbitrary source properties,
but the guarantee we get in the target is given by approximating the guarantee
defined by $\tilde{\tau}$, with a safety property.
This second approximation is optimal by minimality of $\mr{Safe}$, and is
necessary as $\tilde{\tau}$ may not map safety properties to safety
properties \ca{TODO: maybe we should give an example or point to Jeremy instance
where it happens} (in this case, no comparison with $\sctilde$ would be
possible). \ca{I know this can look fuzzy, I will come back to it later (modulo
time)}\jt{I don't understand the point about optimality by minimality}

\ch{It might be that some of the examples from
  \autoref{sec:compiler-correctness} need at least a definition of 
  safety properties already?}

\subsection{Preserving Non-Subset Closed Hyperproperties}
\label{sec:correct-hyper}
\ca{ TODO: It would be interesting to know what is the equivalent
  condition of $\sim$ that gives the insertion/relation.  I think if
  $\sim$ is a total surjective map from target to source then you have
  a reflection (the other way around should give you the insertion) }

Many hyperproperties of interest, including possibilistic
information-flow policy 
are not subset-closed \cite{ClarksonS10}.  $\ccequal$ is not enough to
guarantee the preservation of such hyperproperties, as it only ensures
refinement of the behaviors through compilation but not the other
inclusion, sometimes referred to as \emph{reflection}.
\jt{This is a very abrupt introduction to this subsection}
%
%
It is easy to show that that the following strengthening of $\ccequal$ is equivalent to the preservation of arbitrary hyperproperties. \ca{citation?}
\begin{theorem}[$\hcequal$, $\hyperP$] \label{thm:eq-hp}
  For every compilation chain $\downarrow$, the followings are equivalent
  \MP{
  is the BEH notation defined?
  }
  \begin{align*}
    \hcequal &\equiv  ~\forall \src{W}. ~ \trg{beh(\cmp{W})} = \src{beh(W)}  
    &
    \hyperP   &\equiv  ~\forall \src{W} \forall H \in 2^{\prop}. ~\src{W} ~\satS~ H \Rightarrow \cmp{W} ~\satT~ H
  \end{align*}
\end{theorem}
\MP{ is this coqed? there's no proof in the original, i can't put a
  proof in the proof indx } \ca{this is not Coqed, I can give a proof
  sketch in the appendix but is really immediate stuff.}
The generalization to the trace-relating setting,
$\hctilde \equiv ~\forall \src{W}. ~ \trg{beh(\cmp{W})} =
\tilde{\tau}(\src{beh(W)})$,
does not come with two equivalent formulations in terms of
preservation of hyperproperties.
While for $\hptautilde$ the equivalence holds unconditionally, for
$\hpsigmatilde$ some extra assumptions are required, depending on the
implication one is interested in. It is still possible, and correct, to deduce a
source obligation for a given target hyperproperty $\hT{H}$ when no information
is lost in the the composition $\tilde{\tau} \circ \tilde{\sigma}$ (\IE the two
maps are a Galois \emph{insertion}).
{\it Vice versa}, $\hctilde$ is a consequence of $\hpsigmatilde$ when no information is
lost in composing in the other direction, $\tilde{\sigma} \circ \tilde{\tau}$
(\IE the two maps are a Galois \emph{reflection}).
\jt{Maybe we need a bit more explanation here, because this is not easy to understand
\rb{Although in principle, this recalls the right ideas from the intro, and
there is no earlier place in the section where they could go well.}}

\begin{theorem}[Weak Trinity] \label{thm:wktr}
  Given a trace relation $\sim \subseteq \piS{Trace} \times \trg{Trace_T}$, and
  its induced property mappings $\tilde{\sigma}$ and $\tilde{\tau}$, $\hctilde$
  is equivalent to \hptautilde below. Moreover if
  $\tilde{\tau} \leftrightarrows \tilde{\sigma}$ is a Galois insertion (\IE
  $\tilde{\tau} \circ \tilde{\sigma} = id$), $\hctilde$ implies \hpsigmatilde,
  while if $\tilde{\sigma} \leftrightarrows \tilde{\tau}$ is a Galois reflection
  (\IE $\tilde{\sigma} \circ \tilde{\tau} = id$), \hpsigmatilde implies
  $\hctilde$.

 \begin{minipage}{0.7\textwidth}
  \begin{align*}
     \hptautilde &
        \equiv \forall \src{W} ~\forall \hS{H}. 
            ~\src{W} ~\satS~ \hS{H} \Rightarrow \cmp{W} ~\satT~ \tilde{\tau}(\hS{H})
      \\
     \hpsigmatilde &
        \equiv \forall \src{W} ~\forall \hT{H}. 
          ~\src{W} ~\satS~ \tilde{\sigma}(\hT{H}) \Rightarrow \cmp{W} ~\satT~ \hT{H}
  \end{align*}
\end{minipage}
\begin{minipage}{0.3\textwidth}
  \input{diagram-hc-only}
\end{minipage}
\end{theorem}

\autoref{fig:diagram-preserve-props-cc} sums up our generalizations and orders them according their
strength.

\input{diagram-prop-pres-comp-corr}

%% file: diagram-sc-only.tex
\begin{center}

\begin{tikzpicture}
    \node[right = 3.5 of cctilde](sctilde){
      \phantom{$\sim$}
      \sctilde
    };
    \node[below =1 of sctilde.west,xshift = -1em](scsigma){
      \spsigmatilde
    };
    \node[below =1 of sctilde.east,xshift = 1em](sctau){
      \sptautilde
    };
    \draw[myiff] (sctilde.-60) to (sctau.120);
    \draw[myiff] (sctilde.-120) to  (scsigma.60);
    \draw[myiff] (sctau.180) to (scsigma.0);

\end{tikzpicture}
\end{center}

%% file: diagram-hc-only.tex
  \begin{center} 

    \begin{tikzpicture}[every node/.style={align=center}]

      \node[left = 3.5 of cctilde](hctilde){
        \phantom{$\sim$}
        \hctilde
      };
      \node[below =1 of hctilde.east,xshift = 1em](hctau){
        \hptautilde
      };
      \node[below =1 of hctilde.west,xshift = -1em](hcsigma){
        \hpsigmatilde
      };
      
      \draw[myiff] (hctilde.-60) to (hctau.120);
      \draw[myimpl] (hctilde.-120) to node[midway,sloped,above,font=\tiny] (hpi) {Insertion} (hcsigma.60);
      \draw[myimpl] (hcsigma.0) to node[midway,sloped,above,font=\tiny] (hpr) {Reflection} (hctau.180);

    \end{tikzpicture}


  \end{center}

%% file: diagram-prop-pres-comp-corr.tex
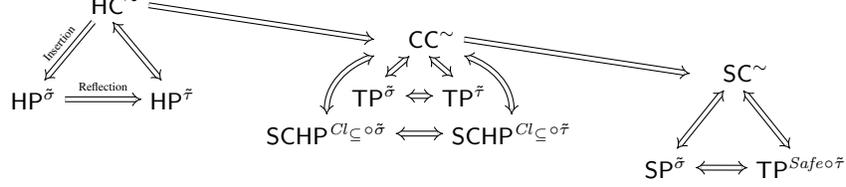
\begin{figure}[!ht]
  \begin{center}


    \begin{tikzpicture}[every node/.style={align=center}]
      \node[](cctilde){
        \phantom{$\sim$}
        \cctilde
      };
      \node[below =.5 of cctilde.east](tctau){
        \tptautilde
      };
      \node[below =.5 of cctilde.west](tcsigma){
        \tpsigmatilde
      };

      \draw[myiff] (cctilde.-60) to (tctau.120);
      \draw[myiff] (cctilde.-120) to (tcsigma.60);
      \draw[myiff] (tctau.180) to (tcsigma.0);


      \node[below =1 of cctilde.east,xshift = 2em](schptau){
        \schptautilde
      };
      \node[below =1 of cctilde.west,xshift = -2em](schpsigma){
        \schpsigmatilde
      };

      \draw[myiff, out = -20, in = 90] (cctilde.-20) to (schptau.90);
      \draw[myiff, out = -160, in = 90] (cctilde.-160) to (schpsigma.90);
      \draw[myiff] (schptau.180) to (schpsigma.0);


       \node[above left = .01 and 3 of cctilde](hctilde){
        \phantom{$\sim$}
        \hctilde
      };
      \node[below =1 of hctilde.east,xshift = 1em](hctau){
        \hptautilde
      };
      \node[below =1 of hctilde.west,xshift = -1em](hcsigma){
        \hpsigmatilde
      };
      
      \draw[myiff] (hctilde.-60) to (hctau.120);
      \draw[,double,-implies,double equal sign distance] (hctilde.-120) to node[midway,sloped,above,] (hpi) {\scalebox{.5}{Insertion}} (hcsigma.60);
      \draw[,double,-implies,double equal sign distance] (hcsigma.0) to node[midway,sloped,above,] (hpr)  {\scalebox{.5}{Reflection}} (hctau.180);

      \draw[,double,-implies,double equal sign distance] (hctilde.0) to (cctilde.180); 

      \node[below right = .01 and 3 of cctilde](sctilde){
      \phantom{$\sim$}
      \sctilde
    };
    \node[below =1 of sctilde.west,xshift = -1em](scsigma){
      \spsigmatilde
    };
    \node[below =1 of sctilde.east,xshift = 1em](sctau){
      \sptautilde
    };
    \draw[myiff] (sctilde.-60) to (sctau.120);
    \draw[myiff] (sctilde.-120) to  (scsigma.60);
    \draw[myiff] (sctau.180) to (scsigma.0);

     \draw[,double,-implies,double equal sign distance] (cctilde.0) to (sctilde.180); 
 
    \end{tikzpicture}

  \end{center}
\caption{Generalization of Compiler Correctness and its trace-relating variations.}
\label{fig:diagram-preserve-props-cc}
\end{figure}


%% file: other_works.tex
\section{Relation to Melton et al. and Sabry and Wadler}
 \label{sec:relatedwork}

 In this section we assume that source and target languages are
 equipped with operational small steps semantics, $\src{\to}$ and
 $\trgb{\to}$ respectively. Let $\piS{State}$ and $\hT{State}$ be the
 set of all possible program states in the source and in the target
 respectively. We define traces as finite or infinite sequences of
 program states, formally
 \begin{align*}
  \piS{Trace} &= \piS{State}^* \cup \piS{State}^{\omega} \\
  \hT{Trace} &= \hT{State}^* \cup \hT{State}^{\omega}.
 \end{align*}
 For both source and target languages, we say that a program produces
 a trace if and only if it can be extracted by an admissible execution
 of the program itself. Formally 
 \begin{align*}
   \src{W \rightsquigarrow} \tS{s} \iff & \exists (\src{W_0 \to W_1 \to \ldots \to ~W_k \to \ldots}). \\
                                        &  ~\exists ~(\src{0 < i_1 < i_2 < \ldots < i_k < \ldots}). \\
                                        & ~\src{W_0} = \tS{s_o} \wedge \forall \src{j}. ~\tS{s_j} = \src{W_{i_j}}.                                        
 \end{align*}
 a similar definition is given for the target, based on the target
 operational semantics $\trgb{\to}$.  The compilation induces a
 relation between source and target states,
\begin{equation*}
\tS{s_k} \sim \tT{t_h} \iff \cmp{s_k} = \tT{t_h}
\end{equation*}
that we lift to traces by requiring the traces to be pointwise related, 
\begin{equation*}
  \tS{s} \sim \tT{t} \iff \forall n. ~\cmp{s_n} = \tT{t_n}
\end{equation*}
Recall the definition of \cctilde

\begin{equation*}
 \cctilde \equiv \forall \src{W} \tT{t}. ~\cmp{W} {\trgb{\rightsquigarrow}} \tT{t} \Rightarrow 
 \exists \tS{s} \sim \tT{t}. ~ \src{W \rightsquigarrow} \tS{s} 
\end{equation*}
that spelled out states that anytime we extract some states from the
execution of $\cmp{W}$, it is possible to extract the same number of
source, and related, states from some execution of $\src{W}$.  In
other words \cctilde requires that whenever $\cmp{W}$ reduces in an
arbitrary (but finite) amount of steps to some target state,
$\cmp{W} \trgb{\to^*} \tT{t_k}$, then $\src{W} \src{\to^*} \tS{s_h}$
for some state $\tS{s_h} \sim \tT{t_k}$, \IE~ $\cmp{s_h} = \tT{t_k}$.

This captures is in general a weaker requirement than the existence of the
decompilation function in \citet{Melton:1986} and
\citet{sabry1997reflection}, but coincide for the case study in
\citet{Melton:1986}, where $\downarrow$ is an injective map. 
\ca{conclusion: their def is in both cases very strong! 
  \begin{itemize}
  \item In \citet{Melton:1986} (they have an insertion) they cannot
    compile two distinct programs to the same target one
  \item In \citet{sabry1997reflection} (they have a reflection) the target language is a subset of the source one  
  \end{itemize}
}

%% file: app-proofs-marco-statements.tex
\section{Elided Formalization Bits}\label{sec:elided}

\subsection{Additional Formalism for \Cref{sec:example-diff-values}}

\paragraph{The Source Language}
Below is the syntax of the source language.
\begin{align*}
	\src{e} \bnfdef
		&\
		\src{n} \mid \src{b} \mid \src{e\ \mathord{op}\ e} \mid \src{\ifte{e}{e}{e}} \mid \src{in_b} \mid \src{in_n} \qquad
        \src{op} \bnfdef
                \ 
                \src{+} \mid \src{\times} \mid \src{\leq} \qquad
        \src{ty} \bnfdef
                \
                \src{\TypeBool} \mid \src{\TypeNat}
	        \\
	\src{r} \bnfdef
		&\
		\src{\nattag n}
                \mid \src{\booltag b} \mid \src{error}
                \qquad
	\src{i} \bnfdef
		\
		\src{\nattag n}
                \mid \src{\booltag b} \qquad
	\src{is} \bnfdef
		\
		\src{i \listconcat is}
                \mid \src{\emptylist} \qquad
	\src{s} \bnfdef
		\
		\src{\pair{is,r}}
\end{align*}

Below are the typing rules for the source language.
\begin{center}
	\typerule{Type-nat}{
	}{
	  \vdash \src{n} : \src{\TypeNat}
	}{}
	\typerule{Type-bool}{
	}{
		\vdash \src{b} : \src{\TypeBool}
	}{}
	\typerule{Type-plus}{
		\vdash \src{e_1} : \src{\TypeNat} &
		\vdash \src{e_2} : \src{\TypeNat}
	}{
		\vdash\src{e_1 + e_2} : \src{\TypeNat}
	}{}
	\typerule{Type-times}{
		\vdash \src{e_1} : \src{\TypeNat} &
		\vdash \src{e_2} : \src{\TypeNat}
	}{
		\vdash\src{e_1 \times e_2} : \src{\TypeNat}
	}{}
	\typerule{Type-ite}{
		\vdash \src{e_1} : \src{\TypeBool}
		&
		\vdash \src{e_2} : \src{ty}
		&
		\vdash \src{e_3} : \src{ty}
	}{
		\vdash\src{\ifte{e_1}{e_2}{e_3}} : \src{ty}
	}{}
	\typerule{Type-le}{
		\vdash \src{e_1} : \src{\TypeNat} &
		\vdash \src{e_2} : \src{\TypeNat}
	}{
		\vdash\src{e_1 \leq e_2} : \src{\TypeBool}
	}{}
	\typerule{Type-in-b}{}{
		\vdash\src{in_b}:\src{\TypeBool}
	}{}
	\typerule{Type-in-n}{}{
		\vdash\src{in_n}:\src{\TypeNat}
	}{}
\end{center}

Well-typed programs do not produce \src{error} (\CoqSymbol, see Theorem type\_soundness in file \texttt{TypeRelationExampleInput.v}).

Below is the big-step semantics of the source language.
\begin{center}
	\typerule{Sem-nat}{
	}{
		\src{n \sem \pair{\emptylist,\nattag n}}
	}{sem-nat}
	\typerule{Sem-in-nat}{
	}{
		\src{in_n \sem \pair{n \listconcat \emptylist,\nattag n}}
	}{sem-in-nat}
	\typerule{Sem-bool}{
	}{
		\src{b \sem \pair{\emptylist,\booltag b}}
	}{}
	\typerule{Sem-in-bool}{
	}{
		\src{in_b \sem \pair{b \listconcat \emptylist,\booltag b}}
	}{}
	\typerule{Sem-op-nat}{
                \src{e_1 \sem \pair{is_1, n_1}} &  
                \src{e_2 \sem \pair{is_2, n_2}} &
                \src{op} \in \src{\{ +, \times \}}
	}{
		\src{e_1 \ \mathord{op}\  e_2 \sem \pair{is_1 \listconcat  is_2, \nattag (n_1 \ \mathord{op}\  n_2)}}
	}{sem-s-nat}
	\typerule{Sem-le}{
                \src{e_1 \sem \pair{is_1, n_1}} &  
                \src{e_2 \sem \pair{is_2, n_2}}  
	}{
		\src{e_1 \leq e_2 \sem \pair{is_1 \listconcat  is_2, \booltag (n_1 \leq  n_2)}}
	}{}
	\typerule{Sem-ite-true}{
                \src{e_1 \sem \pair{is_1, true}} &  
                \src{e_2 \sem \pair{is_2, r_2}}  
	}{
		\src{\ifte{e_1}{e_2}{e_3} \sem \pair{is_1 \listconcat  is_2, r_2}}
	}{}
	\typerule{Sem-ite-false}{
                \src{e_1 \sem \pair{is_1, false}} &  
                \src{e_3 \sem \pair{is_3, r_3}}  
	}{
		\src{\ifte{e_1}{e_2}{e_3} \sem \pair{is_1 \listconcat  is_3, r_3}}
	}{}
\end{center}

\paragraph{The Target Language}
Below is the syntax of the target language.
\begin{align*}
	\trg{e} \bnfdef
		&\
		\trg{n} \mid \trg{e\ \mathord{op}\ e} \mid \trg{\iflete{e}{e}{e}{e}} \mid \trg{in_n} \qquad
        \trg{op} \bnfdef
                \ 
                \trg{+} \mid \trg{\times}
	        \\
	\trg{r} \bnfdef
		&\
		\trg{\nattag n}
                \qquad
	\trg{i} \bnfdef
		\
		\trg{\nattag n} \qquad
	\trg{is} \bnfdef
		\
		\trg{i\listconcat is}
                \mid \trg{\emptylist} \qquad
	\trg{t} \bnfdef
		\
		\trg{\pair{is,r}}
\end{align*}

Below is the big-step semantics of the target language.
\Cref{tr:sem-s-nat,tr:sem-in-nat,tr:sem-nat} are the same as for the source and we present the new rules only:
%
\begin{center}\small
	\typerule{Sem-itele-true}{
                \trg{e_1 \sem \pair{is_1, n_1}} &  
                \trg{e_2 \sem \pair{is_2, n_2}} 
                \\
                n_1 \leq n_2 &
                \trg{e_3 \sem \pair{is_3, n_3}}
	}{
		\trg{\iflete{e_1}{e_2}{e_3}{e_4} \sem \pair{is_1 \listconcat  is_2\listconcat  is_3, n_3}}
	}{}
	\typerule{Sem-ite-false}{
                \trg{e_1 \sem \pair{is_1, n_1}} &  
                \trg{e_2 \sem \pair{is_2, n_2}} 
                \\
                n_1 > 
                n_2 &
                \trg{e_4 \sem \pair{is_4, n_4}}
	}{
		\trg{\iflete{e_1}{e_2}{e_3}{e_4} \sem \pair{is_1 \listconcat  is_2\listconcat  is_4, n_4}}
	}{}
\end{center}

\paragraph{The Compiler}

\begin{align*}
\cmp{\nattag n} =
	&\ \trg{\nattag n}
&
\cmp{\booltag true} =
	&\ \trg{\nattag 1}
&
\cmp{e_1 + e_2} =
	&\ \cmp{e_1} \trgb{+} \cmp{e_2}
\\
\cmp{in_n} =
	&\ \trg{in_n}
&
\cmp{\booltag false} =
	&\ \trg{\nattag 0}
& 
\cmp{e_1 \leq e_2} =
	&\ \trg{\iflete{\cmp{e_1}}{\cmp{e_2}}{1}{0}}
\\
\cmp{in_b} =
	&\ \trg{in_n}
&
\cmp{e_1 \times e_2} =
	&\ \cmp{e_1} \trgb{\times} \cmp{e_2}
& 
\cmp{\ifte{e_1}{e_2}{e_3}} =
	&\ \trg{\iflete{\cmp{e_1}}{0}{\cmp{e_3}}{\cmp{e_2}}}
\end{align*}

\subsection{Additional Formalism for \Cref{SEC:M}}

\paragraph{The Source Language \Sm}
\Sm is a statically-typed language with expressions, commands and simple types (natural numbers and pairs) whose syntax is presented below.
The key aspect of \Sm is the primitive for sending \emph{pairs} over a network interface.
The type system of \Sm is unsurprising.
\begin{align*}
	\src{c} \bnfdef
		&\
		\src{skip} \mid \src{\ifztes{e}{c}{c}} \mid \src{c;c} \mid \sends{e}
	\qquad
	\src{v} \bnfdef
		\
		\src{n} \mid \src{\pair{v,v}}
	\qquad\qquad
	\src{\tau} \bnfdef
		\
		\src{N} \mid \src{\tau\times\tau}
	\\
	\src{e} \bnfdef
		&\
		\src{n} \mid \src{e\op e} 
		\mid \src{\pair{e,e}} \mid \src{\projone{e}} \mid \src{\projtwo{e}}
	\qquad\qquad
	\src{\evalctx} \bnfdef
		\
		\src{\hole{\cdot}} \mid \src{\evalctx\op e} \mid \src{n\op \evalctx}  \mid \src{\projone{\evalctx}} \mid \src{\projtwo{\evalctx}} \mid \src{\pair{\evalctx,e}} \mid \src{\pair{v,\evalctx}}
\end{align*}
\begin{center}
	\typerule{Type-\Sm-skip}{
	}{
		\vdash\src{\skips}
	}{t-sm-skip}
	\typerule{Type-\Sm-seq}{
		\vdash\src{c}
		&
		\vdash\src{c'}
	}{
		\vdash\src{c;c'}
	}{t-sm-seq}
	\typerule{Type-\Sm-send}{
		\vdash\src{e}:\src{\tau\times\tau'}
	}{
		\vdash\src{\sends{e}}
	}{t-sm-send}
	\typerule{Type-\Sm-if}{
		\vdash\src{e}:\src{N}
		&
		\vdash\src{c}
		&
		\vdash\src{c}
	}{
		\vdash\src{\ifztes{e}{c}{c'}}
	}{t-sm-if}
	\typerule{Type-\Sm-n}{}{
		\vdash\src{n}:\src{N}
	}{t-sm-n}
	\typerule{Type-\Sm-op}{
		\vdash\src{e}:\src{N}
		&
		\vdash\src{e'}:\src{N}
	}{
		\src{\Gamma}\vdash\src{e\op e'}:\src{N}
	}{t-sm-op}
	\typerule{Type-\Sm-pair}{
		\vdash\src{e}:\src{\tau}
		&
		\vdash\src{e'}:\src{\tau'}
	}{
		\src{\Gamma}\vdash\src{\pair{e,e'}}:\src{\tau\times\tau'}
	}{t-sm-pair}
	\typerule{Type-\Sm-p1}{
		\vdash\src{e}:\src{\tau\times\tau'}
	}{
		\vdash\src{\projone{e}}:\src{\tau}
	}{t-sm-p1}
	\typerule{Type-\Sm-p2}{
		\vdash\src{e}:\src{\tau\times\tau'}
	}{
		\vdash\src{\projtwo{e}}:\src{\tau'}
	}{t-sm-p2}
\end{center}

\Sm has a contextual, small-step, call by value semantics for expressions ($\src{e \red e'}$) and a big-step semantics for commands ($\src{c \xtos{s} c'}$) that produces traces \src{s}, \IE sequences of messages \src{M} (\IE pairs \src{\pair{v,v}}) sent over the network.
In the following, queuing the empty element is ineffective: $s = \epsilon\listconcat s$.

\begin{center}
	\typerule{Eval-\Sm-Op}{
		\src{n''}=n\op n'
	}{
		\src{n\op n'\red n''}
	}{e-sm-op}
	\typerule{Eval-\Sm-P1}{
	}{
		\src{\projone{\pair{v,v'}} \red v}
	}{e-sm-p1}
	\typerule{Eval-\Sm-P2}{
	}{
		\src{\projtwo{\pair{v,v'}} \red v'}
	}{e-sm-p2}
	\typerule{Eval-\Sm-Ctx}{
		\src{e\red e'}
	}{
		\src{\evalctxs{e} \red \evalctxs{e'}}
	}{e-sm-ctx}


	\typerule{E-\Sm-skip}{ 
	}{
		\src{\skips \xtos{\epsilon} \skips}
	}{es-sm-skip} 
	\typerule{E-\Sm-seq}{
		\src{c\xtos{s}\skips}
		&
		\src{c'\xtos{s'}\skips}
	}{
		\src{c;c'\xtos{s\listconcat s'}\skips}
	}{es-sm-seq}
	\typerule{Eval-\Sm-If-t}{
		\src{e \red\redstar 0}
		&
		\src{c \xtos{s} \skips}
	}{
		\src{\iftes{e}{c}{c'} \xtos{s} \skips}
	}{es-sm-if-t}
	\typerule{Eval-\Sm-If-f}{
		\src{e \red\redstar n}\neq\src{0}
		&
		\src{c' \xtos{s} \skips}
	}{
		\src{\iftes{e}{c}{c'} \xtos{s} \skips}
	}{es-sm-if-f}
	\typerule{Eval-\Sm-Send}{
		\src{e\red\redstar\pair{v,v'}}
	}{
		\src{\sends{e} \xtos{\pair{v,v'}} \skips}
	}{es-sm-send}
	\vrule
	\typerule{Sem-\Sm}{
		\src{c\xtos{s}\skips }
	}{
		\src{c\sem s}
	}{sem-s}
\end{center}

\paragraph{The Target Language \Tm}
\Tm is a statically-typed language with the same types as \Sm, but its primitive for sending over a network interface only sends \emph{natural numbers}.
Nothing changes regarding statements, terms, values, types and evaluation contexts between \Tm and \Sm.
There is only one typing rule that changes compared to \Sm, the one for \sendt{n}.
\Tm has the same dynamic semantics as \Sm, the only changes regard the nature of messages and the reduction of \sendt{n}.
\begin{center}
	$\trg{M} \bnfdef\ \trg{n}$
	\typerule{Type-\Tm-send}{
		\vdash\trg{e}:\trg{N}
	}{
		\vdash\trg{\sendt{e}}
	}{t-tm-send}
	\typerule{Eval-\Tm-Send}{}{
		\trg{\sendt{n} \xtot{n} \skipt}
	}{e-tm-send}
\end{center}

\paragraph{The Compiler from \Sm to \Tm: \compm{\cdot}}
The compiler from \Sm to \Tm is defined inductively on the type derivation of a source statement ($\cmpm{\cdot} :\ \vdash\src{c}\to\trg{c}$) which, in turn, relies on compilation of expressions ($\cmpm{\cdot} :\ \vdash\src{e : \tau} \to \trg{e}$).
The only interesting case is when compiling a \src{\sends{e}}, where we use the source type information concerning the message (\IE a pair) being sent to deconstruct that pair into a sequence of natural numbers, which is what is sent in the target.


The compiler operates on \emph{type derivations} for terms.
Thus, compiling \src{c;c'} would look like the following (using \src{D} as a metavariable to range over derivations).
\begin{align*}
  \compm{
    \AxiomC{\src{D}}
    \UnaryInfC{$\vdash\src{c}$}
    	\AxiomC{\src{D'}}
    	\UnaryInfC{$\vdash\src{c'}$}
    \BinaryInfC{$\vdash\src{c;c'}$}
    \DisplayProof
  } 
  =&\
  \trg{
  	\compm{
  	    \AxiomC{\src{D}}
  	    \UnaryInfC{$\vdash\src{c}$}
  	    \DisplayProof
  	}
    ;
    \compm{
	    \AxiomC{\src{D'}}
	    \UnaryInfC{$\vdash\src{c'}$}
	    \DisplayProof
	}
  }
\end{align*}
However, note that each judgment uniquely identifies which typing rule is being applied and the underlying derivation.
Thus, for compactness, we only write the judgment in the compilation and implicitly apply the related typing rule to obtain the underlying judgments for recursive calls.
\begin{align*}
	\compm{\vdash\src{n:N}}
		&\ =
		\trg{n}
	&
	\compm{\vdash\src{\projone{e}:\tau}}
		&\ =
		\trg{\projone{\compm{\vdash\src{e:\tau\times\tau'}}}}
	\\
	\compm{\vdash\src{e\op e' : N}}
		&\ =
		\trg{\compm{\vdash\src{e : N}}\op\compm{\src{e' : N}}}
	&
	\compm{\vdash\src{\projtwo{e}:\tau'}}
		&\ =
		\trg{\projtwo{\compm{\vdash\src{e:\tau\times\tau'}}}}
	\\
	\compm{\vdash\src{\pair{e,e'}:\tau\times\tau'}}
		&\ =
		\trg{
			\pair{\compm{\vdash\src{e:\tau}},\compm{\vdash\src{e':\tau'}}}
		}
	\\
	\compm{
		\vdash\src{
			\begin{aligned}
				&
				\ifztes{
					\src{e}
					\\&\
				}{
					\src{c}
				}{
					\src{c'}
				}
			\end{aligned}
		}
	}
		&\ =
		\trg{
			\begin{aligned}
				&
				\ifztet{
					\compm{\vdash\src{e:N}}
				\\
				&\
				}{
					\compm{\vdash\src{c}}
				}{
					\compm{\vdash\src{c'}}
				}
			\end{aligned}
		}
	&
	\compm{\vdash\skips}
		&\ =
		\skipt
	\\
	\compm{\vdash\src{\sends{e}}}
		&\ =
		\trg{
			 \gensends{\compm{\vdash\src{e:\tau\times\tau'}}}
		}
	&
	\compm{\vdash\src{c;c'}} 
		&\ = 
		\trg{\compm{\vdash\src{c}};\compm{\vdash\src{c'}}}
\end{align*}

The \gensends{\cdot} function takes a source expression and returns a sequence of instructions sending each element of the expression.
Formally:\ $\gensends{\cdot} :\ \vdash\src{e:\tau} \to \trg{c}$
\begin{align*}
	\gensends{\vdash\src{e:\tau}} =
		\begin{cases}
			\trg{\sendt{{\compm{\vdash\src{{e}}:N }}}} 
			& \ift \src{\tau} = \src{N}
			\\
			\trg{
				\gensends{\vdash\src{\projone{e}:\tau'}};
				\gensends{\vdash\src{\projtwo{e}:\tau''}}
			}
			& \ift \src{\tau} = \src{\tau'\times\tau''}
		\end{cases}
\end{align*}

\subsection{Additional Lemmas and Proofs of \Cref{SEC:M}}\label{sec:proofs-marco}
To prove that the compiler is correct (\Cref{thm:inst-marco-cc}) we need two auxiliary results.
\Cref{thm:inst-marco-gensend} tells us that the way we break down a source send into multiple target ones is correct.
\Cref{thm:inst-marco-cc-expr} tells us that the compilation of expressions is correct.

\begin{lemma}[\gensends{\cdot} works]\label{thm:inst-marco-gensend}
  $\ift
  \trg{\gensends{\compm{\vdash\src{e : \tau\times\tau'}}} \xtot{t} \skipt}
  \thent 
  \src{e\red\redstar v} 
  \andt
  \src{v}\simm\trg{t}$
\end{lemma}
\begin{proof}
	We proceed by induction on \src{\tau} and \src{\tau'}:
	\begin{description}
		\item[$\src{\tau}=\src{N}$ and $\src{\tau'}=\src{N}$]

		By canonicity we have that $\src{e} = \src{\pair{n,n'}}$

		\gensends{\cdot} translates that into \trg{\sendt{\compm{\vdash{\projone{\pair{n,n'}}:N} } };\sendt{\compm{\vdash\projtwo{\pair{n,n'}}:N}}}.

		By the target semantics and \Cref{tr:e-tm-send}, that produces $\trg{t}=\trg{n;n'}$.

		We need to prove that 
		\begin{itemize}
			\item $\src{e\equiv\pair{n,n'}\red\redstar\pair{n,n;}}$ which is trivially true;
			\item $\src{\pair{n,n'}}\sim\trg{n;n'}$, which holds by \Cref{tr:tr-rel-n}.
		\end{itemize}

		\item[$\src{\tau}=\src{\tau_1\times\tau_1'}$ and $\src{\tau'}=\src{\tau_2\times\tau_2'}$]

		So by canonicity $\src{e}=\src{\pair{\pair{e_1,e_1'},\pair{e_2,e_2'}}}$.

		By definition of \gensends{\cdot}:
		\begin{align*}
			&
			\gensends{\compm{\vdash\src{\pair{\pair{e_1,e_1'},\pair{e_2,e_2'}} : \tau\times\tau'}}}
			\\
			=&\
			\trg{
				\gensends{\compm{\vdash\src{\projone{\pair{\pair{e_1,e_1'}},\pair{e_2,e_2'}} : \tau}}};
				\gensends{\compm{\vdash\src{\projtwo{\pair{\pair{e_1,e_1'}},\pair{e_2,e_2'}} : \tau'}}}
			}
		\end{align*}

		By assumption we have HP1
			\trg{
				\begin{aligned}
					&
					\gensends{\compm{\vdash\src{\projone{\pair{\pair{e_1,e_1'}},\pair{e_2,e_2'}} : \tau}}};
					\\
					&
					\gensends{\compm{\vdash\src{\projtwo{\pair{\pair{e_1,e_1'}},\pair{e_2,e_2'}} : \tau'}}} 
				\end{aligned}
				\xtot{t_1;t_2} \skipt
			}

		We apply the induction hypothesis and we get
		\begin{description}
			\item[IH-SV1] $\src{\projone{\pair{\pair{e_1,e_1'},\pair{e_2,e_2'}}} \red\redstar \pair{v_1,v_1'}}$
			\item[IH-SV2] $\src{\projtwo{\pair{\pair{e_1,e_1'},\pair{e_2,e_2'}}} \red\redstar \pair{v_2,v_2'}}$
			\item[IH-SR1] $\src{\pair{v_1,v_1'}}\sim\trg{t_1}$
			\item[IH-SR1] $\src{\pair{v_2,v_2'}}\sim\trg{t_2}$ 
		\end{description}

		We need to prove
		\begin{itemize}
			\item $\src{\pair{\pair{e_1,e_1'},\pair{e_2,e_2'}} \red\redstars \pair{\pair{v_1,v_1'},\pair{v_2,v_2'}}}$

			By IH-SV1, IH-SV2 and determinism of the semantics.
			
			\item $\src{\pair{\pair{v_1,v_1'},\pair{v_2,v_2'}}}\sim\trg{t_1;t_2}$
			
			This holds by \Cref{tr:tr-rel-m-m} with IH-SR1 and IH-SR2, for $\src{M}=\src{\pair{v_1,v_1'}}$ and $\src{M'}=\src{\pair{v_2,v_2'}}$.
		\end{itemize}

		\item[$\src{\tau}=\src{N}$ and $\src{\tau'}=\src{\tau_1\times\tau_2}$]

		Analogous to the other cases, by IH and \Cref{tr:tr-rel-n-m}.

		\item[$\src{\tau}=\src{\tau_1\times\tau_2}$ and $\src{\tau'}=\src{N}$]

		Analogous to other cases, by IH and \Cref{tr:tr-rel-m-n}.
		\qedhere
	\end{description}
\end{proof}

\begin{lemma}[\compm{\cdot} is correct for expressions]\label{thm:inst-marco-cc-expr}
$\ift 
  \trg{\compm{\vdash e:\tau}\red\redstar\compm{\vdash v:\tau}}
\thent
  \src{e\red\redstar v}
$ 
\end{lemma}
\begin{proof}
	Trivial induction on the typing derivation of \src{e}.
\end{proof}

\hrule

\begin{proof}[of \autoref{thm:inst-marco-cc}]
	By definition we need to prove that
	$$\forall\src{c},\trg{t}. \ift \trg{\compm{\src{\vdash c}}\sem\trg{t}} \thent \exists\src{s}\sim\trg{t} \andt \src{c\sem Q}$$
	We proceed by induction on the typing derivation of \src{c}.
	\begin{description}
		\item[Base.]
		\begin{description}
			\item[$\src{c}=\skips$]
			Trivial, as no trace is emitted.
		\end{description}
		\item[Inductive.]
		\begin{description}
			\item[$\src{c}=\src{c;c'}$]
			By IH.
			\item[$\src{c}=\src{\iftes{e}{c}{c'}}$]
			
			By \autoref{thm:inst-marco-cc-expr} and IH.

			\item[$\src{c}=\src{\sends{e}}$]

			By HP we have that HPQ \trg{\compm{\src{\vdash \sends{e}}}\xto{t} \skipt}.

			We can then apply \autoref{thm:inst-marco-gensend} with HPQ and we have HPV $\src{v}\sim\trg{t}$ and HPS \src{e\red\redstar v}.

			By definition of \compm{\cdot} and of $\trg{\sem}$ we need to prove that
			$\src{\sends{t}\xtos{v}\skips}$ and $\src{v}\sim\trg{t}$.

			The former holds by HPS and \Cref{tr:es-sm-send}.

			The latter holds by HPV.
			\qedhere
		\end{description}
	\end{description}
\end{proof}

\subsection{Additional Formalism for \Cref{sec:sec-comp-traces}}

Syntax of $\Sr$ and $\Tr$
\begin{align*}
  &
  \src{e} \bnfdef \cdots \mid \src{e_1; e_2} \mid \src{f(e)} \mid \SrOutL\ \src{n}
  \mid \src{arg}
  \qquad
  \src{fs} \bnfdef \src{\pair{f_1, e_1},} \ldots \src{, \pair{f_n, e_n}} 
    &
  \trg{e} \bnfdef \cdots \mid \TrOutH\ \trg{n}
  \\
  &
  \SrPar \bnfdef \pair{\src{fs}, \src{e}}
  \qquad
  \SrCtx \bnfdef \src{fs}
  \qquad
  \SrPrg \bnfdef \SrCtx\src{[}\SrPar\src{]}
  \qquad
  \src{i} \bnfdef \cdots 
  \mid \SrOutL\ \src{n} 
  &
  \trg{i} \bnfdef \cdots \mid \TrOutH\ \trg{n} 
\end{align*}

%% file: composition-appendix.tex
\section{Composition results} 
\label{sec:composition} 

In this appendix we deduce some insights from the composition of trace-relations and property mappings.
%
We describe the correctness theorem obtained for a multistep compiler, where each step
is proved correct for a possibly different trace relation.
We obtain the expected result: it is indeed possible to compose individual
compiler correctness proofs for each step, to prove the correctness of the whole
chain.

Assume $\downarrow^\src{s}_\ooth{i}$ is a compilation chain from a source to an intermediate 
language and $\downarrow^\ooth{i}_\trg{t}$ from the intermediate to a target language.%
\footnote{For the intermediate
language we use a \ooth{verbatim}, \ooth{emerald} font.
  \MP{
    just added the macro OTH, if you need a third colour.
    if you don't typeset strange math, you can also use OOTH, which is verbatim and nice but works poorly with math (unless on mac)
  }
} 
Let $\sim_{\src{s},\ooth{i}} \subseteq \src{Trace_S} \times \ooth{Trace_I}$ and 
$\sim_{\ooth{i},\trg{t}} \subseteq \ooth{Trace_I} \times \trg{Trace_T}$, such that $\downarrow^\src{s}_\ooth{i} \in \cctildei{\sim_{\src{s},\ooth{i}}}$
and $\downarrow^\ooth{i}_\trg{t} \in \cctildei{\sim_{\ooth{i},\trg{t}}}$. It is straightforward to show that compiling from source to target through $\downarrow^\src{s}_\trg{t} = \downarrow^\src{s}_\ooth{i} ; \downarrow^\ooth{i}_\trg{t} $ satisfies the following 

\begin{equation*}
  \cctildei{(\sim_{\src{s},\ooth{i}}; \sim_{\ooth{i},\trg{t}})} \equiv \forall \src{W} \forall \trg{t}.
    ~\src{W} \downarrow^\src{s}_\trg{t} \trg{\rightsquigarrow t} \Rightarrow
    \exists \src{s} \sim_{\src{s},\ooth{i}}; \sim_{\ooth{i},\trg{t}} \trg{t}. ~ \wsmkt{W}{s}
\end{equation*}

where $\src{s} \sim_{\src{s},\ooth{i}}; \sim_{\ooth{i},\trg{t}} \trg{t} \iff \exists \ooth{w} \in \ooth{Trace_I}. ~\src{s} \sim_{\src{s},\ooth{i}} \ooth{w} \wedge \ooth{w} \sim_{\ooth{i},\trg{t}} \trg{t}$. 

\medskip
It follows that target guarantees, as well as source obligations, can
be established like in \Cref{sec:trinity}, using the composition of
the two relation $\sim_{\src{s},\ooth{i}}; \sim_{\ooth{i},\trg{t}}$,
or just by composition of the property mappings.  \ca{Write it down?
  Painful notation and not great insights...}  Similarly for
$\downarrow^\src{s}_\ooth{i} \in \rtctildei{\sim_{\src{s},\ooth{i}}}$
and
$\downarrow^\ooth{i}_\trg{t} \in \rtctildei{\sim_{\ooth{i},\trg{t}}}$,
their composition
$\downarrow^\src{s}_\trg{t} \in \rtctildei{(\sim_{\src{s},\ooth{i}};
  \sim_{\ooth{i},\trg{t}})}$.

%% file: rob-pres-safety.tex

I/O events are not the only instance of events that compilers consider.
\MP{
  indeed one may say that the compiler that erases such events is correct and secure, trivially by definition since there are no target traces ... 
}
Especially in the setting of secure compilation, where a compartmentalized partial program interacts with a context, \emph{interaction traces} are often used~\cite{patrignani_thesis,PatrignaniG18,AbateBGHPT19,JuglaretHAEP16}.
Consider a language analogous to that of the previous section, where the context \com{C} defines a set of functions \com{F_c} and the program defines a different set \com{F_p}.
Interaction traces (generally) record the control flow of calls between these two sets via actions that are \com{call~f~v} and \com{ret~v}~\cite{JeffreyR05}.
These actions indicate a call to function \com{f} with parameter \com{v} and a return with return value \com{v}.
In case the context calls a function in \com{F_p} (or returns to a function in \com{F_p}), the action is decorated with a \com{?} (\IE those actions are \com{call~f~v?} and \com{ret~v?}).
Dually, the program calling a function in \com{F_c} (or returning to it) generates an action decorated with a \com{!} (\IE those actions are \com{call~f~v!} and \com{ret~v!}).

\citet{PatrignaniG18} consider precisely such a setting.
Moreover, they define a compiler that preserves safety properties of source programs  (\IE it is \ii{RSC} in the sense of \autoref{fig:diagram-preserve-props-sec}) by relying on capability machines.
The interesting point, however, is that they also consider source and target traces to be distinct since the source has \src{bools} and \src{nats} 
and the target only has \trg{nats}.%
\footnote{%
  Technically, their difference on values also encompasses heap locations and capabilities, but we elide that here since this is sufficient to draw our analogy.
} 
Thus, to prove \ii{RSC}, they rely on a cross-language relation on values, which is scaled to trace actions, and then scaled point-wise to traces (analogously to what we have done in \Cref{sec:example-diff-values,sec:instance-marco,sec:sec-comp-traces}).

Besides defining a relation on traces (which is an instance of $\sim$), they also define a relation between source and target safety properties that supports concurrent programs.%
\footnote{%
  They call those safety properties monitors since they focus on safety~\cite{Schneider00} and indicate \src{s} with \src{M} and \trg{t} with \trg{M}.
}
Thus, they really provide an instantiation of $\tau$
\MP{
  Jeremy: right? this is not $\tilde{\tau}$.
}
that maps all safe source traces to the related target ones.
This ensures that no additional target trace is introduced in the target property, and source safety property are mapped to target safety ones  by $\tau$.

Their compiler is then proven to generate code that respects $\tau$, so they achieve a variation of $\ii{RSC}^{\tau}$ \ca{What do you mean? \rsctilde or \rsptautilde or even $\ii{RSP}^{\tau}$?}
 \ca{In that paper do you prove RSCtilde or RSPtau?? In the first case our framework tells you for free what are the guarantees also on arbitrary source properties. 
     In the second case you are proving something a bit weaker than \rsctilde itself }
Their proofs are based on standard techniques either for secure compilation (\IE trace-based backtranslation~\cite{MarcosSurvey}) and for correct compilation (\IE forward/backward simulation~\cite{Leroy09b}).


%% file: instance-marcodeepak.tex
\label{sec:example-more-events}
\citet{PatrignaniG17} study the preservation of hypersafety
from the perspective of secure compilation. Again, 
their result can be interpreted in our setting. They consider reactive systems, where trace alphabets are partitioned in input actions \com{\alpha?} and output actions \com{\alpha!}, whose concatenation generate traces \com{\OB{\alpha?\alpha!}}. We use the same notation as before and indicate such sequences as \src{s} and \trg{t} respectively.
The set of target output actions \trgb{\alpha!} includes an action $\trg{\tick}$ that has no source counterpart (\IE $\nexists\src{\alpha?}\sim\trg{\tick}$), and whose output does not depend on internal state (thus it cannot leak secrets).%
\footnote{%
	Technically, they assume a set of \trg{\tick} actions, but for this analogy a single action suffices.
}
By emitting \trg{\tick} whenever undesired inputs are fed to a compiled program (\EG passing a \trg{nat} when a \trg{bool} is expected), 
hypersafety is preserved (as \tick\ does not leak secrets)~\cite{PatrignaniG17}.

More formally, they assume a relation on actions $\sim$ that is total on the source actions and injective.
From there, they define \trg{TPC}---which here corresponds to an instance of $\tau$---%
that maps the set of valid source traces to the set of valid target traces (that now mention \trg{\tick}) as follows:
\begin{align*}
	\fun{\trg{TPC}}{\piS{\pi}} =
		&\ 
		\myset{
			\trg{t}
		}{ 
			\trg{t} \in \bigcup_{n\in\mb{N}} \fun{int_n}{\piS{\pi}} 
		}
			\;\text{ where}
	\;
	\fun{int_0}{\piS{\pi}} =
		\ 
		\myset{
			\trg{t}
		}{ 
			\exists \src{s}\in\piS{\pi} \wedge \src{s}\sim\trg{t}}
	\\
	\fun{int_{n+1}}{\piS{\pi}} = 
		&\ 
		\myset{
			\trg{t}
		}{
				(i)\
				\trg{t}\equiv\trg{\trg{t}_1}\trgb{\alpha?}\tick\trg{t_2} 
				\wedge 
				(ii)\
				\trg{t_1\trg{t}_2}\in\fun{int_n}{\piS{\pi}} 
				\wedge 
				(iii)\
				\fun{undesired}{\trgb{\alpha?}}
	}
\end{align*}
where \fun{undesired}{\trgb{\alpha?}} indicates that \trgb{\alpha?} is an undesired input (intuitively, this is an information that can be derived from the set of source traces~\cite{PatrignaniG17}).

Informally, given a set of source traces \piS{\pi}, \trg{TPC} generates all target traces that are related (pointwise) to a source trace (case \funname{int_0}).
Then (case \funname{int_{n+1}}), it adds all traces (\trg{t}) with interleavings of undesired input \trgb{\alpha?} (point iii) followed by \trg{\tick} (point i) as long as the interleavings split a trace $\trg{t_1t_2}$ that has already been mapped (point ii).

\trg{TPC} is an instance of $\tau$ that maps  
source hypersafety to target hypersafety (and therefore, safety to safety),
thus our theory can be instantiated for the preservation of these classes of hyperproperties as well.